\documentclass{article}

\usepackage{epsfig}
\usepackage{graphics}
\usepackage{graphicx}
\usepackage{url}
\usepackage{subfigure}
\usepackage{hyperref}

\usepackage[centertags]{amsmath}
\usepackage{amsfonts}
\usepackage{amsthm}
\usepackage{amssymb}

\begin{document}

\title{Designing Complex Dynamics in Cellular Automata with Memory}

\author{Genaro J. Mart\'{\i}nez$^{1-9}$ \\ Andrew Adamatzky$^{1,2}$ \\ Ramon Alonso-Sanz$^{1,2,10}$}

\date{July 3, 2013}

\maketitle

\noindent {\large Paper published at {\it International Journal of Bifurcation and Chaos} 23(10), 1330035 (131 pages), 2013. \url{http://www.worldscientific.com/doi/abs/10.1142/S0218127413300358}} \\ \\

\begin{centering}
$^1$ International Center of Unconventional Computing, University of the West of England, BS16 1QY Bristol, United Kingdom \\
\url{http://uncomp.uwe.ac.uk/} \\
$^2$ Laboratorio de Ciencias de la Computaci\'on, Escuela Superior de C\'omputo, Instituto Polit\'ecnico Nacional, M\'exico, D.F. \\
\url{http://uncomp.uwe.ac.uk/LCCOMP/} \\
$^3$ Foundation of Computer Science Laboratory, Hiroshima University, Higashi-Hiroshima, Japan \\
\url{http://www.iec.hiroshima-u.ac.jp/} \\
$^4$ Centre for Chaos and Complex Networks, City University of Hong Kong, Hong Kong, P. R. China \\
\url{http://www.ee.cityu.edu.hk/~cccs/} \\
$^5$ School of Science, Hangzhou Dianzi University, Hangzhou, P. R. China \\
$^6$ Centre for Complex Systems, Instituto Polit\'ecnico Nacional, M\'exico, D.F. \\
\url{http://www.isc.escom.ipn.mx/sistemascomplejos/} \\
$^7$ Centro de Ciencias de la Complejidad, Universidad Nacional Aut\'onoma de M\'exico, M\'exico, D.F. \\
\url{http://c3.fisica.unam.mx/} \\
$^8$ Laboratoire de Recherche Scientifique, Paris, France \\
\url{http://labores.eu/} \\
$^9$ Institut des Syst\`emes Complexes en Normandie, Normandie, France \\
\url{http://iscn.univ-lehavre.fr/} \\
$^{10}$ ETSI Agronomos Estad{\'i}stica, Universidad Polit\'ecnica de Madrid, Madrid, Espa\~na \\
\end{centering}

\begin{abstract}
Since their inception at {\it Macy conferences} in later 1940s  complex systems remain the most controversial topic of inter-disciplinary sciences. The term `complex system' is the most vague and liberally used scientific term. Using elementary cellular automata (ECA), and exploiting the CA classification, we demonstrate elusiveness of `complexity' by shifting space-time dynamics of the automata from simple to complex by enriching cells with {\it memory}. This way, we can transform any ECA class to another ECA class --- without changing skeleton of cell-state transition function --- and vice versa by just selecting a right kind of memory. A systematic analysis display that memory helps `discover' hidden information and behaviour on trivial --- uniform, periodic, and non-trivial --- chaotic, complex --- dynamical systems. \\

\noindent {\bf keywords:} elementary cellular automata, classification, memory, computability, gliders, collisions, complex systems.
\end{abstract}

\newpage

\tableofcontents

\newpage

\listoffigures

\newpage

\section{Introduction}
\label{intro}

A complexity theory emerged from studies of computable problems in computer science and mathematical foundations of computation, when a need came to compare performance and resource-efficiency of algorithms. Typically time complexity (number of steps) and space complexity (memory of a single processor and number of processors) are expressed in terms of a Turing machine or an equivalent mathematical device. Each specific kind of a Turing machine represents a certain class of complexity [Minsky, 1967], [Arbib, 1969], [Hopcroft \& Ullman, 1987]. When related to complex systems meaning of the word `complexity' is different and heavily depends on its context. Complexity of a system is almost never quantified but often related to unpredictability. 

Theory of cellular automata (CA) refers to complexity its entire life [von Neumann, 1966], [Adamatzky \& Bull, 2009], [Boccara, 2004], [Chopard \& Droz, 1998], [Hoekstra et al., 2010], [Kauffman, 1993], [Margenstern, 2007], [McIntosh, 2009], [Mainzer \& Chua, 2012], [Mitchell, 2002], [Morita, 1998], [Margolus et al., 1986], [Poundstone, 1985], [Park et al., 1986], [Toffoli \& Margolus, 1987], [Schiff, 2008], [Sipper, 1997], [Wolfram, 1986], [Mart{\'i}nez et al., 2013a], [Mart{\'i}nez et al., 2013b]. Due to transparency of cellular automata structures their complexity can be measured and analysed [Wolfram, 1984a], [Culik II \& Yu, 1988].

An elementary cellular automaton (ECA) is a one-dimensional array of finite automata, each automaton takes two states and updates its state in discrete times according to its own state and states of its two closest neighbours, all cells update their state synchronously. Thus in 1980s Wolfram subdivided ECA onto four complexity classes [Wolfram, 1984a]:

\begin{itemize}
\item {class I.} CA evolving uniformly.
\item {class II.} CA evolving periodically.
\item {class III.} CA evolving chaotically.
\item {class IV.} Include all previous cases, known as the class {\it complex}.
\end{itemize}

Also these classes can be defined in terms of CA evolution as follows:

\begin{itemize}
\item If the evolution is dominated by a unique state of alphabet from any random initial condition, it belongs to {\it class I}.
\item If the evolution is dominated by blocks of cells which are periodically repeated from any random initial condition, hence it belongs to {\it class II}.
\item If the evolution is dominated by sets of cells without some defined pattern for a long time from any random initial condition, hence it belongs to {\it class III}.
\item If the evolution is dominated by non-trivial structures emerging and travelling along of the evolution space where also uniform, periodic, or chaotic regions can coexist with these structures, it belongs to {\it class IV}. This class is named frequently as: {\it complex behaviour}, {\it complex dynamics}, or simply {\it complex}.
\end{itemize} 

\begin{figure}
\begin{center}
\subfigure[]{\scalebox{0.55}{\includegraphics{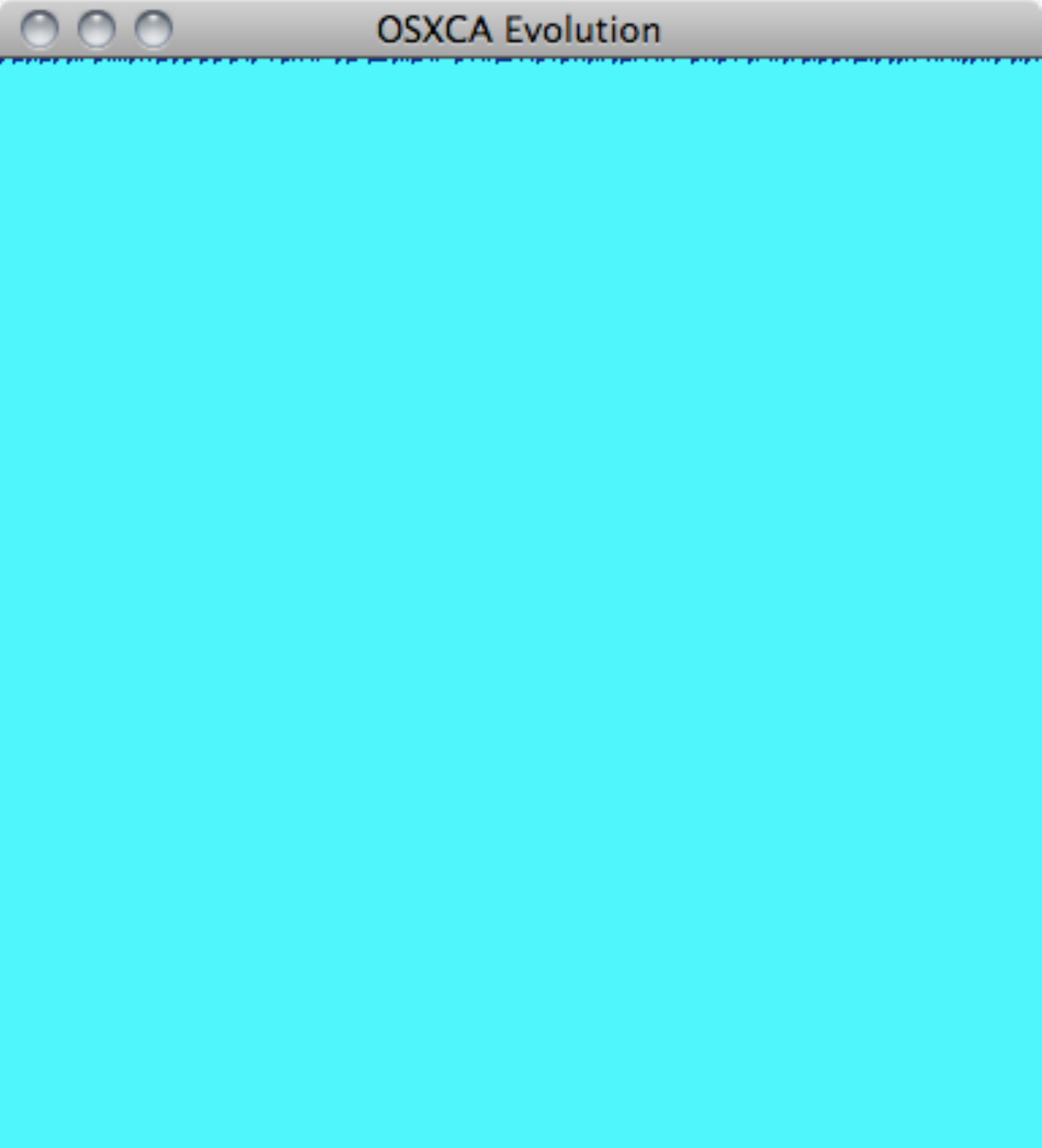}}} \hspace{0.8cm}
\subfigure[]{\scalebox{0.55}{\includegraphics{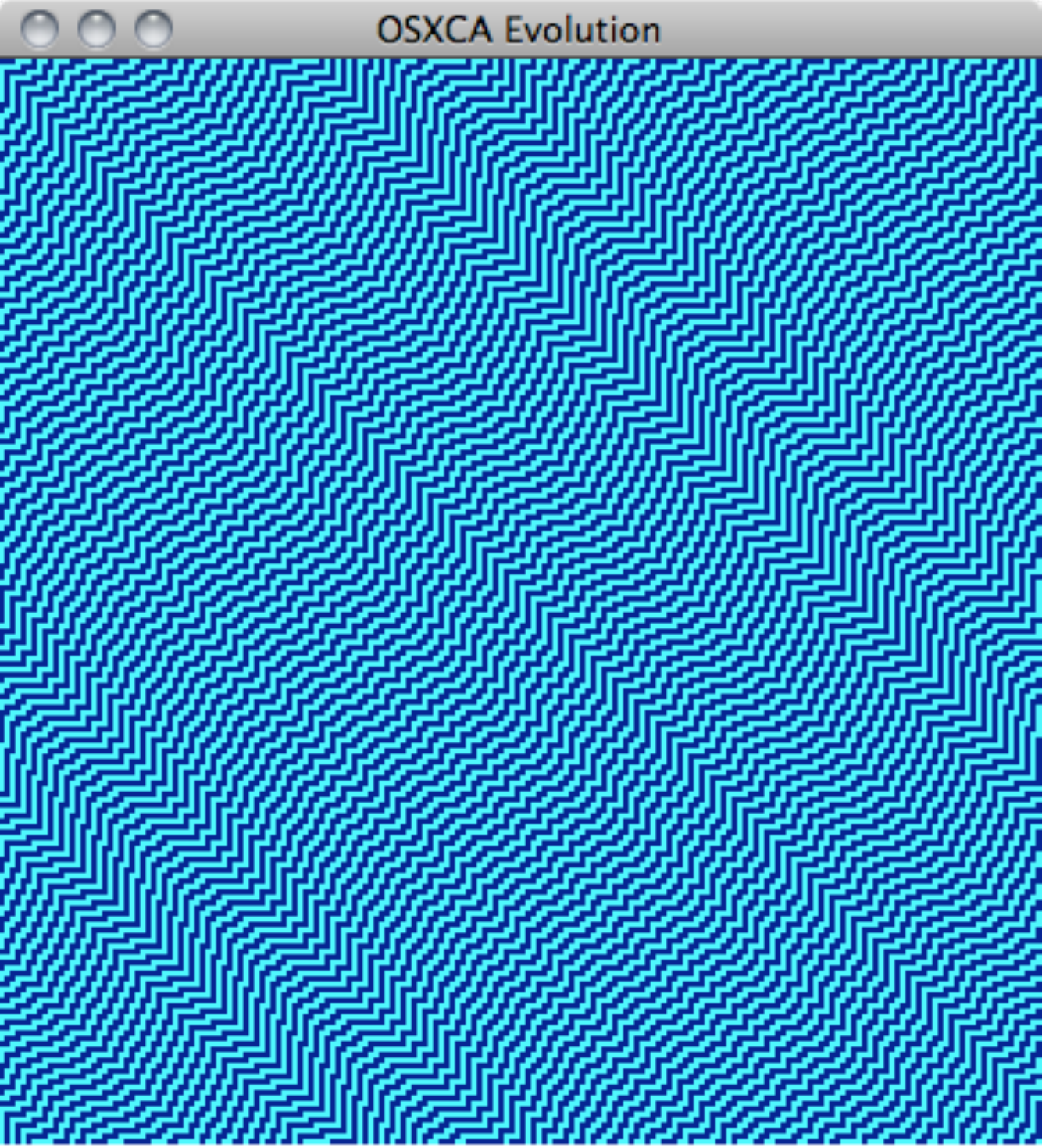}}} 
\subfigure[]{\scalebox{0.55}{\includegraphics{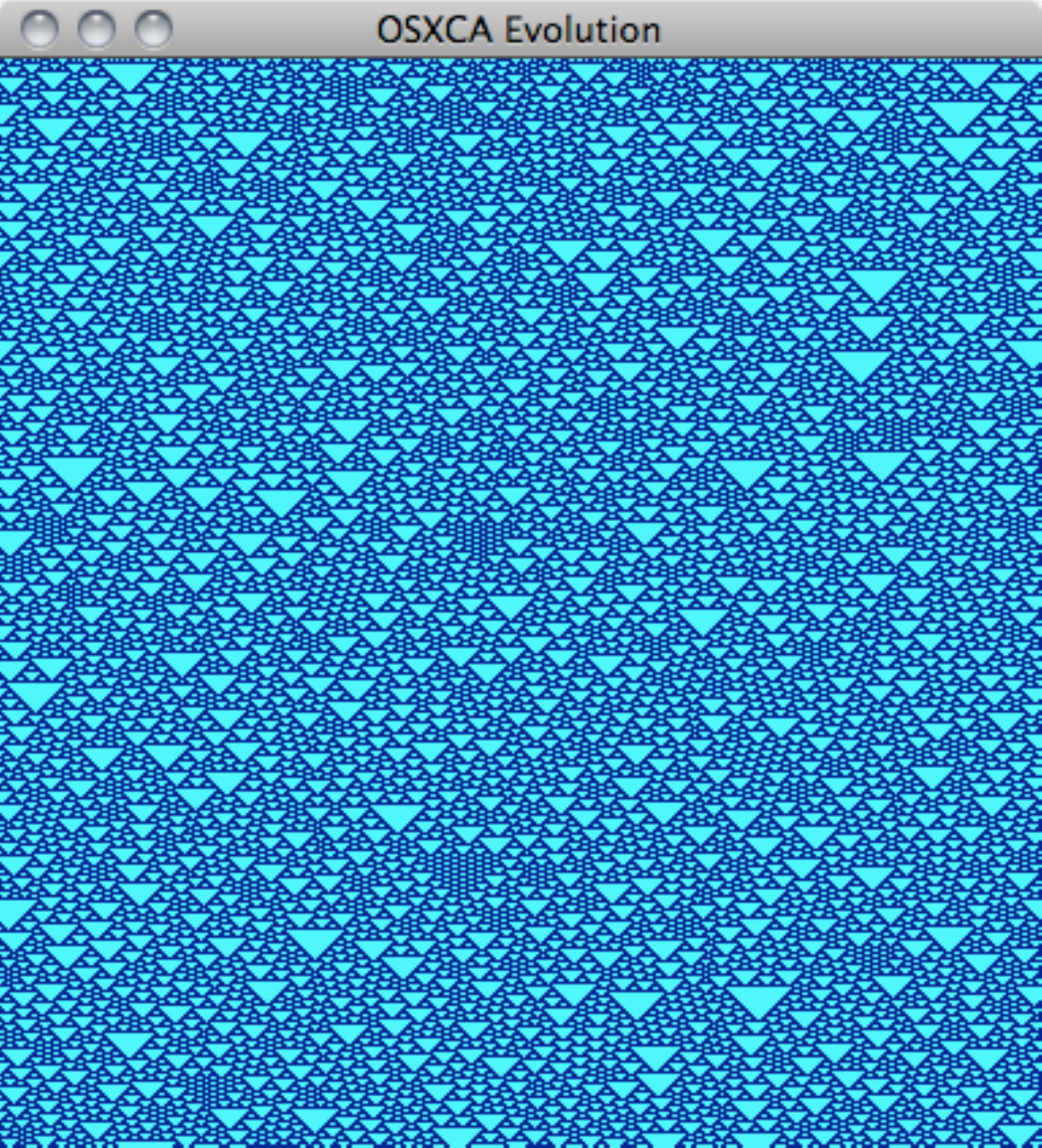}}} \hspace{0.8cm}
\subfigure[]{\scalebox{0.55}{\includegraphics{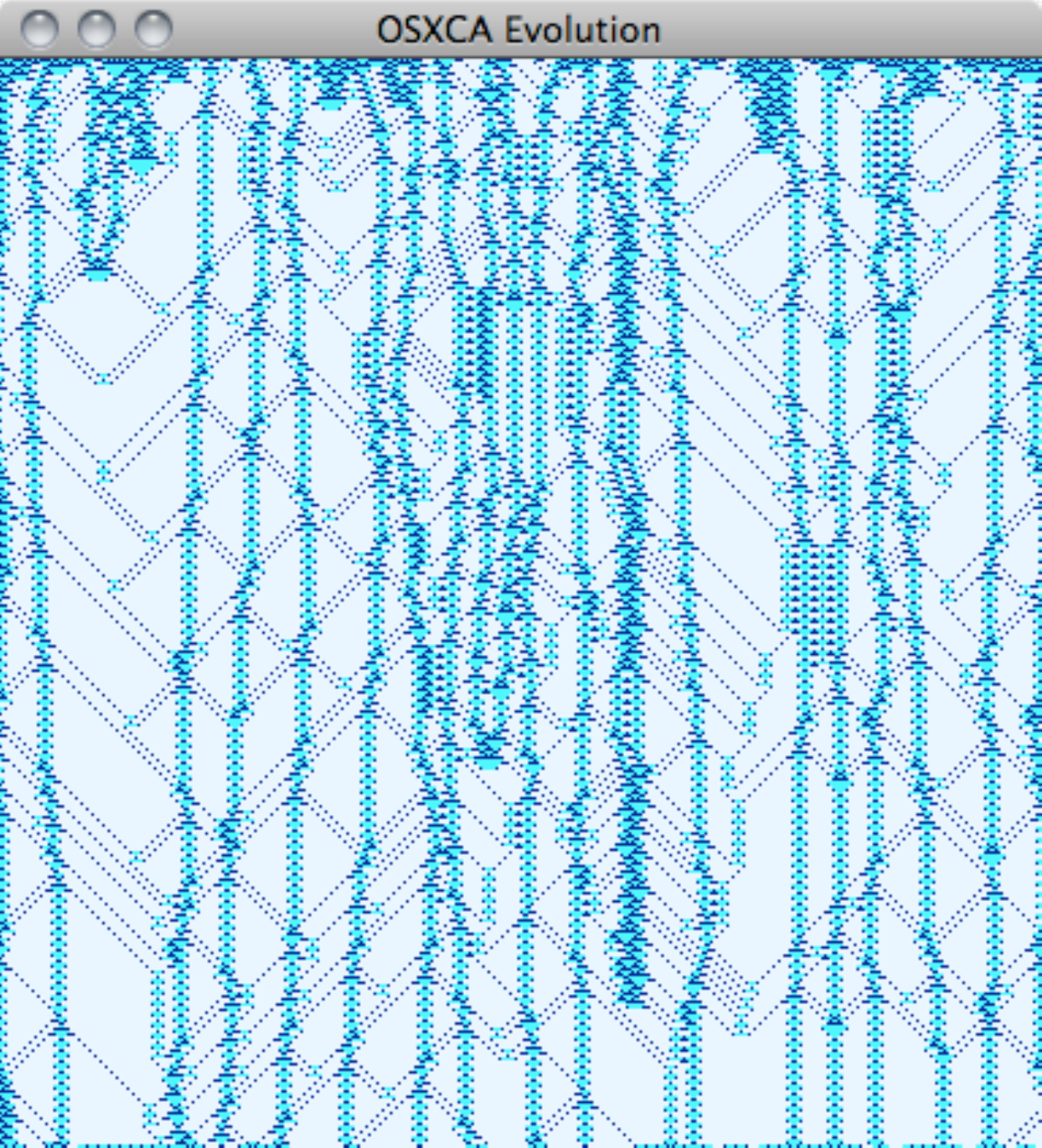}}}
\end{center}
\caption{Examples of space-time evolution of ECA rules: 
(a)~class I, ECA rule 8, 
(b)~class II, ECA rule 15, 
(c)~class III, ECA rule 126,
(d)~class IV, ECA rule 54 (a periodic background is filtered). 
All automata illustrated start their development at the same random initial condition with a density of 50\% of states 0, light (light blue) dots, and states 1, dark (dark blue) dots. Each automaton is a horizontal ring of 385 cells evolved for 400 time steps.}
\label{WolframClasses}
\end{figure}

Figure~\ref{WolframClasses} illustrates the Wolfram's classes by a selected ECA rule (following the Wolfram's notation for ECA [Wolfram, 1983]), and all evolutions begin with the same random initial condition. Fig.~\ref{WolframClasses}a shows  ECA rule 8 converging quickly to a homogeneous state, the class I. Figure~\ref{WolframClasses}b displays blocks of cells which evolve periodically exhibiting a right shift, this is an interesting reversible ECA rule 15, the class II. Figure~\ref{WolframClasses}c displays a typical chaotic evolution with ECA rule 126, where no regular patterns detected  or no limit point can be identified, the class III. Finally, Fig.~\ref{WolframClasses}d displays the so-called complex class or class IV with ECA rule 54. There we can see non-trivial patterns emerging in the evolution space, and such patterns conserve their form and travel along of the evolution space. The patterns collide with each other and annihilate or fuse, or undergo soliton-like transformations or produce new structures. These patterns are referred to as {\it gliders} in the CA literature (glider is a concept widely accepted and popularised by Conway from its famous 2D CA {\it Game of Life} [Gardner, 1970]). In space-time configurations developed by functions from  class IV we can see regions with periodic configurations, fragments of chaos, and well-defined non-trivial patterns. Frequently in complex rules the background is dominated by a stable state, such as happens in Conway's Game of Life. In this case, particularly the complex ECA rule 54 and 110 can evolve with a periodic background (called ether) where these gliders emerge and live. Gliders in GoL and other CAs as the 2D Brian's Brain CA [Toffoli \& Margolus, 1987] caught the attention of Langton and thus contributed to development of Artificial Life field [Langton, 1984], [Langton, 1986].

Since the publication of the paper ``Universality and complexity in cellular automata'' in 1984 [Wolfram, 1984a] there have been a number of disputes about validity of the classification. Wolfram selected certain ECA rules to illustrate each class. Although, he commented textually that: {\it $k=2, r=1$ cellular automata are too simple to support universal computation} [Wolfram, 1984a] (page 31). Nevertheless, in his book ``Cellular Automata and Complexity'' [Wolfram, 1994] ECA rule 110 was awarded its own appendix (Table 15, Structures in Rule 110, pages 575--577). It contains specimens of evolution including a list of thirteen gliders compiled by Lind, and also presents the conjecture that the rule could be universal. Wolfram wrote:  {\it One may speculate that the behaviour of rule 110 is sophisticated enough to support universal computation}. Finally, in [Cook, 2004], [Wolfram, 2002] it was proved that ECA rule 110 is computationally universal because it simulates a novel cyclic tag system with package of gliders and collisions on millions of cells.\footnote{Large snapshots of this large machine working in ECA rule 110 are available in \url{http://uncomp.uwe.ac.uk/genaro/rule110/ctsRule110.html}.}

The paper written by Culick II and Yu titled ``Undecidability of CA Classification Schemes'' [Culik II \& Yu, 1988], [Sutner, 1989] discussed the properties of Wolfram ECA classes and stated that {\it it is undecidable to which class a given cellular automaton belongs} (page 177).

Further attempts of ECA classification have been made in [Gutowitz et al., 1987], [Li \& Packard, 1990], [Aizawa \& Nishikawa, 1986], Ada94, [Sutner, 2009]. Gutowitz developed a statistical analysis in ``Local structure theory for cellular automata'' [Gutowitz et al., 1987]. An extended classification of ECA classes with mean field theory was proposed by McIntosh in ``Wolfram's Class IV and a Good Life'' [McIntosh, 1990]. An interesting schematic diagram conceptualising  classes in CA was made by Li and Packard in ``The Structure of the Elementary Cellular Automata Rule Space'' [Li \& Packard, 1990]. Patterns recognition and classification was presented in ``Toward the classification of the patterns generated by one-dimensional cellular automata'' [Aizawa \& Nishikawa, 1986]. An extended analysis of CA was presented in  ``Identification of Cellular Automata'' by Adamatzky in [Adamatzky, 1994] relating to the problem of given a sequence of configurations of an unknown CA hence how to  reconstruct the cell-state transition rule.  Sutner has been discussed this classification and also the principle of equivalence computation in ``Classification of Cellular Automata'' [Sutner, 2009], with emphasis in class IV or computable CA. A fruitful approach with additive 2D CA was suggested by Eppstein [Eppstein, 1999].\footnote{You can see such discussion from Tim Tyler's CA FAQ in \url{http://cafaq.com/classify/index.php}.}

In this classification, class IV (called complex) is of particular interest because such rules present non-trivial behaviour with a rich diversity of patterns (gliders) emerging and non-trivial interactions between them, gliders are referred as well as  mobile self-localizations, particles, or fragments of waves. This feature was relevant to implementation of a register machine in GoL [Berlekamp et al., 1982] to determine its universality. Thus Rendell has developed an elaborated Turing machines in GoL with thousands of thousands of cells [Rendell, 2011a], [Rendell, 2011b]. Although across of the history, these bridges connection between {\it complexity of a CA} (or any other dynamical system) and their {\it universality} is not always obvious [Adamatzky, 2002], [Mills, 2008].

Other recommendable reference sources to mention include Mitchell's  ``Complexity: A Guided Tour'' [Mitchell, 2009], Wolfram's ``A New Kind of Science'' [Wolfram, 2002], Bar-Yam's ``Dynamics of Complex Systems'' [Bar-Yam, 1997], and ``The Universe as Automaton: From Simplicity and Symmetry to Complexity'' [Mainzer \& Chua, 2012] by Mainzer and Chua.

\section{One-dimensional cellular automata}
\label{ca}

CA are discrete dynamical systems, with a finite alphabet that evolve on a regular lattice in parallel. In the paper we deal with one-dimensional cellular automata.

\subsection{Elementary cellular automata (ECA)}
\label{eca}

A CA is a tuple $ \langle \Sigma,\varphi,\mu, c_0 \rangle$ where $d$ is a dimensional lattice and each cell $x_i$, $i \in N$, takes a state from a finite alphabet $\Sigma$ such that $x \in \Sigma$. A sequence $s \in \Sigma^n$ of $n$ cell-states represents a string or a global configuration $c$ on $\Sigma$. We write a set of finite configurations as $\Sigma^n$. Cells update their states by an evolution rule $\varphi: \Sigma^{\mu} \rightarrow \Sigma$, such that $\mu=2r+1$ represents a cell neighbourhood that consists of a central cell and a number of $r$-neighbours connected locally. If $k=|\Sigma|$ hence there are $k^{2r+1}$ neighbourhoods and  $k^{k^{2r+1}}$ evolution rules.

\begin{figure}[th]
\centerline{\includegraphics[width=3in]{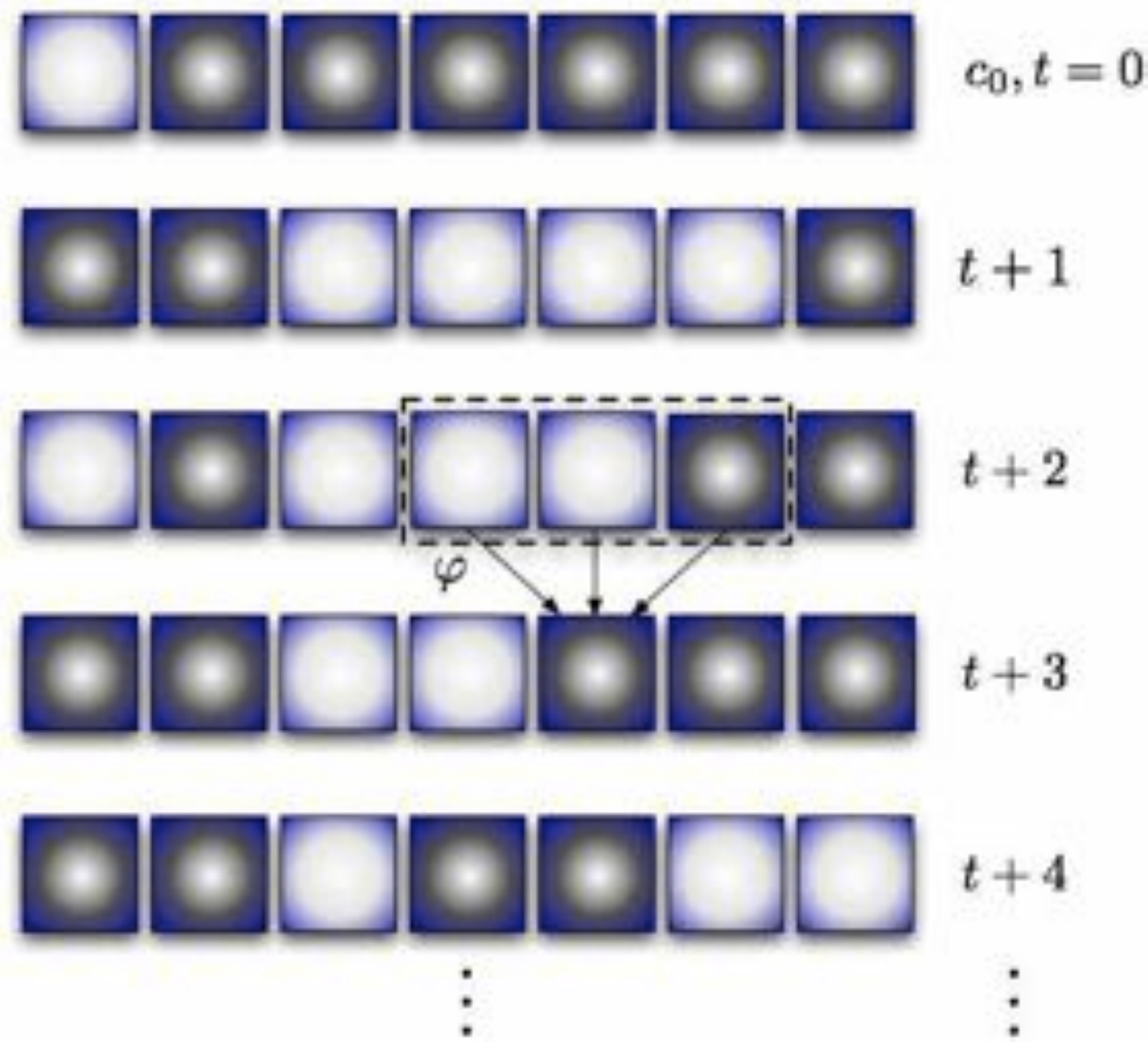}}
\caption{Dynamics in ECA on an arbitrary one-dimensional array transformed for a specific evolution rule $\varphi$.}
\label{evolECA}
\end{figure}

An evolution diagram for a CA is represented by a sequence of configurations $\{c_i\}$ generated by the global mapping $\Phi:\Sigma^n \rightarrow \Sigma^n$, where a global relation is given as $\Phi(c^t) \rightarrow c^{t+1}$. Thus $c_0$ is the initial configuration. Cell states of a configuration $c^t$ are updated simultaneously by the local rule, as follows:

\begin{equation}
\varphi(x_{i-r}^t, \ldots, x_{i}^t, \ldots, x_{i+r}^t) \rightarrow x_i^{t+1}
\end{equation}

\noindent  where $i$ indicates cell position and $r$ is the radius of neighbourhood in $\mu$. Thus, the {\it elementary} CA represents a system of order $(k=2,\ r=1)$ (in Wolfram's notation [Wolfram, 1983]), the well-known {\it ECA}.

To represent a specific ECA evolution rule we will write the evolution rule in a decimal notation, e.g. $\varphi_{R54}$ represents the evolution rule 54. Thus Fig.~\ref{evolECA} illustrates how an evolution dynamics works for ECA.

\subsection{Elementary cellular automata with memory (ECAM)}
\label{ecam}

Conventional CA are memoryless:  new state of a cell depends on the neighbourhood configuration solely at the preceding time step of $\varphi$. CA with memory are an extension of CA in such a way that every cell $x_i$ is allowed to remember its states during some fixed period of its evolution. CA with memory have been proposed originally by Alonso-Sanz in [Alonso-Sanz \& Martin, 2003], [Alonso-Sanz, 2006], [Alonso-Sanz, 2009a], [Alonso-Sanz, 2009b], [Alonso-Sanz, 2011].

Hence we implement a memory function $\phi$, as follows:

\begin{equation}
s_{i}^{(t)} = \phi (x^{t-\tau+1}_{i}, \ldots, x^{t-1}_{i}, x^{t}_{i})
\end{equation}

\noindent where $1\le \tau \le t$ determines the {\it degree of memory}. Thus, $\tau=1$ means no memory (or conventional evolution), whereas $\tau=t$ means unlimited trailing memory. Each cell trait $s_{i} \in \Sigma$ is a state function of the series of states of the cell $i$ with memory backward up to a specific value $\tau$. In the memory implementations run here, commences to act as soon as $t$ reaches the $\tau$ time-step. Initially, i.e., $t<\tau$, the automaton evolves in the conventional way. Later, to proceed in the dynamics, the original rule is applied on the cell states $s$ as:

\begin{equation}
\varphi(\ldots, s_{i-1}^{(t)}, s_{i}^{(t)}, s_{i+1}^{(t)}, \ldots) \rightarrow x^{t+1}_i
\end{equation}

\noindent to get an evolution with memory. Thus in CA with memory,  while the mapping $\varphi$ remains unaltered, historic memory of all past iterations is retained by featuring each cell as a summary of its past states from $\phi$. We can say that cells canalise memory to the map $\varphi$ [Alonso-Sanz, 2009a].

\begin{figure}[th]
\centerline{\includegraphics[width=6in]{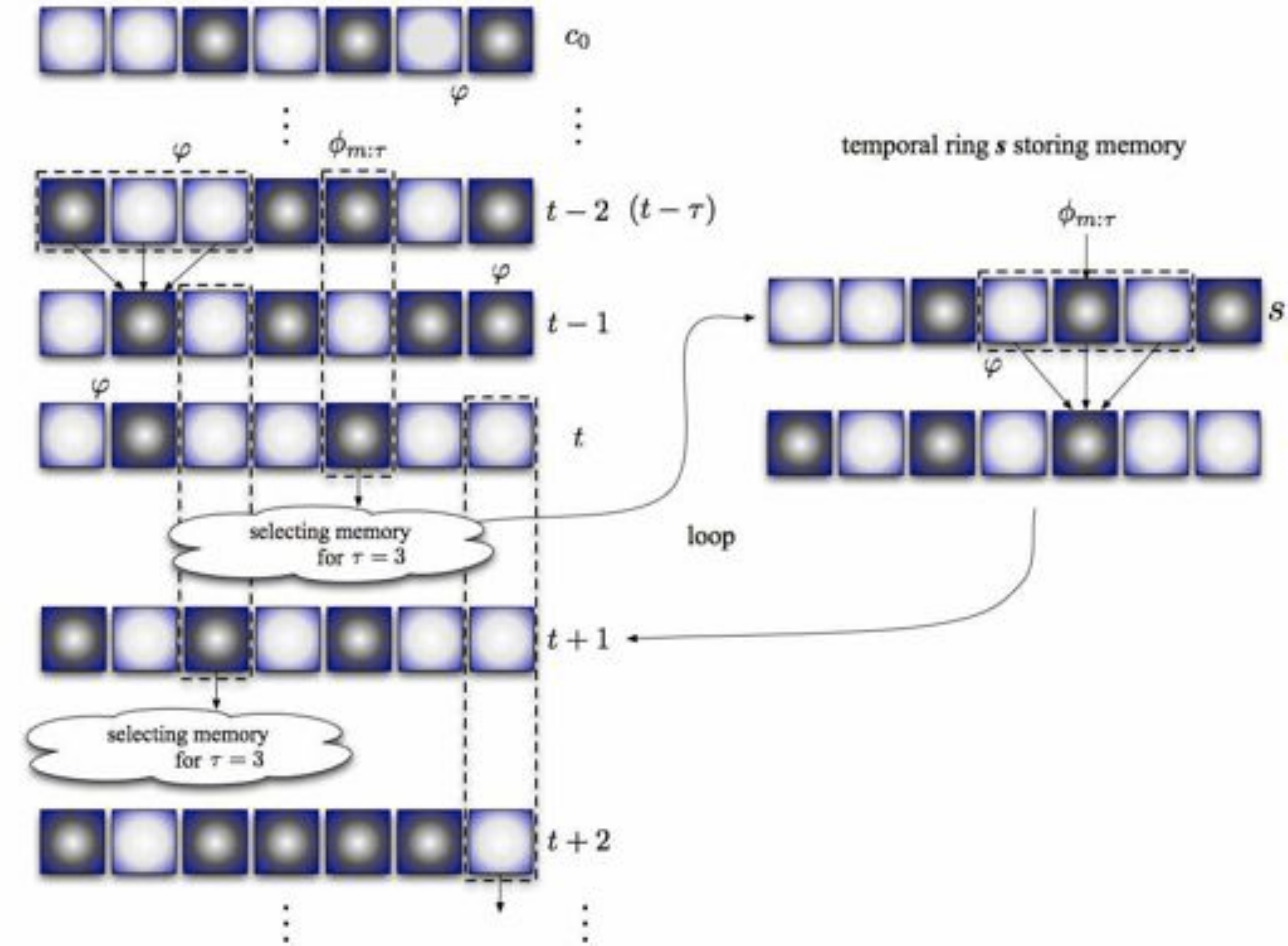}}
\caption{Dynamics in ECAM on an arbitrary one-dimensional array and hypothetical evolution rule $\varphi$ and memory function $\phi_{m:\tau}$ with $\tau=3$.}
\label{evolECAM}
\end{figure}

Let us consider the {\it memory function} $\phi$ in a form of {\it majority memory}, 

$$\phi_{maj} \rightarrow s_{i},$$

\noindent where in case of a tie, i.e., same number of 1s and 0s in past configurations, the last value $x_i^{t}$ is to be adopted as $s_i^{(t)}$, which implies no memory effect. These $\#1=\#0$ ties are only feasible when $\tau$ is even, in which case the effect of memory may appear as somehow \emph{weaker}, or simply \emph{different}, compared to the effect of the odd $\tau-1$ or $\tau+1$ close lengths of memory. Thus, $\phi_{maj}$ function represents the classic majority function. For three values [Minsky, 1967], then we have that:

\begin{equation}
\phi_{maj}(a,b,c) : (a \wedge b) \vee (b \wedge c) \vee (c \wedge a)
\label{majMem}
\end{equation}

Any map of previous states may act as memory (not only majority). Thus, minority, parity, alpha, $\ldots$, or any CA rule acting as memory, weighted memory, $\ldots$, etc. (for full details please see [Alonso-Sanz, 2009a], [Alonso-Sanz, 2011]).

Evolution rules representation for ECAM in this paper is given in [Mart{\'i}nez et al., 2010a], [Mart{\'i}nez et al., 2010b], [Mart{\'i}nez et al., 2011], [Mart{\'i}nez et al., 2012a], [Mart{\'i}nez et al., 2012b], as follows:

\begin{equation}
\phi_{CARm:\tau}
\end{equation}

\noindent where $CAR$ is the decimal notation of a particular ECA rule and $m$ is the kind of memory used with a specific value of $\tau$. This way, for example, the majority memory ($maj$) incorporated in ECA rule 30 employing five steps of a cell's history ($\tau=5$) is denoted simply as: $\phi_{R30maj:5}$. The memory is functional as the CA itself, see schematic explanation in Fig.~\ref{evolECAM}.

\section{Chaos moving to complexity when endowing the dynamics with memory: A case study}
\label{chaostocomplex}

In this section, we consider a particular case to illustrate the effect of memory, deriving in complex dynamics from a chaotic rule [Mart{\'i}nez et al., 2010b]. Here we deal with a chaotic ECA (class III), the evolution rule 126. This is a special chaotic rule because such evolution yield sets of regular languages [Wolfram, 1984b], [McIntosh, 2009]. We can deduce from previous analysis that ECA rule 126 could contain another kind of interesting information. Selecting a kind of memory we will see that particularly ECAM $\phi_{R126maj:4}$ displays a large number of glider guns emerging from random initial conditions, and emergence of a number of non-trivial patterns colliding constantly [Mart{\'i}nez et al., 2010b].

\subsection{ECA rule 126}

The local-state transition function $\varphi$ corresponding to ECA rule 126 is represented as follows:

\[
\varphi_{R126} = \left\{
	\begin{array}{lcl}
		1 & \mbox{if} & 110, 101, 100, 011, 010, 001 \\
		0 & \mbox{if} & 111, 000
	\end{array} \right. .
\]

\begin{figure}[th]
\begin{center}
\subfigure[]{\scalebox{0.6}{\includegraphics{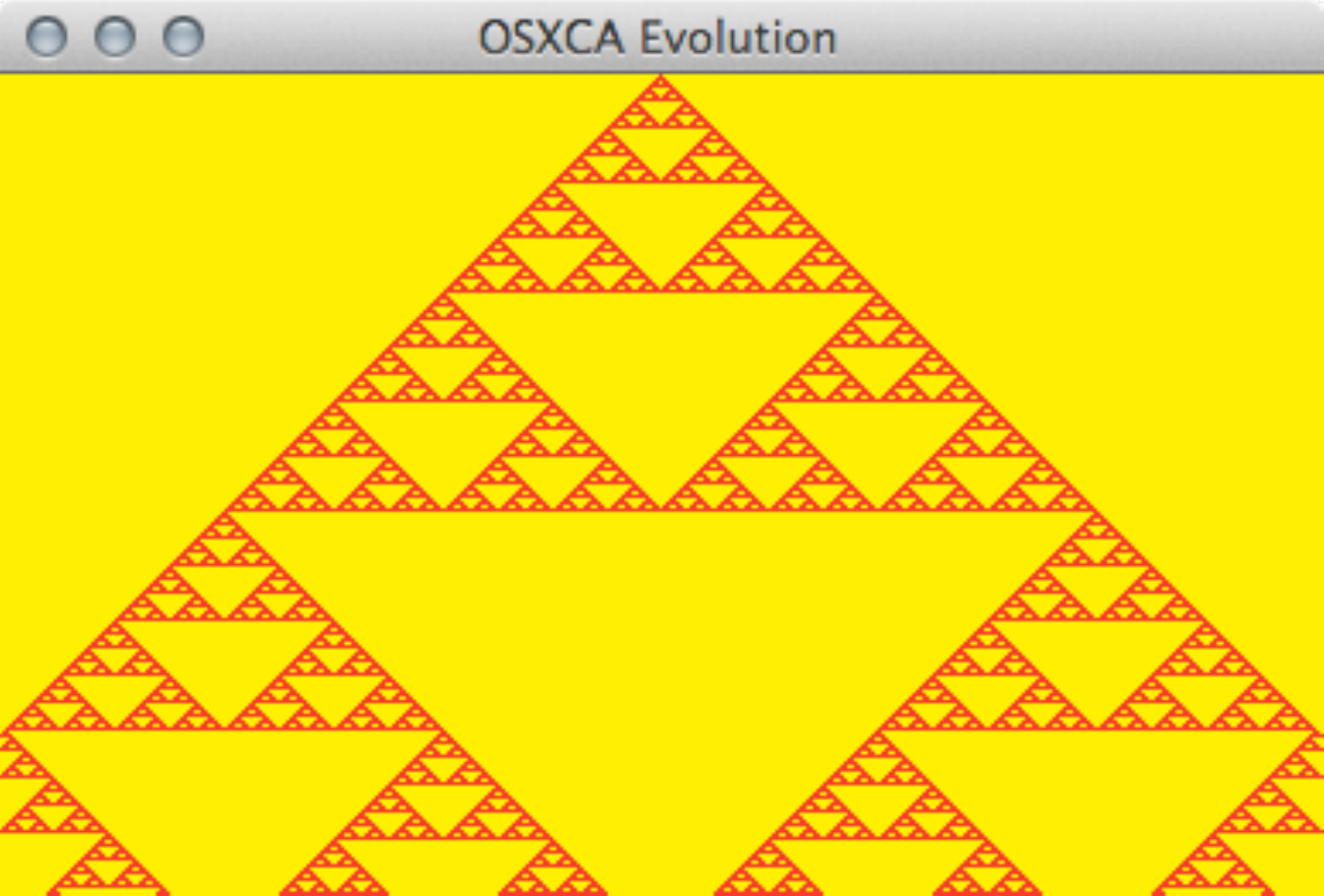}}} 
\subfigure[]{\scalebox{0.6}{\includegraphics{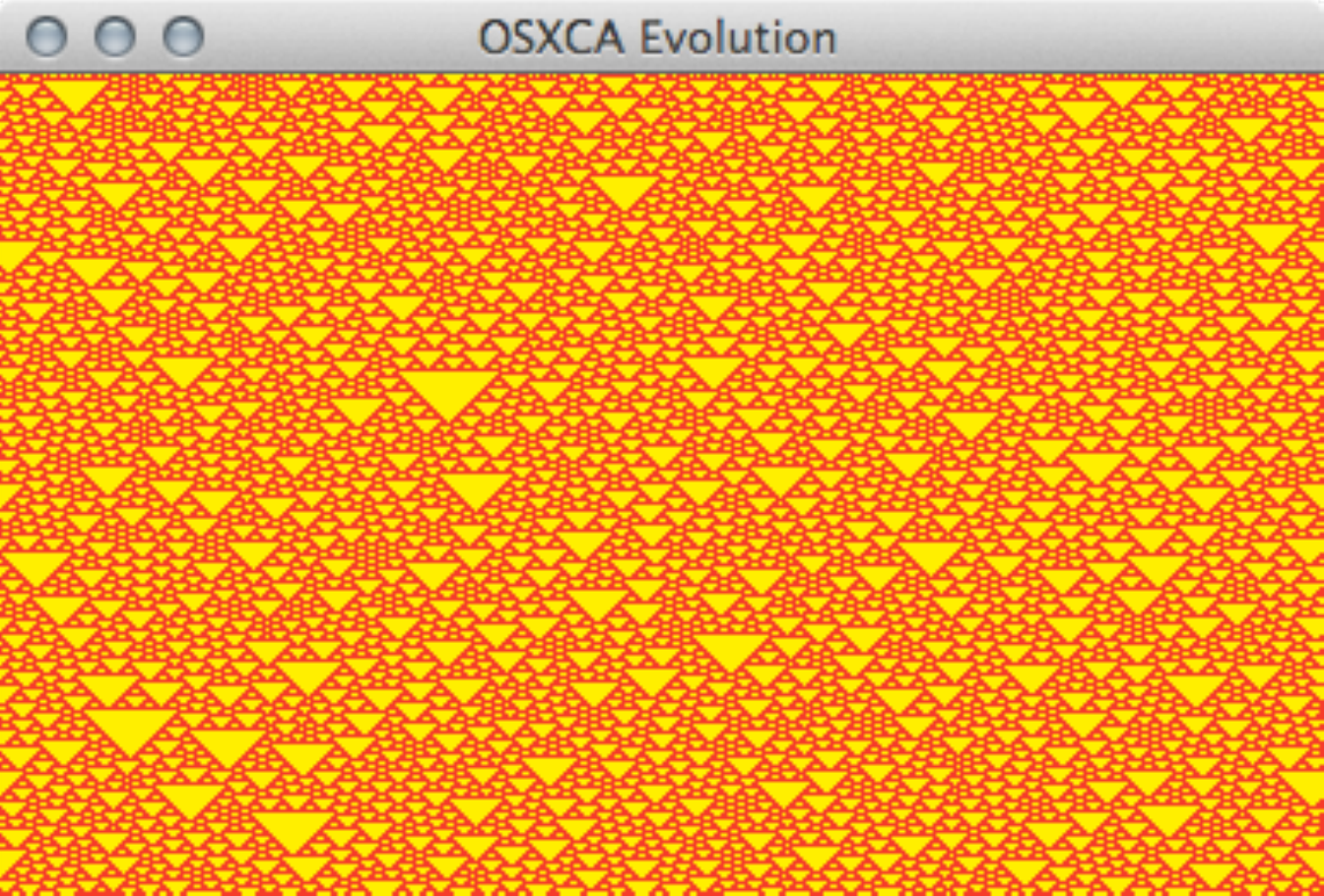}}} \end{center}
\caption{(a) Typical fractal and (b) chaotic global evolution of ECA rule 126. (a)~initially all cells in `0' but one in state `1,' (b)~evolution from random initial configuration with 50\% of `0' and `1' states. Evolution on a horizontal ring of 387 cells with time going down up to 240 time steps).}
\label{randomEvol}
\end{figure}

ECA rule 126 has a chaotic global behaviour typical from Class III in Wolfram's classification [Wolfram, 1994] (Fig.~\ref{WolframClasses}). In $\varphi_{R126}$ we can easily recognize an initial high probability of alive cells, i.e. cells in state `1'; with a 75\% to appear in the next time and, complement of only 25\% to get state 0. It will be always a new alive cell iff $\varphi_{R126}$ has one or two alive cells such that the equilibrium reached when there is an overpopulation condition. Figure~\ref{randomEvol} shows these cases in typical evolutions of ECA rule 126, both evolving from a single cell in state `1' (Fig.~\ref{randomEvol}a) and from a random initial configuration (Fig.~\ref{randomEvol}b) where a high density of 1's is evidently in the evolution.

While looking on chaotic space-time configuration in Fig.~\ref{randomEvol} we understand the difficulty for analysing the rule's behaviour and selecting any coherent activity among periodic structures without special tools.

\subsection{Mean field approximation}

In this section we use a probabilistic analysis with mean field theory to uncover basic properties of $\varphi_{R126}$ evolution space and its related chaotic behaviour. Such analysis we help us to explore the evolution space with specific initial conditions, that might lead to discoveries of non-trivial behaviour.

Mean field theory is a established technique for discovering statistical properties of CA without analysing evolution spaces of individual rules [McIntosh, 2009]. The method assumes that states in $\Sigma$ are independent and do not correlate with each other in the local function $\varphi_{R126}$. Thus we can study probabilities of states in a neighbourhood in terms  of the probability of a single state (the state in which the neighbourhood evolves), and probability of the neighbourhood as a product of the probabilities of each cell in it. McIntosh in [McIntosh, 1990] presents an explanation of Wolfram's classes with a mixture of probability theory and de Bruijn diagrams, resulting in a classification based on mean field theory curve, as follows:

\begin{itemize}
\item class I: monotonic, entirely on one side of diagonal;
\item class II: horizontal tangency, never reaches diagonal;
\item class IV: horizontal plus diagonal tangency, no crossing;
\item class III: no tangences, curve crosses diagonal.
\end{itemize}

For one dimensional case, all neighbourhoods are considered as follows:

\begin{equation}
p_{t+1}=\sum_{j=0}^{k^{2r+1}-1}\varphi_{j}(X)p_{t}^{v}(1-p_{t})^{n-v}
\label{MFp1D}
\end{equation}

\noindent such that $j$ is an index relating each neighbourhood and $X$ are cells $x_{i-r}, \ldots, x_{i},$ $\ldots, x_{i+r}$. Thus $n$ is the number of cells into every neighbourhood, $v$ indicates how often state `1' occurs in $X$, $n-v$ shows how often state `0' occurs in the neighbourhood $X$, $p_{t}$ is the probability of cell being in state `1' while $q_{t}$ is the probability of cell being in state `0' i.e., $q=1-p$. The polynomial for ECA rule 126 is defined as follows:

\begin{equation}
p_{t+1}=3p_{t}q_{t}.
\label{pR126}
\end{equation}

Because $\varphi_{R126}$ is classified as a chaotic rule, we expect no tangencies and its curve must cross the identity; recall that $\varphi_{R126}$ has a 75\% of probability to produce a state one. 

\begin{figure}[th]
\centerline{\includegraphics[width=3.8in]{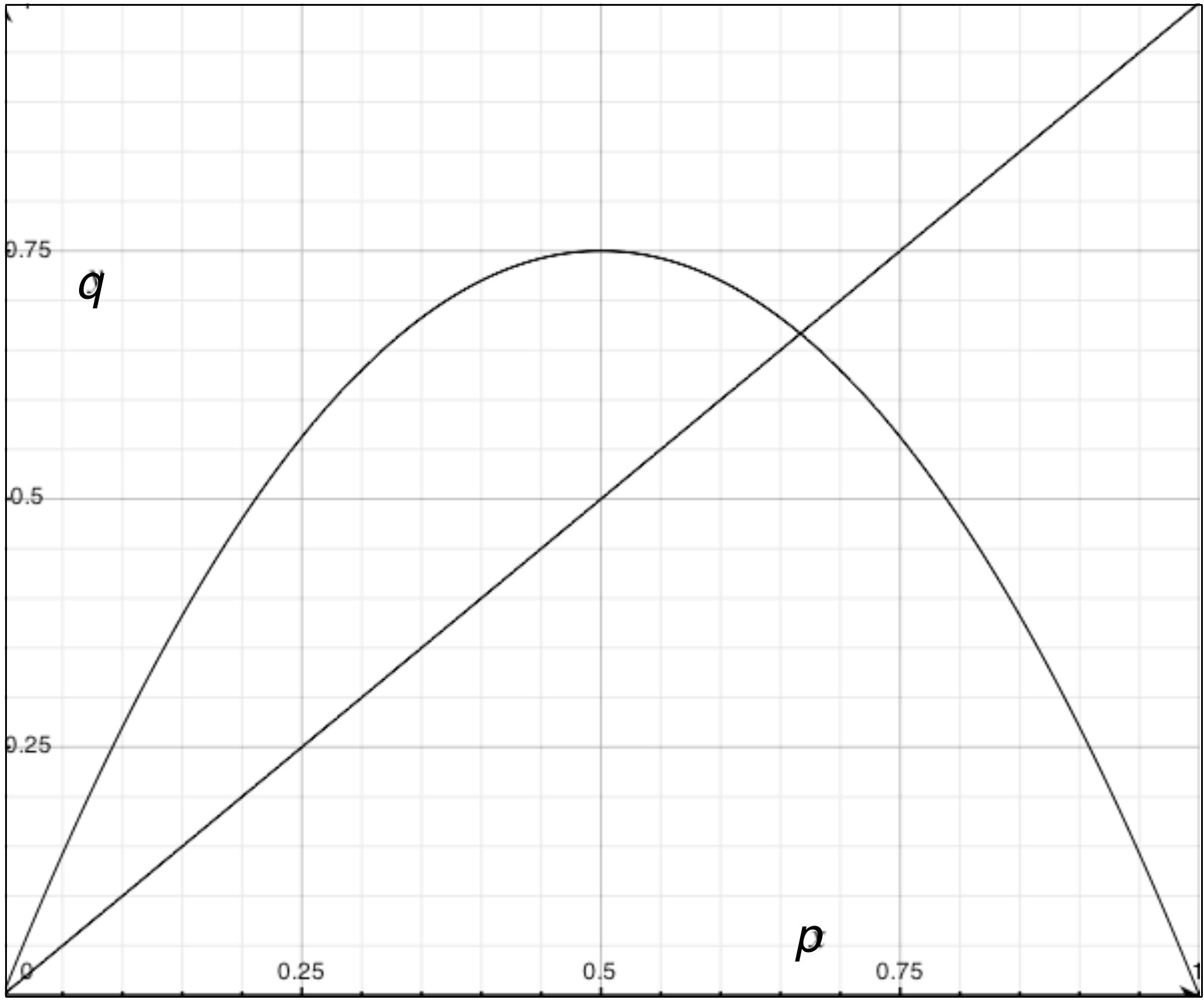}}
\caption{Mean field curve for ECA rule 126. }
\label{meanField}
\end{figure}

Mean field curve (Fig.~\ref{meanField}) confirms that probability of state `1' in space-time configurations of $\varphi_{R126}$ is 0.75 for high densities related to big populations of 1's. The curve demonstrates also that $\varphi_{R126}$ is chaotic because the curve cross the identity with a first fixed point at the origin $f=0$ and the absence of unstable fixed points inducing non stable regions in the evolution. Nevertheless, the stable fixed point is $f=0.6683$, which represents a `concentration' of `1's  diminishing during the automaton evolution.

So the initial inspection indicates no evidence of complex behaviour emerging in $\varphi_{R126}$. Of course a deeper analysis is necessary for obtaining more features from a chaotic rule, so the next sections explain other techniques to study in particular periodic structures.

\subsection{Basins of attraction}
A basin (of attraction) field of a finite CA is the set of basins of attraction into which all possible states and trajectories are driven by the local function $\varphi$. The topology of a single basin of attraction may be represented by a diagram, the {\it state transition graph}. Thus the set of graphs composing the field specifies the global behaviour of the system [Wuensche \& Lesser, 1992].

Generally a basin can also recognise CA with chaotic or complex behaviour following previous results on attractors [Wuensche \& Lesser, 1992]. Thus we have  Wolfram's classes  represented as a basin classifications, following the Wuensche's characterisation:

\begin{itemize}
\item class I: very short transients, mainly point attractors (but possibly also periodic attractors) very high in-degree, very high leaf density (very ordered dynamics);
\item class II: very short transients, mainly short periodic attractors (but also point attractors), high in-degree, very high leaf density;
\item class IV: moderate transients, moderate-length periodic attractors, moderate in-degree, very moderate leaf density (possibly complex dynamics);
\item class III: very long transients, very long periodic attractors, low in-degree, low leaf density (chaotic dynamics).
\end{itemize}

\begin{figure}
\centerline{\includegraphics[width=5.5in]{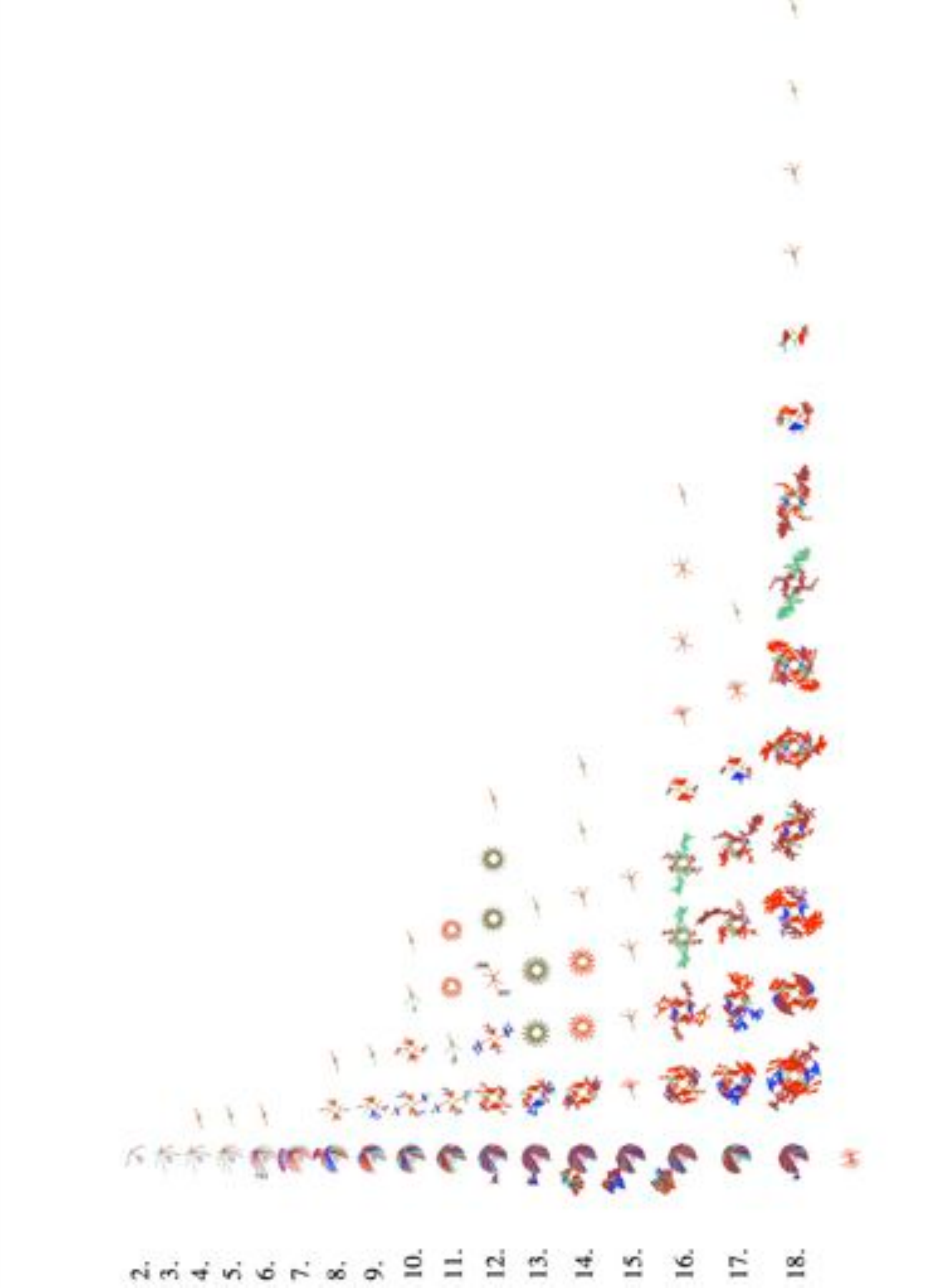}}
\caption{The whole set of non-equivalent basins in ECA rule 126 from $l=2$ to $l=18$.}
\label{r126_2-18}
\end{figure}

The basins depicted in Fig.~\ref{r126_2-18} show the whole set of non-equivalent basins in ECA rule 126 from $l=2$ to $l=18$ ($l$ means length of array) attractors, all they display not high densities from an attractor of mass one and attractors of mass 14.\footnote{Basins and attractors were calculated with {\it Discrete Dynamical System} DDLab [Wuensche, 2011] available from \url{http://www.ddlab.org/}} This way, ECA rule 126 displays some non symmetric basins and some of them have long transients that induce a relation with chaotic rules.

\begin{figure}[th]
\begin{center}
\subfigure[]{\scalebox{0.45}{\includegraphics{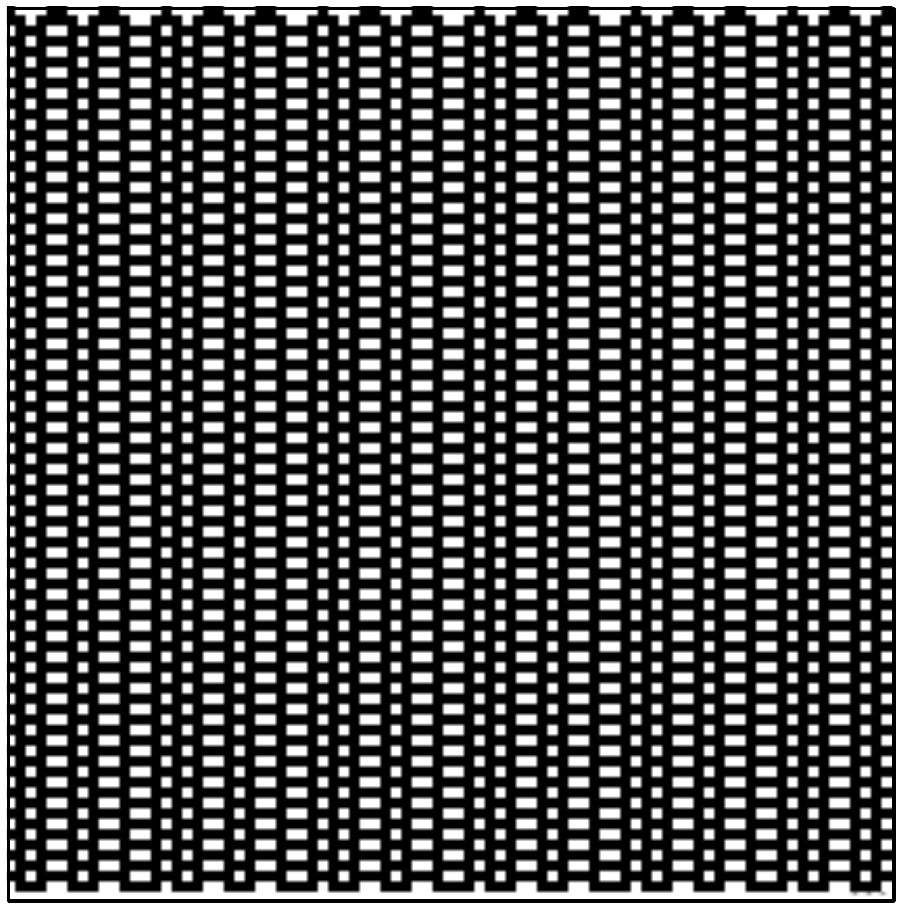}}} 
\subfigure[]{\scalebox{0.45}{\includegraphics{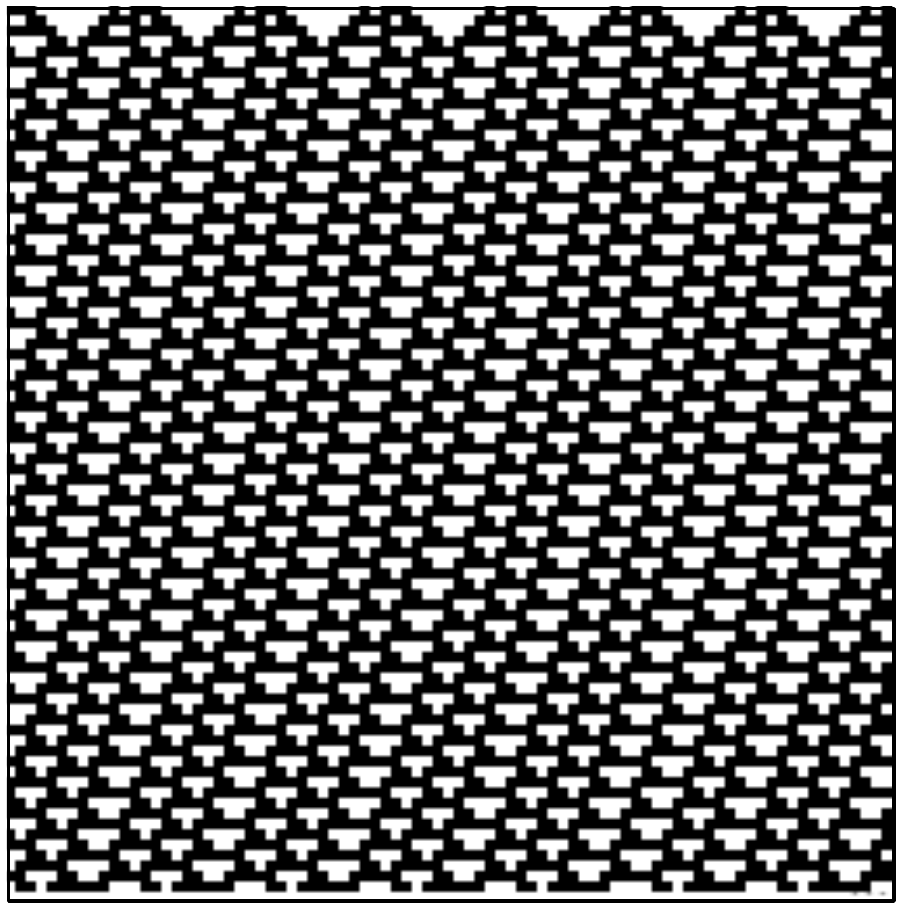}}}
\subfigure[]{\scalebox{0.45}{\includegraphics{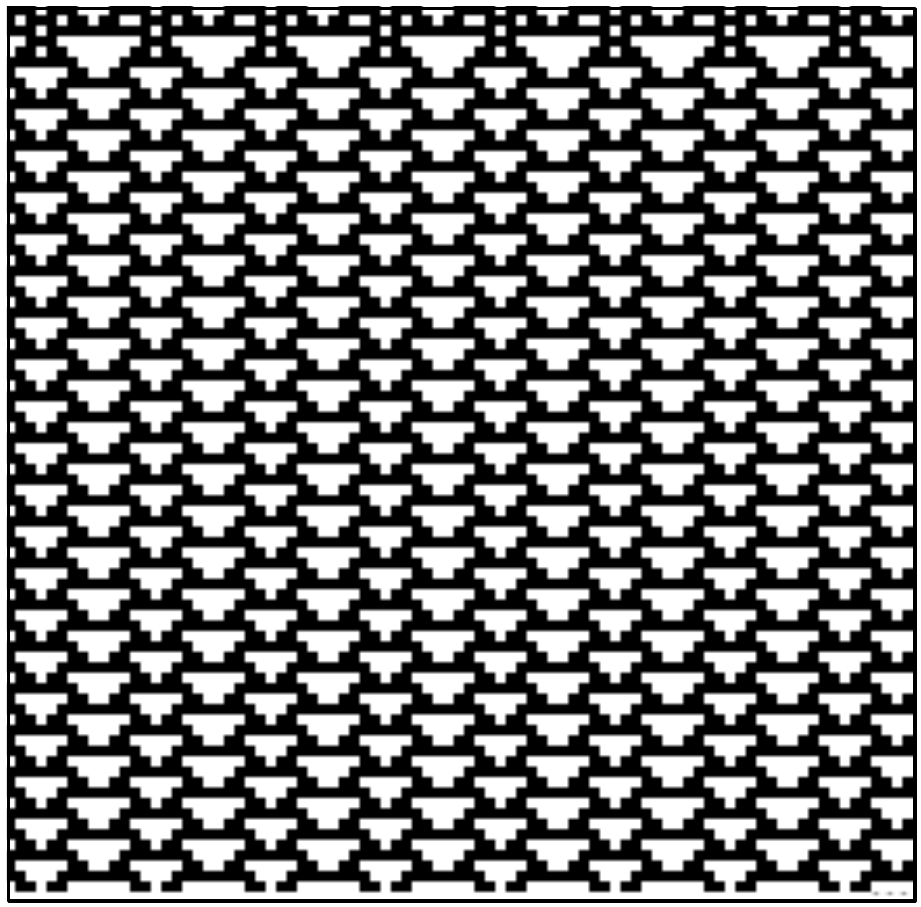}}}
\subfigure[]{\scalebox{0.45}{\includegraphics{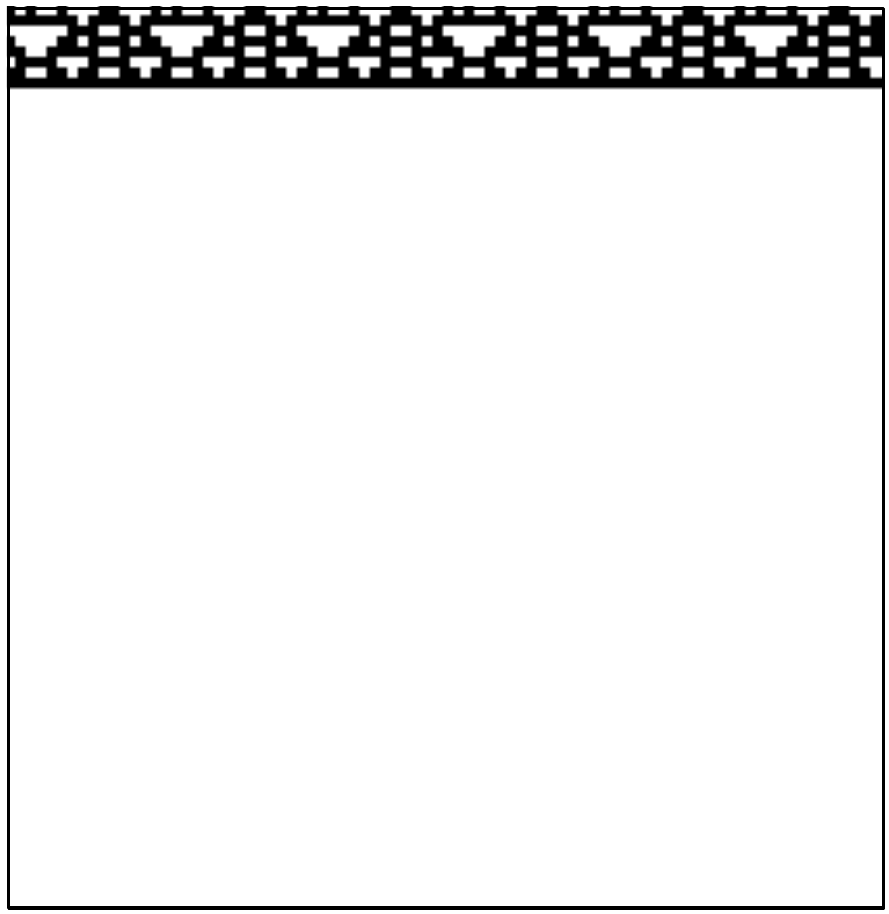}}}
\end{center}
\caption{Periodic patterns calculated from some exemplar attractors.}
\label{cyclesR126}
\end{figure}

Particularly we can see specific cycles in Fig.~\ref{cyclesR126} where the following structures could be found:

\begin{itemize}
\item[(a)] static configurations as still life patterns ($l=8$);
\item[(b)] traveling configurations as gliders ($l=15$);
\item[(c)] meshes ($l=12$);
\item[(d)] or empty universes ($l=14$).
\end{itemize}

The cycle diagrams expose only displacements to the left, and this empty universe evolving to the stable state 0 is constructed all times on the first basin for each cycle, see Fig.~\ref{r126_2-18}.

This way some cycles could induce a non trivial activity in rule 126, but the associated initial conditions are not generally predominant. However some information could be derived from periodic patterns that have a high frequency inside this evolution space. This can be done by using filters.

\subsection{De Bruijn diagrams}

De Bruijn diagrams [McIntosh, 2009], [Voorhees, 1996] are proven to be an adequate tool for describing evolution rules in one dimension CA, although originally they were used in shift-register theory (the treatment of sequences where their elements overlap each other). Paths in a de Bruijn diagram may represent chains, configurations or classes of configurations in the evolution space. 

For a one-dimensional CA of order $(k,r)$, the de Bruijn diagram is defined as a directed graph with $k^{2r}$ vertices and $k^{2r+1}$ edges. The vertices are labeled with elements of an alphabet of length $2r$. An edge is directed from vertex $i$ to vertex $j$, if and only if, the $2r-1$ final symbols of $i$ are the same that the $2r-1$ initial ones in $j$ forming a neighbourhood of $2r+1$ states represented by $i \diamond j$. In this case, the edge connecting $i$ to $j$ is labeled with $\varphi(i \diamond j)$ (the value of the neighbourhood defined by the local function) [Voorhees, 2008].

\begin{figure}[th]
\centerline{\includegraphics[width=2in]{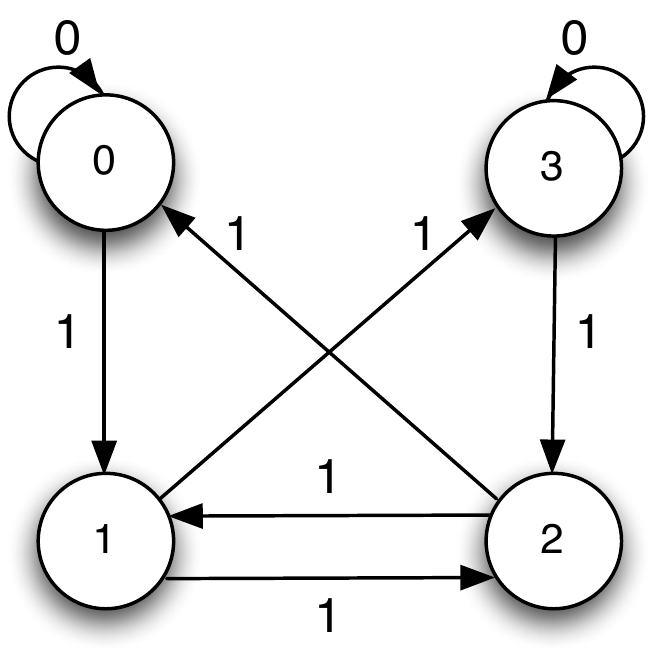}}
\caption{De Bruijn diagram for the ECA rule 126.}
\label{dB126}
\end{figure}

The extended de Bruijn diagrams [McIntosh, 2009] are useful for calculating all periodic sequences by the cycles defined in the diagram. These ones also show the {\it shift} of a sequence for a certain number of {\it generations}. Thus we can get de Bruijn diagrams describing periodic sequences for ECA rule 126.

\begin{figure}[th]
\begin{center}
\subfigure[]{\scalebox{0.55}{\includegraphics{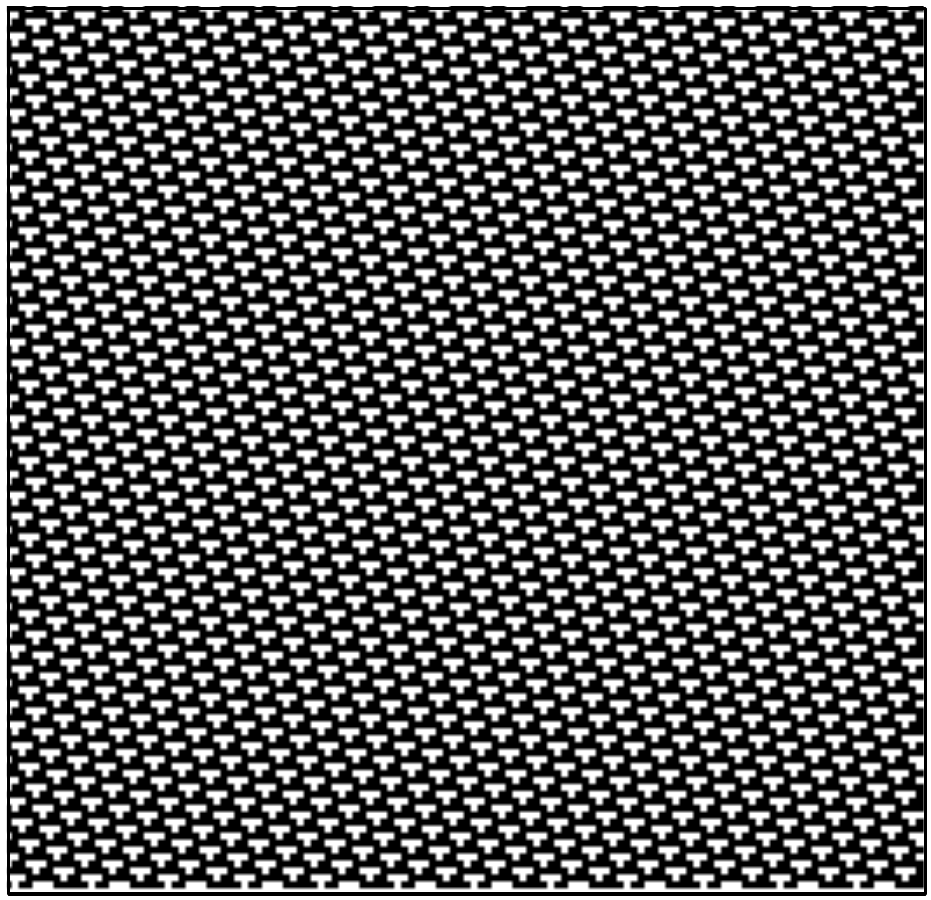}}} \hspace{0.1cm}
\subfigure[]{\scalebox{0.55}{\includegraphics{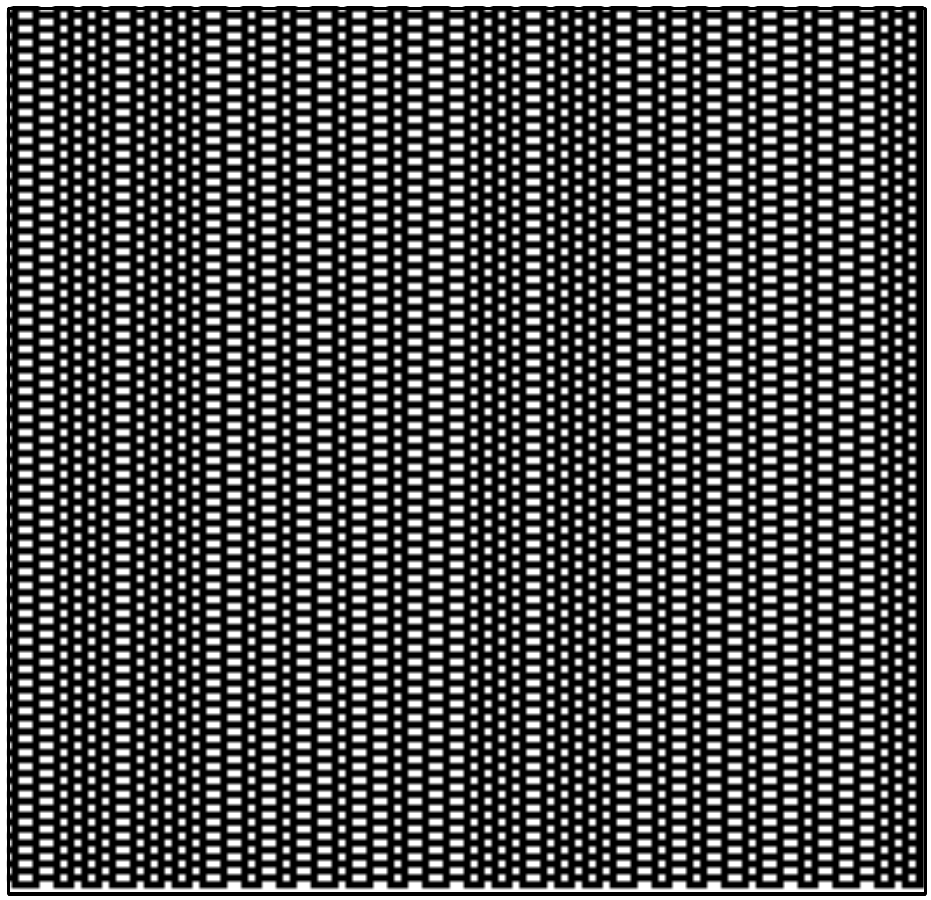}}} \hspace{0.1cm}
\subfigure[]{\scalebox{0.55}{\includegraphics{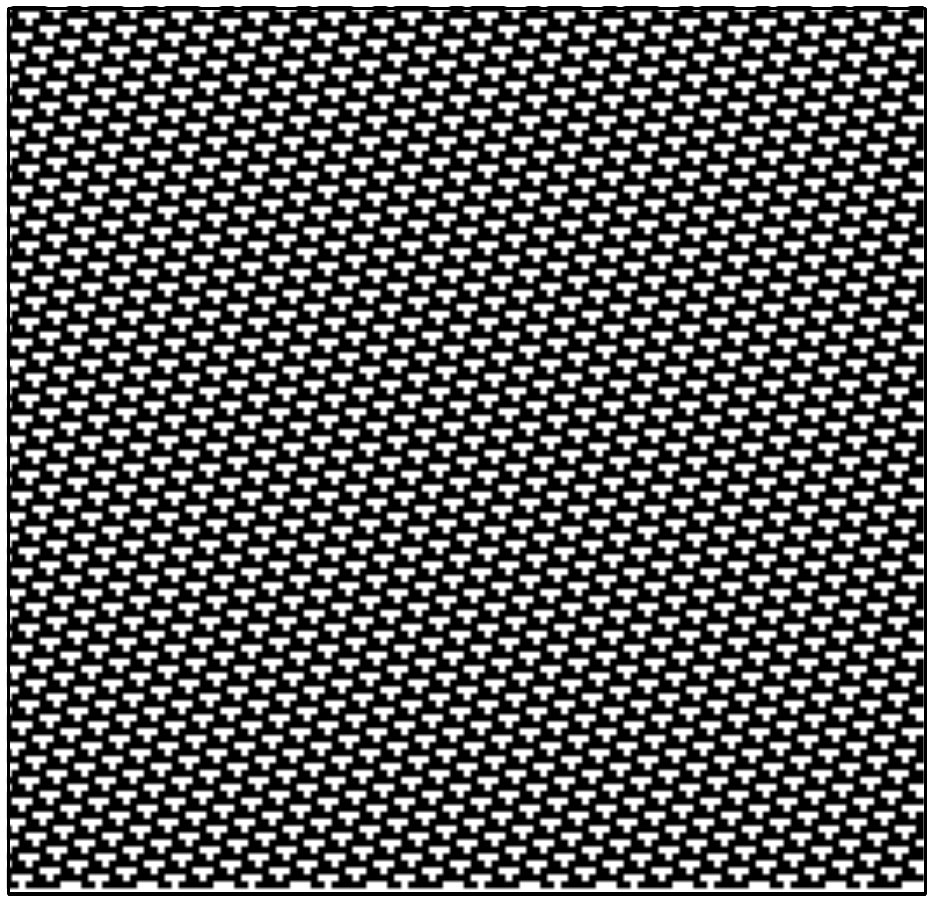}}}
\end{center}
\caption{Patterns calculated with extended de Bruijn diagrams, in particular from cycles of order $(x,2)$ (that means $x$-shift  in 2-generations).}
\label{dB126-1}
\end{figure}

The de Bruijn diagram associated to ECA rule 126 is depicted in Fig. \ref{dB126}.\footnote{De Bruijn diagrams were calculated using NXLCAU21 designed by McIntosh; available in \url{http://delta.cs.cinvestav.mx/~mcintosh/cellularautomata/SOFTWARE.html}} Figure~\ref{dB126} shows that there are two neighbourhoods evolving into $0$ and six neighbourhoods into $1$. State 1 has higher frequency. This indicates  a possibility that the local transition function is injective and {\it Garden of Eden} configurations [Amoroso \& Cooper, 1970] exist. These are configurations that cannot be constructed from other configurations, i.e., configurations without  ancestors. In one dimension, the {\it subset} diagram can calculate quickly the Garden of Eden configurations, and the {\it pair diagram} can calculate configurations with multiple ancestors [McIntosh, 1990]. Classical analysis in graph theory has been applied to de Bruijn diagrams for studying topics such as reversibility [Nasu, 1978], [Mora et al., 2005]; in other sense, cycles in the diagram indicate periodic constructions in the evolution of the automaton if the label of the cycle agrees with the sequence defined by its nodes, taking periodic boundary conditions. Let us take the equivalent construction of a de Bruijn diagram in order to describe the evolution in two steps of ECA rule 126 (having now nodes composed by sequences of four symbols); the cycles of this new diagram are presented in Fig.~\ref{dB126-1}.

Cycles inside de Bruijn diagrams can be used for obtaining regular expressions representing a periodic pattern. Figure~\ref{dB126-1} displays three patterns calculated as: (a) {\sf shift $-3$ in 2 generations} representing a pattern with displacement to the left, (b) {\sf shift 0 in 2 generations} describing a static pattern traveling without displacement, and (c) {\sf shift $+3$ in 2 generations} is exactly the symmetric pattern given in the first evolution.

So, we can also see in Fig.~\ref{dB126-1} that it is possible to find patterns traveling in both directions, as gliders or mobile structures. But generally these constructions (strings) cannot live in combination with others structures and therefore it is really hard to have this kind of objects with such characteristics. Although, ECA rule 126 has at least one glider! This will be explained in the next section.

\subsection{Filters help for discover hidden dynamics}

Filters are essential tools for discovering hidden order in chaotic or complex rules. Filters were introduced in CA studies by Wuensche who employed them to automatically classify cell-state transition functions, see [Wuensche, 1999]. Also filters related to tiles were successfully applied and deduced in analysing space-time behaviour of ECA governed by rules 110 and 54 [Mart{\'i}nez, 2006b], [Mart{\'i}nez et al., 2006].

\begin{figure}[th]
\centerline{\includegraphics[width=6.5in]{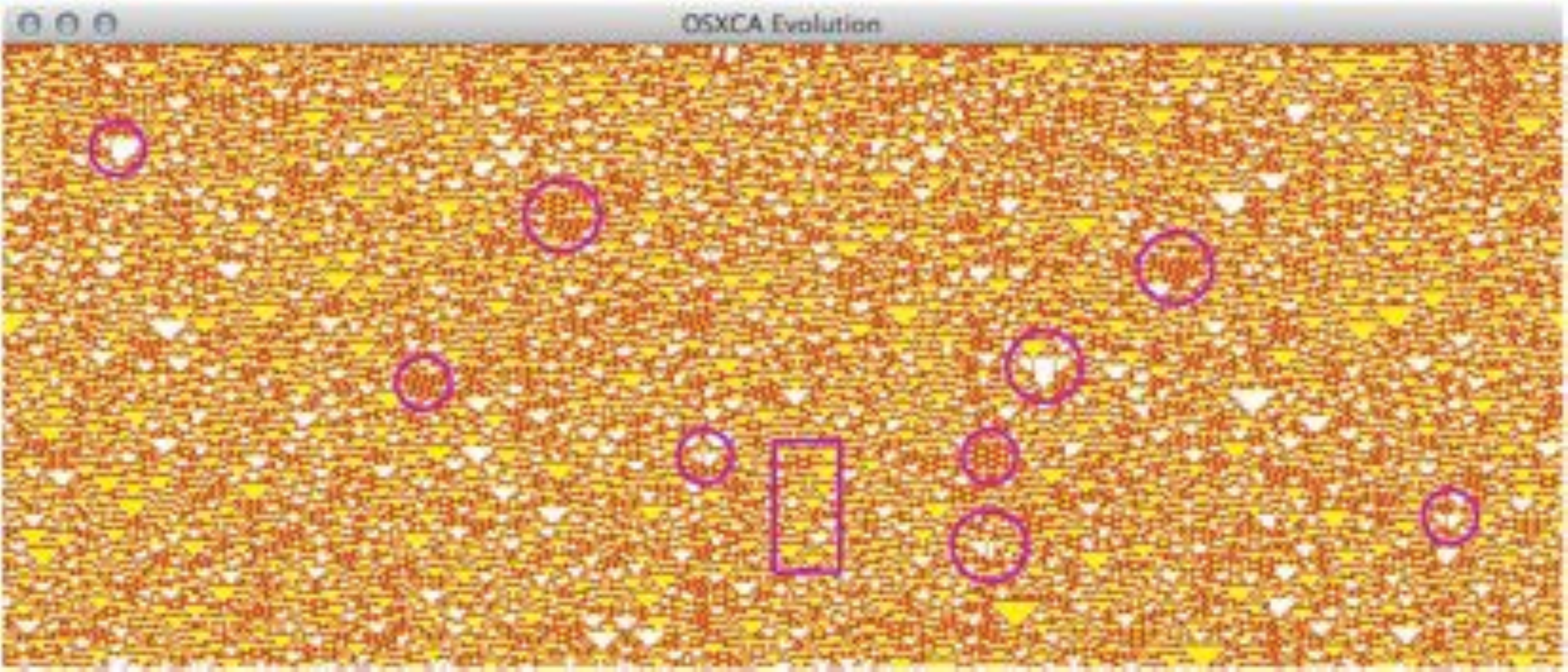}}
\caption{Filtered space-time configuration in ECA rule 126.}
\label{R126filtered}
\end{figure}

This way, we have found that ECA rule 126 has two types of two dimension tiles (which together work as filters over $\varphi_{R126}$):

\begin{itemize} 
\item the tile {\footnotesize $t_1 = \begin{bmatrix} 1111 \\ 1001 \end{bmatrix}$}, and
\item the tile {\footnotesize $t_2 = \begin{bmatrix} 0000000 \\ 0111110  \\ 1100011 \\ 0110110 \\ 1111111 \end{bmatrix}$}.
\end{itemize}

Filter $t_1$ works more significantly on configurations generated by $\varphi_{R126}$, the second one is not frequently found although it is exploited when ECA rule 126 is altered with memory (as we will see in the next section).

The application of the first filter is effective to discover gaps with little patterns traveling on triangles of `1' states in the evolution space. Although even in this case it may be unclear how a dynamics would be interpreted, a careful inspection on the evolution brings to light very small localisations (as still life), as shown in Fig.~\ref{R126filtered}.

This localisation emerging in ECA rule 126 and pinpointed by a filter is the periodic pattern calculated with the basin (Fig.~\ref{cyclesR126}a), and with the de Bruijn diagram (Fig.~\ref{dB126-1}b). The last one offers more information because such cycles allow to classify the whole phases when this glider is coded in the initial condition. Circles in Fig.~\ref{R126filtered} show some interesting regions that now are more clear with filters working. Some of them display very simple gliders (stationary), periodic meshes, and non-periodic structures emerging and existing inside chaotic patterns in several generations.

\subsection{Dynamics emerging in ECA rule 126 with memory}
\label{ECAM126section}

CA with memory had open a new family of evolution rules with different and interesting dynamics [Alonso-Sanz, 2009a], [Alonso-Sanz, 2011]. In this paper we explore three types of memory: {\sf minority}, {\sf majority}, and {\sf parity}. In the latter case, $s_{i}^{(t)} = x^{t-\tau+1}_{i} \oplus \ldots\oplus x^{t-1}_{i} \oplus x^{t}_{i}$.

\begin{figure}
\begin{center}
\subfigure[]{\scalebox{0.45}{\includegraphics{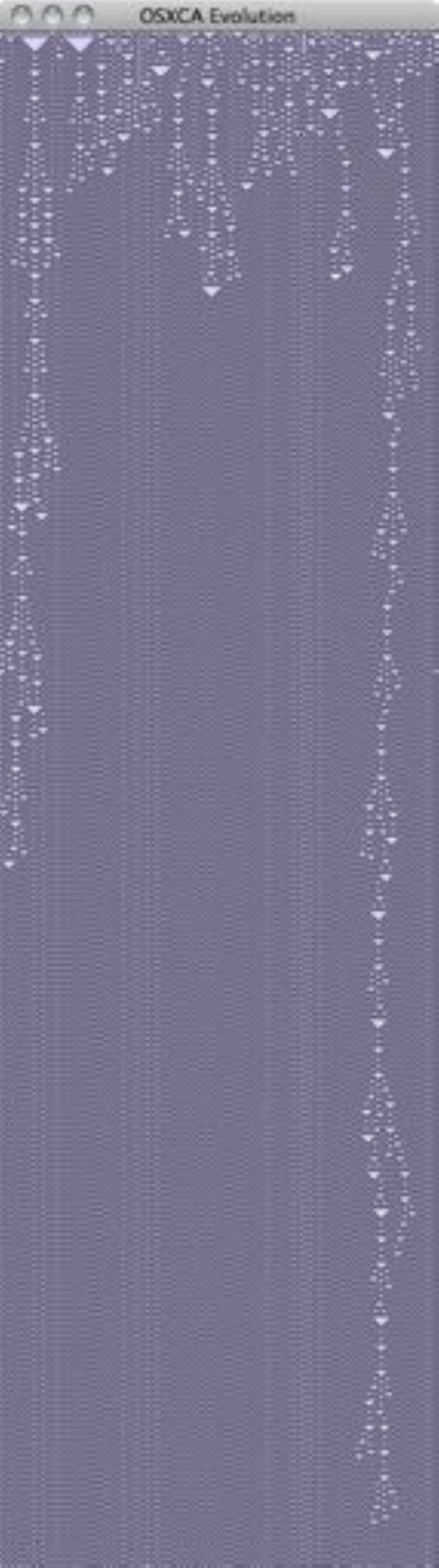}}} \hspace{0.6cm}
\subfigure[]{\scalebox{0.45}{\includegraphics{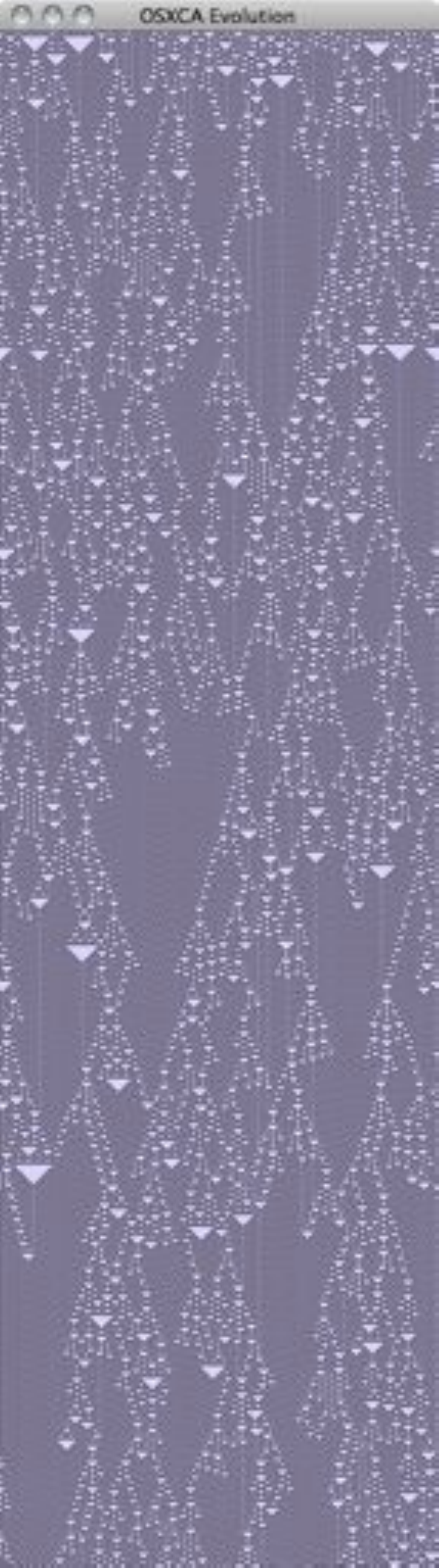}}} \hspace{0.6cm}
\subfigure[]{\scalebox{0.45}{\includegraphics{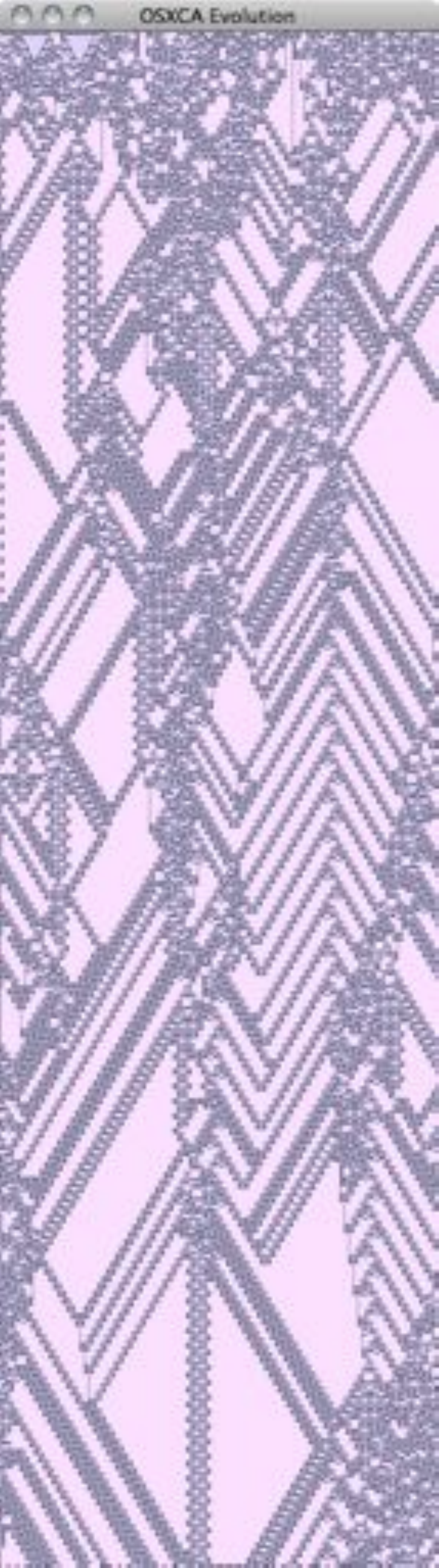}}}
\end{center}
\caption{(a) $\phi_{R126min:3}$ displays a typical evolution of ECAM rule 126 with minority memory $\tau=3$, 
(b) $\phi_{R126par:2}$ displays an evolution but now evolving with parity memory, and (c) the most interesting evolution with ECAM rule $\phi_{R126maj:4}$, where we can see the emergence of complex patterns as gliders and glider guns. In this case a filter is selected for a best view of complex patterns and their interactions. Snapshots start with same random initial conditions on a ring of 296 cells evolving in 1036 generations.}
\label{majMemory}
\end{figure}

\begin{figure}
\centerline{\includegraphics[width=5.5in]{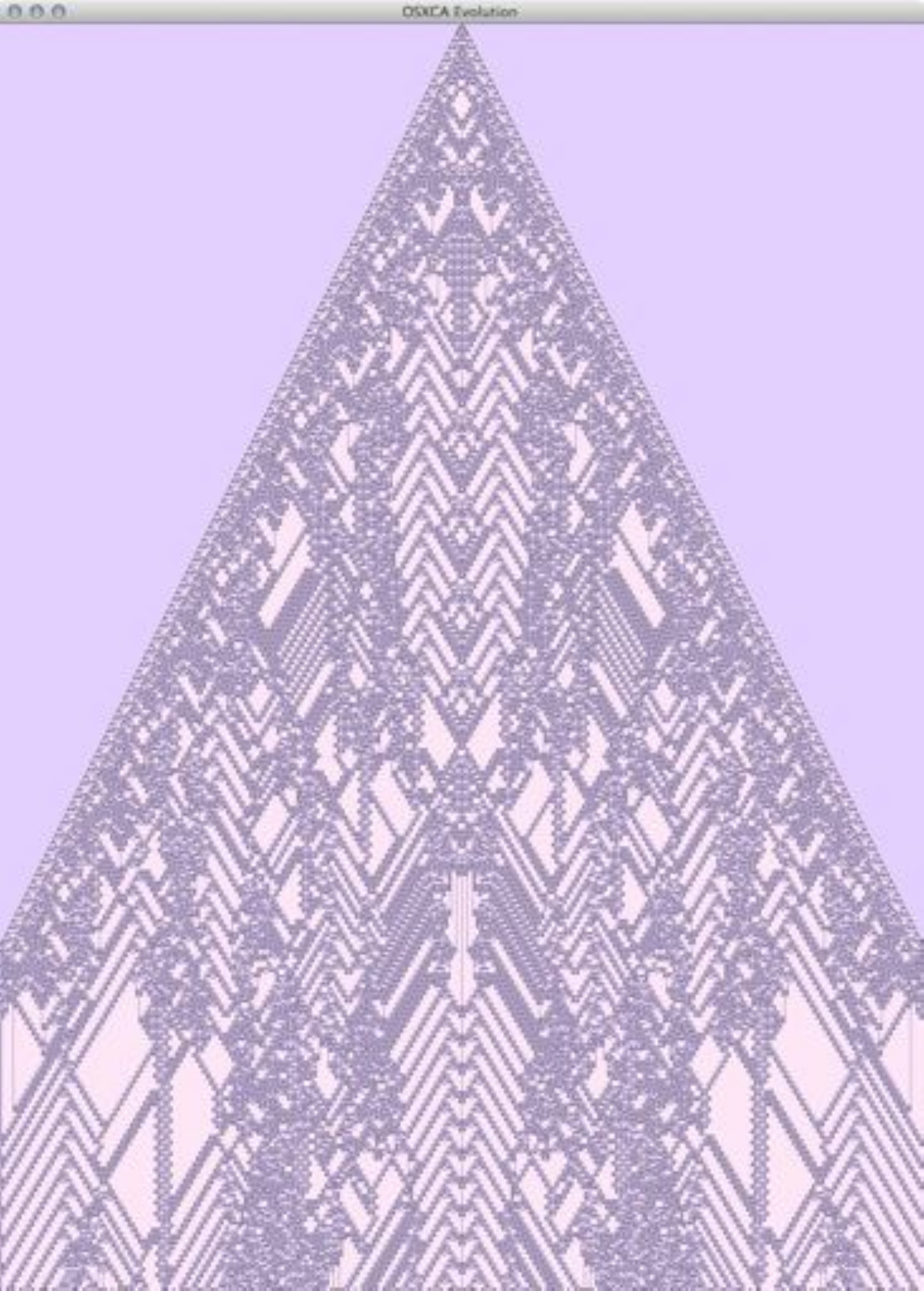}}
\caption{Filtered space-time configuration of ECAM $\phi_{R126maj:4}$ evolving with a ring of 843 cells, periodic boundaries, starting just from one non-quiescent cell and running for 1156 steps.}
\label{126maj4_843cells_1156gen}
\end{figure}

\begin{figure}
\centerline{\includegraphics[width=5.5in]{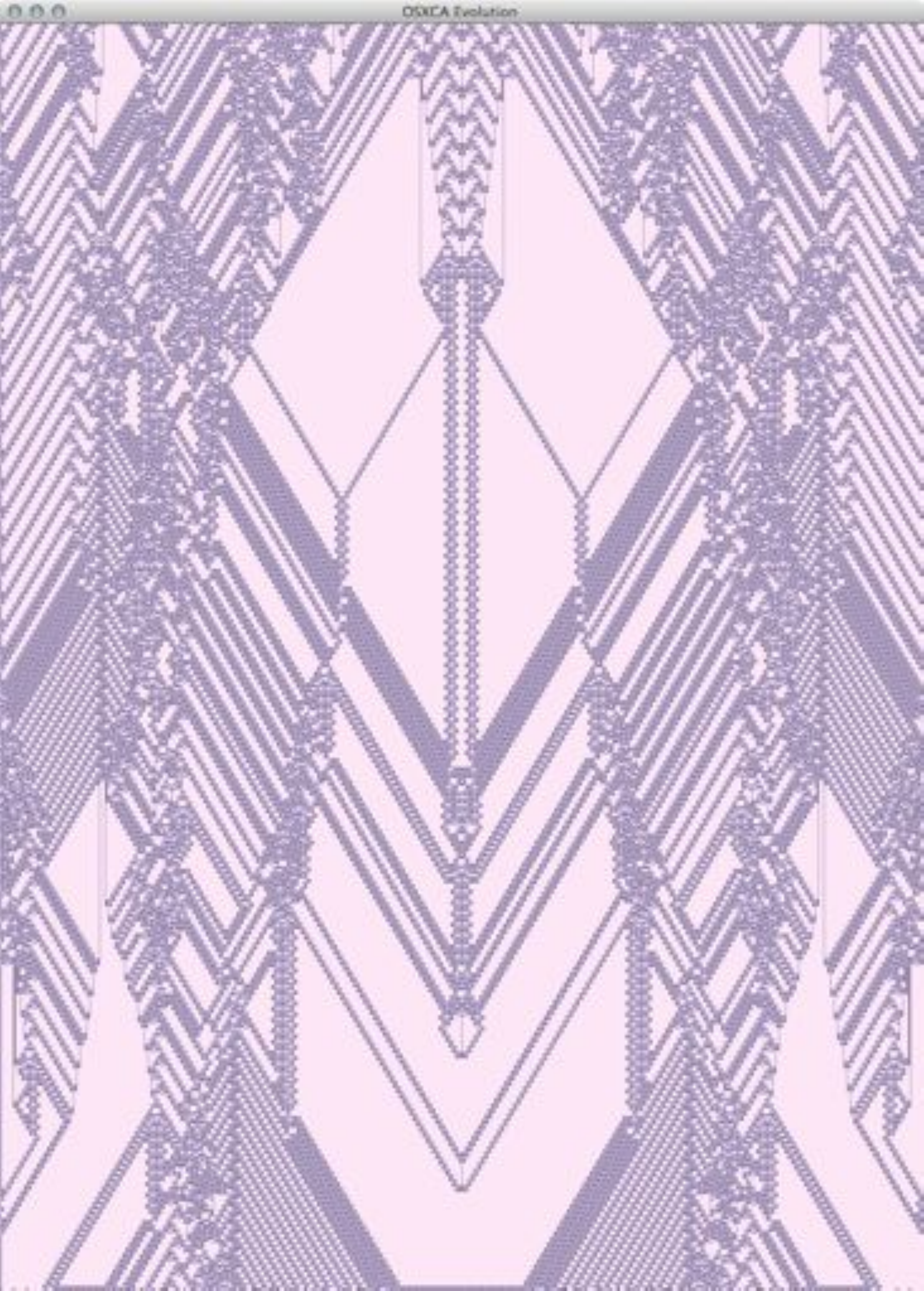}}
\caption{Continued evolution to 2312 steps.}
\label{126maj4_843cells_2312gen}
\end{figure}

\begin{figure}
\centerline{\includegraphics[width=5.5in]{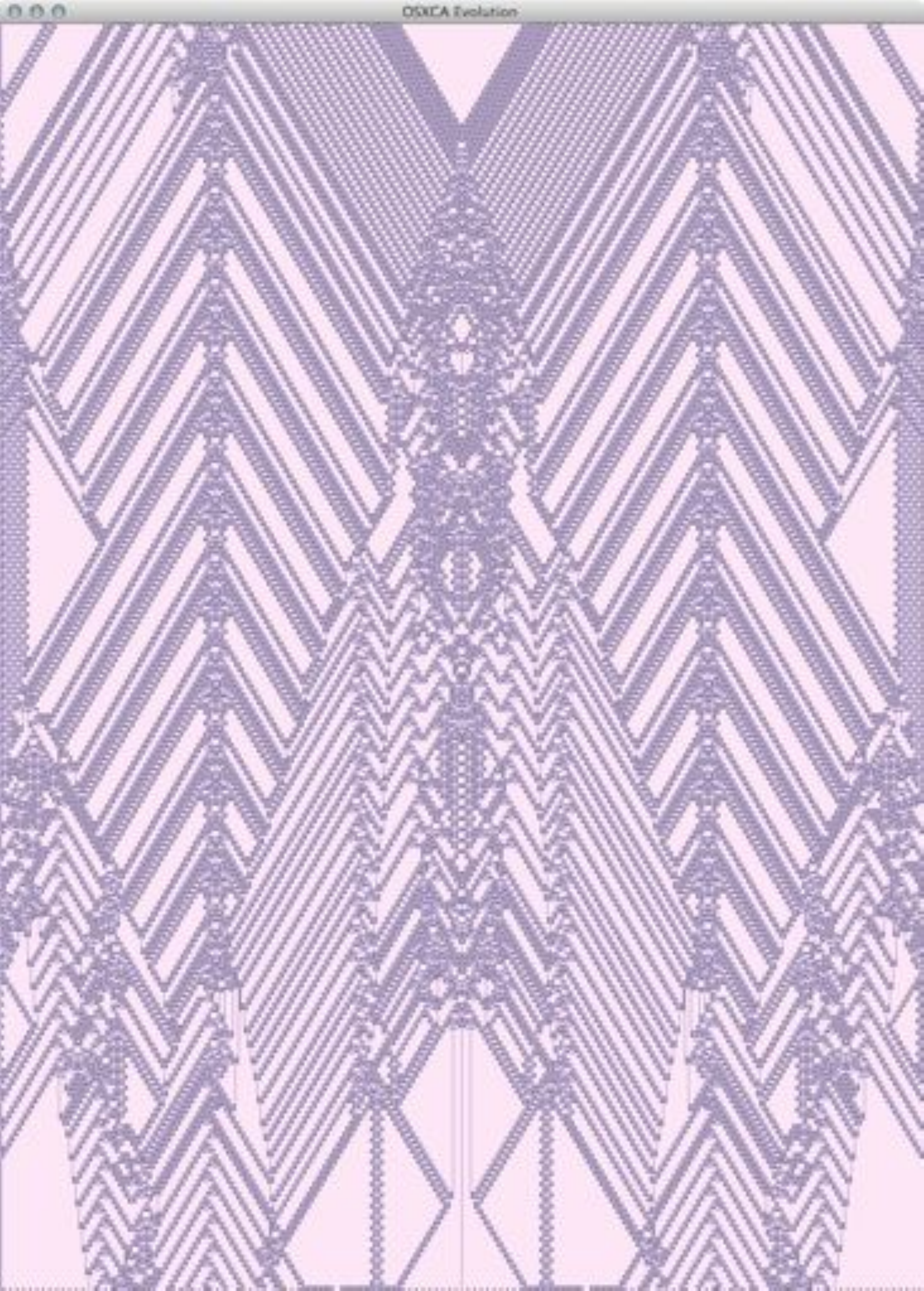}}
\caption{Continued evolution to 3468 steps.}
\label{126maj4_843cells_3468gen}
\end{figure}

\begin{figure}
\centerline{\includegraphics[width=5.5in]{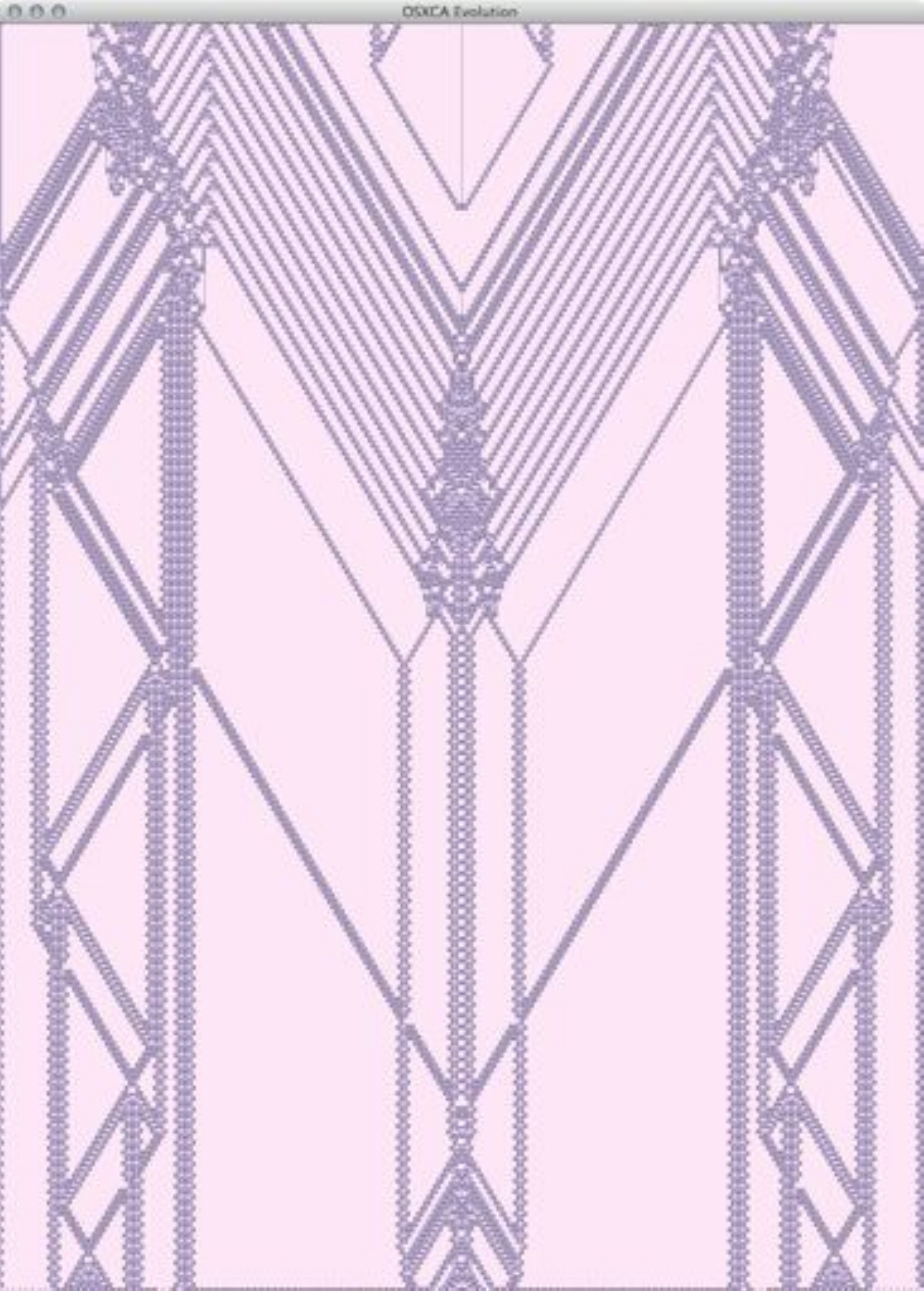}}
\caption{Continuee evolution to 4624 steps.}
\label{126maj4_843cells_4624gen}
\end{figure}

\begin{figure}
\centerline{\includegraphics[width=5.5in]{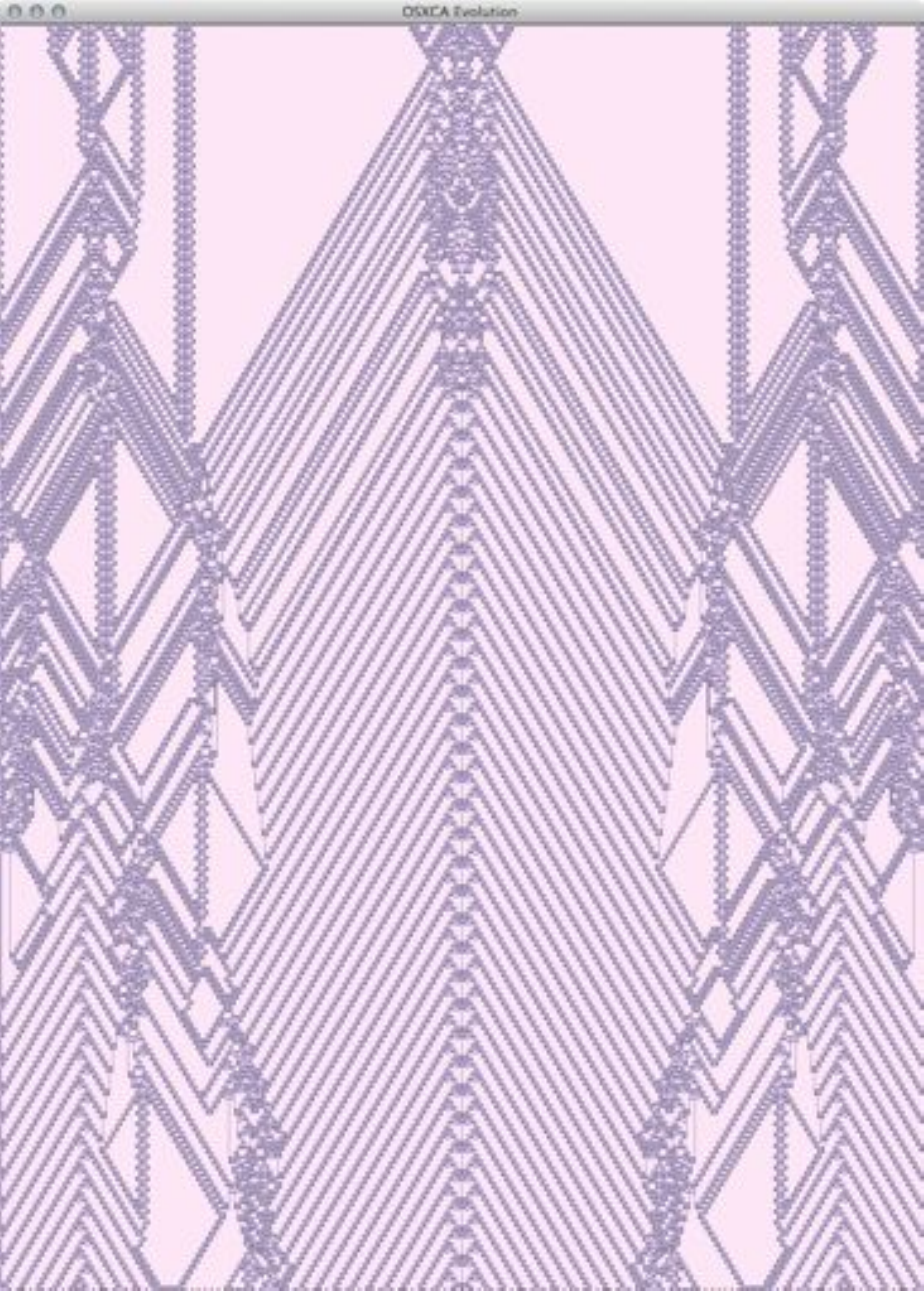}}
\caption{Continued evolution to 5780 steps.}
\label{126maj4_843cells_5780gen}
\end{figure}

Figure~\ref{majMemory} illustrates three different kinds of dynamics emerging in ECAM rule 126, for some values of $\tau$.\footnote{Evolutions of $\phi_{R126maj:\tau}$ were calculated with {\it OSXLCAU21 system} available in \url{http://uncomp.uwe.ac.uk/genaro/OSXCASystems.html}} Exploring different values of $\tau$, we found that large odd values of $\tau$ tend to define {\it macrocells}-like patterns [Wolfram, 1994], [McIntosh, 2009], while even values are responsible for a mixture of periodic and chaotic dynamics. Figure~\ref{majMemory}(a)i illustrates large periodic regions with few complex patterns traveling isolation  developed by function $\phi_{R126min:3}$. Figure~\ref{majMemory}(b) shows the function $\phi_{R126par:2}$, its evolution is more interesting because we can see the emergence of some complex patterns than also interact producing other types of complex structures, including mobile self-localizations or gliders. By exploring systematically distinct values of $\tau$, we found that $\phi_{R126maj:4}$ produces an impressive and non-trivial emergence of patterns traveling and colliding. Fig.~\ref{majMemory}(c) shows the most interesting evolution with well defined complex patterns, not just mobile self-localisations but also the emergence of glider guns, they are complex patterns which travel on the evolution space emitting periodically another kind of gliders.

An interesting evolution is starting with a single non-quiescent cell. Particularly, $\phi_{R126maj:4}$ displays  a growth complex behaviour. An example of this space-time configuration is given in Fig.~\ref{126maj4_843cells_1156gen} showing the first 1152 steps, where in this case the automaton needed other 30,000 steps to reach a stationary configuration. Filter is convenient to eliminate the non relevant information about gliders. At the same figure we can see a number of gliders, glider guns, still-life configurations, and a wide number of combinations of such patterns colliding and traveling with different velocities and densities. Consequently, we can classify a number of periodic structures, objects, and interesting reactions. For full details about ECAM $\phi_{R126maj:4}$ please see the paper [Mart{\'i}nez et al., 2010b]. Another case was presented with the ECA rule 30 in [Mart{\'i}nez et al., 2010a], and others ECA rules in [Mart{\'i}nez et al., 2012a].

By selecting a majority memory function on the chaotic ECA rule 126 we can transform its dynamics to complex dynamics. Thus, for some $CA$ rule with a memory $m$ function $\phi$ and value $\tau$ we can derive a complex system from a chaotic system or vice versa, transform a chaotic system to complex.

Further we explore systematically --- on the 88 equivalent ECA rules --- if memory functions are able to cover the Wolfram's classes transforming each class. This way, we prove experimentally that each class may {\it jump} to another class with a kind of memory. We  will also show that by selecting a memory we can reach any other class starting from any class. The full exploration is showed in the appendix~\ref{ECAMappendix}.

\section{Programming dynamics using memory}

There are varieties of CA classifications, including Wolfram's classes [Wolfram, 1983], intra- and inter-class connection probabilities [Li \& Packard, 1990], $\lambda$-parameter [Langton, 1986], classification by patterns [Aizawa \& Nishikawa, 1986], $Z$-parameter and attractors basin [Wuensche \& Lesser, 1992], [Wuensche, 1999], local structure approximation [Gutowitz et al., 1987], mean field and de Bruijn approximation [McIntosh, 1990], non-trivial collective behaviours [Chat\'e \& Manneville, 1992], glider classification [Eppstein, 1999], equivalence computation [Sutner, 2009], morphology-bases classification [Adamatzky et al., 2006], nonlinear dynamics [Chua, 2006], [Chua, 2007], [Chua, 2009], [Chua, 2011], [Chua, 2012], [Mainzer \& Chua, 2012], communication complexity [D\"urr et al., 2004], generative morphological diversity [Adamatzky \& Mart{\'i}nez, 2010], basis of lattice analysis [Gunji, 2010], genetic algorithms [Das et al., 1994], compression-based approach [Zenil, 2010], expressiveness (biodiversity) [Redeker et al., 2013], evolutionary computation [Wolz \& de Oliveira, 2008].

The present study with {\it memory function} opens a new and complementary properties on CA classes, producing a number of interesting properties.

\subsection{ECAM classification}
In this section, we propose a classification based in memory functions. This tables are published in [Mart{\'i}nez].

A ECAM is a ECA composed with a memory function, the new rule opens new and extended domain of rules based in the ECA domain [Mart{\'i}nez et al., 2010a].

To derive a new rule from a basic ECA rule one should select an ECA rule and compose this rule with a memory function (in our analysis we have considered three basic functions\,: majority, minority, and parity). Therefore, the memory function will determine if the original ECA rule preserves the same class (respective to Wolfram's classes) or if it changes to another class.

Following this simple principle, we know now that ECA rules composed with memory can be classified as follows:

\begin{itemize}
\item[] {\bf \em strong}, because the memory functions are unable to transform one class to another;
\item[] {\bf \em moderate}, because the memory function can transform the rule to another class and conserve the same class as well;
\item[] {\bf \em weak}, because the memory functions do most transformations and the rule changes to another different class quickly.
\end{itemize}

Table~\ref{ECAM_classes} presents the ECA classification based in memory functions. \\

\begin{table}[th]
\centering
\begin{tabular}{lcl}
  \hline\noalign{\smallskip}
  \multicolumn{3}{p{2cm}}{classification} \\
  \cline{1-2}\noalign{\smallskip}
  type & num. & rules \\
  \noalign{\smallskip}\hline\noalign{\smallskip}
  {\sf strong} & 39 & 2, 7, 9, 10, 11, 15, 18, 22, 24, 25, 26, 30, 34, \\
                   &      & 35, 41, 42, 45, 46, 54, 56, 57, 58, 62, 94, 106, \\
                   &      & 108, 110, 122, 126, 128, 130, 138, 146, 152, \\
                   &      & 154, 162, 170, 178, 184. \\
  {\sf moderate} & 34 & 1, 3, 4, 5, 6, 8, 13, 14, 27, 28, 29, 32, 33, 37, \\
                   &      & 38, 40, 43, 44, 72, 73, 74, 77, 78, 104, 132, \\
                   &      & 134, 136, 140, 142, 156, 160, 164, 168, 172. \\
  {\sf weak} & 15 & 0, 12, 19, 23, 36, 50, 51, 60, 76, 90, 105, 150, \\
                   &      & 200, 204, 232. \\
  \noalign{\smallskip}\hline
\end{tabular}
\caption{ECAM classification.}
\label{ECAM_classes}
\end{table}

\noindent {\bf Proposition 1.} Dynamics of CA from a `strong class' can be changed by {\bf any} memory function. \\

\noindent {\bf Proposition 2.} Dynamics of CA from `moderate class' can be changed by {\bf at least one} or more memory functions. \\

\noindent {\bf Proposition 3.} Dynamics of CA from `weak class' cannot be affected by {\bf none} kind of memories studied in present paper. \\

Memory classification presents a number of interesting properties.

We have ECA rules which composed with a particular kind of memory  are able to reach another class including the original dynamic. The main feature is that, at least, this rule with memory is able to reach every different class. Rules with this property are called {\it universal ECAM} (5 rules).

\begin{center}
    \begin{tabular}{ r p{9cm}}
    {\sf universal ECAM:} & 22, 54, 130, 146, 152.
    \end{tabular}
\end{center}

Particularly, all these UECAM are classified as {\sf strong} in ECAM's classification.

\begin{center}
    \begin{tabular}{ r p{9cm}}
    {\sf strong:} & 22, 54, 130, 146, 152. \\
    {\sf moderate:} & - \\
    {\sf weak:} & -
    \end{tabular}
\end{center}

On the other hand, we have ECA that when composed with memory are able to yield a complex ECAM but with elements of the original ECA rule. They are called {\it complex ECAM} (44 rules). Several of these complex rules are illustrated in Appendix~A.

\begin{center}
    \begin{tabular}{ r p{9cm}}
    {\sf complex ECAM:} & 6, 9, 10, 11, 13, 15, 22, 24, 25, 26, 27, 30, 33, 35, 38, 40, 41, 42, 44, 46, 54, 57, 58, 62, 72, 77, 78, 106, 108, 110, 122, 126, 130, 132, 138, 142, 146, 152, 156, 162, 170, 172, 178, 184.
    \end{tabular}
\end{center}

\noindent and they can be particularised in terms of ECAM's classification, as follows:

\begin{center}
    \begin{tabular}{ r p{9cm}}
    {\sf strong:} & 9, 10, 11, 15, 22, 24, 25, 26, 30, 35, 41, 42, 46, 54, 57, 58, 62, 106, 108, 110, 122, 126, 130, 138, 146, 152, 162, 170, 178, 184. \\
    {\sf moderate:} & 6, 13, 27, 33, 38, 40, 44, 72, 77, 78, 132, 142, 156, 172. \\
    {\sf weak:} & -
    \end{tabular}
\end{center}

It is remarkable that none of the rules classified in {\sf weak} class is able to reach complex behaviour. These set of rules are strongly robust to any perturbation in terms of ECAM's classification.

\subsection{ECA classifications versus ECAM classification}
In this instance, we will compare several ECA classifications reported in CA literature all along the CA-history versus memory classification.

\subsubsection{Wolfram's classification (1984)}

Wolfram's classification in ``Universality and complexity in cellular automata'', establishes four classes:

\begin{center}
\{{\sf uniform (class I), periodic (class II), chaotic (class III), complex (class IV)}\}
\end{center}

See details in [Wolfram, 1994], [Martin et al., 1984], [Wolfram, 2002].

\begin{center}
    \begin{tabular}{ r p{9cm}}
    {\bf class I:} & 0, 8, 32, 40, 128, 136, 160, 168. \\
    & \\
    {\sf strong:} & 128. \\
    {\sf moderate:} & 8, 32, 40, 136, 160, 168. \\
    {\sf weak:} & 0.
    \end{tabular}
\end{center}

\begin{center}
    \begin{tabular}{ r p{9cm}}
    {\bf class II:} & 1, 2, 3, 4, 5, 6, 7, 9, 10, 11, 12, 13, 14, 15, 19, 23, 24, 25, 26, 27, 28, 29, 33, 34, 35, 36, 37, 38, 42, 43, 44, 46, 50, 51, 56, 57, 58, 62, 72, 73, 74, 76, 77, 78, 94, 104, 108, 130, 132, 134, 138, 140, 142, 152, 154, 156, 162, 164, 170, 172, 178, 184, 200, 204, 232. \\
    & \\
    {\sf strong:} & 2, 7, 9, 10, 11, 15, 24, 25, 26, 34, 35, 42, 46, 56, 57, 58, 62, 94, 108, 130, 138, 152, 154, 162, 170, 178, 184. \\
    {\sf moderate:} & 1, 3, 4, 5, 6, 13, 14, 27, 28, 29, 33, 37, 38, 43, 44, 72, 73, 74, 77, 78, 104, 132, 134, 140, 142, 156, 164, 172. \\
    {\sf weak:} & 12, 19, 23, 36, 50, 51, 76, 200, 204, 232.
    \end{tabular}
\end{center}

\begin{center}
    \begin{tabular}{ r p{9cm}}
    {\bf class III:} & 18, 22, 30, 45, 60, 90, 105, 122, 126, 146, 150. \\
    & \\
    {\sf strong:} & 18, 22, 30, 45, 122, 126, 146. \\
    {\sf moderate:} & - \\
    {\sf weak:} & 60, 90, 105, 150.
    \end{tabular}
\end{center}

\begin{center}
    \begin{tabular}{ r p{9cm}}
    {\bf class IV:} & 41, 54, 106, 110. \\
    & \\
    {\sf strong:} & 41, 54, 106, 110. \\
    {\sf moderate:} & - \\
    {\sf weak:} & -
    \end{tabular}
\end{center}

\subsubsection{Li and Packard's classification (1990)}

Li and Packard's classification in ``The Structure of the Elementary Cellular Automata Rule Space'', establishes five ECA classes:

\begin{center}
\{{\sf null, fixed point, periodic, locally chaotic, chaotic}\}.
\end{center}

For details please see [Li \& Packard, 1990].

\begin{center}
    \begin{tabular}{ r p{9cm}}
    {\bf null:} & 0, 8, 32, 40, 128, 136, 160, 168. \\
    & \\
    {\sf strong:} & 128. \\
    {\sf moderate:} & 8, 32, 40, 136, 160, 168. \\
    {\sf weak:} & 0.
    \end{tabular}
\end{center}

\begin{center}
    \begin{tabular}{ r p{9cm}}
    {\bf fixed point:} & 2, 4, 10, 12, 13, 24, 34, 36, 42, 44, 46, 56, 57, 58, 72, 76, 77, 78, 104, 130, 132, 138, 140, 152, 162, 164, 170, 172, 184, 200, 204, 232. \\
    & \\
    {\sf strong:} & 2, 10, 24, 34, 42, 46, 56, 57, 58, 130, 138, 152, 162, 170, 184. \\
    {\sf moderate:} & 4, 13, 44, 72, 77, 78, 104, 132, 140, 164, 172. \\
    {\sf weak:} & 12, 36, 76, 200, 204, 232.
    \end{tabular}
\end{center}

\begin{center}
    \begin{tabular}{ r p{9cm}}
    {\bf periodic:} & 1, 3, 5, 6, 7, 9, 11, 14, 15, 19, 23, 25, 27, 28, 29, 33, 35, 37, 38, 41, 43, 50, 51, 74, 94, 108, 131(62), 134, 142, 156, 178. \\
    & \\
    {\sf strong:} & 7, 9, 11, 15, 25, 35, 41, 62, 94, 108, 178. \\
    {\sf moderate:} & 1, 3, 5, 6, 14, 27, 28, 29, 33, 37, 38, 43, 74, 134, 142, 156. \\
    {\sf weak:} & 19, 23, 50, 51.
    \end{tabular}
\end{center}

\begin{center}
    \begin{tabular}{ r p{9cm}}
    {\bf locally chaotic:} & 26, 73, 154. \\
    & \\
    {\sf strong:} & 26, 154. \\
    {\sf moderate:} & 73. \\
    {\sf weak:} & -
    \end{tabular}
\end{center}

\begin{center}
    \begin{tabular}{ r p{9cm}}
    {\bf chaotic:} & 18, 22, 30, 45, 54, 60, 90, 105, 106, 132, 129(126), 137(110), 146, 150, 161(122). \\
    & \\
    {\sf strong:} & 18, 22, 30, 45, 54, 106, 122, 126, 110, 122, 146. \\
    {\sf moderate:} & - \\
    {\sf weak:} & 60, 90, 105, 150.
    \end{tabular}
\end{center}

\subsubsection{Wuensche's equivalences (1992)}

Wuensche in ``The Global Dynamics of Cellular Automata'', establishes three ECA kinds of symmetries:

\begin{center}
\{{\sf symmetric, semi-asymmetric, full-asymmetric}\}.
\end{center}

For details please see [Wuensche \& Lesser, 1992].

\begin{center}
    \begin{tabular}{ r p{9cm}}
    {\bf symmetric:} & 0, 1, 4, 5, 18, 19, 22, 23, 32, 33, 36, 37, 50, 51, 54, 72, 73, 76, 77, 90, 94, 104, 105, 108, 122, 126, 128, 132, 146, 150, 160, 164, 178, 200, 204, 232. \\
    & \\
    {\sf strong:} & 18, 22, 54, 108, 122, 126, 128, 146, 178. \\
    {\sf moderate:} & 1, 4, 5, 32, 33, 72, 73, 77, 104, 132, 160, 164. \\
    {\sf weak:} & 0, 19, 23, 36, 50, 51, 76, 90, 105, 150, 200, 204.
    \end{tabular}
\end{center}

\begin{center}
    \begin{tabular}{ r p{8cm}}
    {\bf semi-asymmetric:} & 2, 3, 6, 7, 8, 9, 12, 13, 26, 27, 30, 34, 35, 38, 40, 41, 44, 45, 58, 62, 74, 78, 106, 110, 130, 134, 136, 140, 154, 162, 168, 172. \\
    & \\
    {\sf strong:} & 2, 7, 9, 26, 30, 34, 35, 41, 45, 58, 62, 106, 110, 130, 154, 162. \\
    {\sf moderate:} & 3, 6, 8, 13, 27, 38, 40, 44, 74, 78, 134, 136, 140, 168, 172. \\
    {\sf weak:} & 12.
    \end{tabular}
\end{center}

\begin{center}
    \begin{tabular}{ r p{8cm}}
    {\bf full-asymmetric:} & 10, 11, 14, 15, 24, 25, 28, 29, 42, 43, 46, 57, 60, 138, 142, 152, 156, 170, 184. \\
    & \\
    {\sf strong:} & 10, 11, 15, 24, 25, 42, 46, 57, 138, 152, 170, 184. \\
    {\sf moderate:} & 14, 28, 29, 43, 142, 156. \\
    {\sf weak:} & 60.
    \end{tabular}
\end{center}

Also, Wuensche classifies a set of ``maximally chaotic'' rules or known as ``chain rules'' (for details please see [Wuensche, 1999]).

\begin{center}
    \begin{tabular}{ r p{9cm}}
    {\bf chain rules:} & 30, 45, 106, 154. \\
    & \\
    {\sf strong:} & 30, 45, 106, 154. \\
    {\sf moderate:} & - \\
    {\sf weak:} & -
    \end{tabular}
\end{center}

\subsubsection{Index complexity classification (2002)}

Index complexity in ``A Nonlinear Dynamics Perspective of WolframÕs New Kind of Science. Part I: Threshold of Complexity'', establishes three ECA classes:

\begin{center}
\{{\sf red ($k=1$), blue ($k=2$), green ($k=3$)}\}.
\end{center}

For details please see [Chua et al., 2002].

\begin{center}
    \begin{tabular}{ r p{9cm}}
    {\bf red ($k=1$):} & 0, 1, 2, 3, 4, 5, 7, 8, 10, 11, 12, 13, 14, 15, 19, 23, 32, 34, 35, 42, 43, 50, 51, 76, 77, 128, 136, 138, 140, 142, 160, 162, 168, 170, 178, 200, 204, 232. \\
    & \\
    {\sf strong:} & 2, 7, 10, 11, 15, 34, 35, 42, 128, 138, 162, 170, 178. \\
    {\sf moderate:} & 1, 3, 4, 5, 8, 13, 14, 32, 43, 77, 136, 140, 142, 160, 168. \\
    {\sf weak:} & 0, 12, 19, 23, 50, 51, 76, 200, 204, 232.
    \end{tabular}
\end{center}

\begin{center}
    \begin{tabular}{ r p{9cm}}
    {\bf blue ($k=2$):} & 6, 9, 18, 22, 24, 25, 26, 28, 30, 33, 36, 37, 38, 40, 41, 44, 45, 54, 56, 57, 60, 62, 72, 73, 74, 90, 94, 104, 106, 108, 110, 122, 126, 130, 132, 134, 146, 152, 154, 156, 164. \\
    & \\
    {\sf strong:} & 9, 18, 22, 24, 25, 26, 30, 41, 45, 54, 56, 57, 62, 94, 106, 108, 110, 122, 126, 130, 146, 152, 154. \\
    {\sf moderate:} & 6, 28, 33, 37, 38, 40, 44, 72, 73, 74, 104, 132, 134, 156, 164. \\
    {\sf weak:} & 36, 60, 90.
    \end{tabular}
\end{center}

\begin{center}
    \begin{tabular}{ r p{9cm}}
    {\bf green ($k=3$):} & 27, 29, 46, 58, 78, 105, 150, 172, 184. \\
    & \\
    {\sf strong:} & 46, 58, 184. \\
    {\sf moderate:} & 27, 29, 78, 172. \\
    {\sf weak:} & 105, 150.
    \end{tabular}
\end{center}

\subsubsection{Density parameter with $d$-spectrum classification (2003)}

Density parameter with $d$-spectrum in ``Experimental Study of Elementary Cellular Automata Dynamics Using the Density Parameter'', establishes three ECA classes:

\begin{center}
\{{\sf P, H, C}\}.
\end{center}

For details please see [Fat\`{e}s, 2003].

\begin{center}
    \begin{tabular}{ r p{9cm}}
    {\bf P:} & 0, 1, 2, 3, 4, 5, 6, 7, 8, 9, 10, 11, 12, 13, 14, 15, 19, 23, 24, 25, 27, 28, 29, 32, 33, 34, 35, 36, 37, 38, 40, 42, 43, 44, 50, 51, 56, 57, 58, 62, 72, 74, 76, 77, 78, 104, 108, 128, 130, 132, 134, 136, 138, 140, 142, 152, 156, 160, 162, 164, 168, 170, 172, 178, 184, 200, 204, 232. \\
    & \\
    {\sf strong:} & 2, 7, 9, 10, 11, 15, 24, 25, 34, 35, 42, 56, 57, 58, 62, 108, 128, 130, 138, 152, 162, 170, 178, 184. \\
    {\sf moderate:} & 1, 3, 4, 5, 6, 8, 13, 14, 27, 28, 29, 32, 33, 37, 38, 40, 43, 44, 72, 74, 77, 78, 104, 132, 134, 136, 140, 142, 156, 160, 164, 168, 172. \\
    {\sf weak:} & 0, 12, 19, 23, 36, 50, 51, 76, 200, 204, 232.
    \end{tabular}
\end{center}

\begin{center}
    \begin{tabular}{ r p{9cm}}
    {\bf H:} & 26, 41, 54, 73, 94, 110, 154. \\
    & \\
    {\sf strong:} & 26, 41, 94, 110, 154. \\
    {\sf moderate:} & 73. \\
    {\sf weak:} & -
    \end{tabular}
\end{center}

\begin{center}
    \begin{tabular}{ r p{9cm}}
    {\bf C:} & 18, 22, 30, 45, 60, 90, 105, 106, 122, 126, 146, 150. \\
    & \\
    {\sf strong:} & 18, 22, 30, 45, 106, 122, 126, 146. \\
    {\sf moderate:} & - \\
    {\sf weak:} & 60, 90, 105, 150.
    \end{tabular}
\end{center}

\subsubsection{Communication complexity classification (2004)}

Communication complexity classification in ``Cellular Automata and Communication Complexity'' establishes three ECA classes:

\begin{center}
\{{\sf bounded, linear, other}\}.
\end{center}

For details see [D\"urr et al., 2004].

\begin{center}
    \begin{tabular}{ r p{9cm}}
    {\bf bounded:} & 0, 1, 2, 3, 4, 5, 7, 8, 10, 12, 13, 15, 19, 24, 27, 28, 29, 32, 34, 36, 38, 42, 46, 51, 60, 71(29), 72, 76, 78, 90, 105, 108, 128, 130, 136, 138, 140, 150, 154, 156, 160, 162(missing), 170, 172, 200, 204. \\
    & \\
    {\sf strong:} & 2, 7, 10, 24, 34, 42, 46, 108, 128, 130, 138, 154, 162, 170. \\
    {\sf moderate:} & 1, 3, 4, 5, 8, 13, 15, 27, 28, 29, 32, 38, 72, 78, 136, 140, 156, 160, 172. \\
    {\sf weak:} & 0, 12, 19, 36, 51, 60, 76, 90, 105, 150, 200, 204.
    \end{tabular}
\end{center}

\begin{center}
    \begin{tabular}{ r p{9cm}}
    {\bf linear:} & 11, 14, 23, 33, 35, 43, 44, 50, 56, 58, 77, 132, 142, 152, 168, 178, 184, 232. \\
    & \\
    {\sf strong:} & 11, 35, 56, 58, 152, 178, 184. \\
    {\sf moderate:} & 14, 33, 43, 44, 77, 132, 142, 168. \\
    {\sf weak:} & 23, 50, 232.
    \end{tabular}
\end{center}

\begin{center}
    \begin{tabular}{ r p{9cm}}
    {\bf other:} & 6, 9, 18, 22, 25, 26, 30, 37, 40, 41, 45, 54, 57, 62, 73, 74, 94, 104, 106, 110, 122, 126, 134, 146, 164. \\
    & \\
    {\sf strong:} & 9, 18, 22, 25, 26, 30, 41, 45, 54, 57, 62, 94, 106, 110, 122, 126, 146. \\
    {\sf moderate:} & 6, 37, 40, 73, 74, 104, 134, 164. \\
    {\sf weak:} & -
    \end{tabular}
\end{center}

Additionally, bound class can be refined in other four subclasses.

\begin{center}
    \begin{tabular}{ r p{6cm}}
    {\bf bounded by additivity:} & 15, 51, 60, 90, 105, 108, 128, 136, 150, 160, 170, 204. \\
    & \\
    {\sf strong:} & 15, 51, 108, 128, 170. \\
    {\sf moderate:} & 136, 160. \\
    {\sf weak:} & 60, 90, 105, 150, 204.
    \end{tabular}
\end{center}

\begin{center}
    \begin{tabular}{ r p{4cm}}
    {\bf bounded by limited sensibility:} & 0, 1, 2, 3, 4, 5, 8, 10, 12, 19, 24, 29, 34, 36, 38, 42, 46, 72, 76, 78, 108, 138, 200. \\
    & \\
    {\sf strong:} & 2, 10, 24, 34, 42, 46, 108, 138. \\
    {\sf moderate:} & 1, 3, 4, 5, 8, 29, 38, 72, 78. \\
    {\sf weak:} & 0, 12, 19, 36, 76, 200.
    \end{tabular}
\end{center}

\begin{center}
    \begin{tabular}{ r p{6cm}}
    {\bf bounded by half-limited sensibility:} & 7, 13, 28, 140, 172. \\
    & \\
    {\sf strong:} & 7. \\
    {\sf moderate:} & 13, 28, 140, 172. \\
    {\sf weak:} & -
    \end{tabular}
\end{center}

\begin{center}
    \begin{tabular}{ r p{9cm}}
    {\bf bounded for any other reason:} & 27, 32, 130, 156, 162. \\
    & \\
    {\sf strong:} & 130, 162. \\
    {\sf moderate:} & 27, 32, 156. \\
    {\sf weak:} & -
    \end{tabular}
\end{center}

\subsubsection{Topological classification (2007)}

Topological classification in ``A Nonlinear Dynamics Perspective of Wolfram's New Kind of Science. Part VII: Isles of Eden'', establishes six ECA classes: 

\begin{center}
\{{\sf period-1, period-2, period-3, Bernoulli $\sigma_t$-shift, complex Bernoulli-shift, hyper Bernoully-shift}\}.
\end{center}

For details please see [Chua et al., 2007].

\begin{center}
    \begin{tabular}{ r p{9cm}}
    {\bf period-1:} & 0, 4, 8, 12, 13, 32, 36, 40, 44, 72, 76, 77, 78, 94, 104, 128, 132, 136, 140, 160, 164, 168, 172, 200, 204, 232. \\
    & \\
    {\sf strong:} & 94, 128. \\
    {\sf moderate:} & 4, 8, 13, 32, 40, 44, 72, 77, 78, 104, 132, 136, 140, 160, 164, 168, 172. \\
    {\sf weak:} & 0, 12, 36, 76, 200, 204, 232.
    \end{tabular}
\end{center}

\begin{center}
    \begin{tabular}{ r p{9cm}}
    {\bf period-2:} & 1, 5, 19, 23, 28, 29, 33, 37, 50, 51, 108, 156, 178. \\
    & \\
    {\sf strong:} & 108, 178. \\
    {\sf moderate:} & 1, 5, 28, 29, 33, 37, 156. \\
    {\sf weak:} & 19, 23, 50, 51.
    \end{tabular}
\end{center}

\begin{center}
    \begin{tabular}{ r p{9cm}}
    {\bf period-3:} & 62. \\
    & \\
    {\sf strong:} & 62. \\
    {\sf moderate:} & - \\
    {\sf weak:} & -
    \end{tabular}
\end{center}

\begin{center}
    \begin{tabular}{ r p{7cm}}
    {\bf Bernoulli $\sigma_t$-shift:} & 2, 3, 6 , 7, 9, 10, 11, 14, 15, 24, 25, 27, 34, 35, 38, 42, 43, 46, 56, 57, 58, 74, 130, 134, 138, 142, 152, 162, 170, 184. \\
    & \\
    {\sf strong:} & 2, 7, 9, 10, 11, 15, 24, 25, 34, 35, 42, 46, 56, 57, 58, 130, 138, 152, 162, 170, 184. \\
    {\sf moderate:} & 3, 6, 14, 27, 38, 43, 74, 134, 142. \\
    {\sf weak:} & -
    \end{tabular}
\end{center}

\begin{center}
    \begin{tabular}{ r p{9cm}}
    {\bf complex Bernoulli-shift:} & 18, 22, 54, 73, 90, 105, 122, 126, 146, 150. \\
    & \\
    {\sf strong:} & 18, 22, 122, 126, 146. \\
    {\sf moderate:} & 73. \\
    {\sf weak:} & 90, 105, 150.
    \end{tabular}
\end{center}

\begin{center}
    \begin{tabular}{ r p{9cm}}
    {\bf hyper Bernoully-shift:} & 26, 30, 41, 45, 60, 106, 110, 154. \\
    & \\
    {\sf strong:} & 26, 30, 41, 45, 110, 154. \\
    {\sf moderate:} & - \\
    {\sf weak:} & 60.
    \end{tabular}
\end{center}

\subsubsection{Power spectral classification (2008)}

Power spectral classification in ``Power Spectral Analysis of Elementary Cellular Automata'', establishes four ECA classes: 

\begin{center}
\{{\sf category 1: extremely low power density, category 2: broad-band noise, category 3: power law spectrum, exceptional rules}\}.
\end{center}

For details please see [Ninagawa, 2008].

\begin{center}
    \begin{tabular}{ r p{4cm}}
    {\bf category 1  extremely low power density:} & 0, 1, 4, 5, 8, 12, 13, 19, 23, 26, 28, 29, 33, 37, 40, 44, 50, 51, 72, 76, 77, 78, 104, 128, 132, 133(94), 136, 140, 156, 160, 164, 168, 172, 178, 200, 232. \\
    & \\
    {\sf strong:} & 26, 94, 128, 178. \\
    {\sf moderate:} & 1, 4, 5, 8, 13, 28, 29, 33, 37, 40, 44, 72, 77, 78, 104, 132, 136, 140, 156, 160, 164, 168, 172. \\
    {\sf weak:} & 0, 12, 19, 23, 50, 51, 76, 200, 232.
    \end{tabular}
\end{center}

\begin{center}
    \begin{tabular}{ r p{6cm}}
    {\bf category 2 broad-band noise:} & 2, 3, 6, 7, 9, 10, 11, 14, 15, 18, 22, 24, 25, 26, 27, 30, 34, 35, 38, 41, 42, 43, 45, 46, 56, 57, 58, 60, 74, 90, 105, 106, 129(126), 130, 134, 138, 142, 146, 150, 152, 154, 161(122), 162, 170, 184. \\
    & \\
    {\sf strong:} & 2, 7, 9, 10, 11, 15, 18, 22, 24, 25, 26, 30, 34, 35, 41, 42, 45, 46, 56, 57, 58, 106, 122, 126, 130, 138, 146, 152, 154, 162, 170, 184. \\
    {\sf moderate:} & 3, 6, 14, 27, 38, 43, 74, 134, 142, . \\
    {\sf weak:} & 60, 90, 105, 150.
    \end{tabular}
\end{center}

\begin{center}
    \begin{tabular}{ r p{9cm}}
    {\bf category 3 power law spectrum:} & 54, 62, 110. \\
    & \\
    {\sf strong:} & 54, 62, 110. \\
    {\sf moderate:} & . \\
    {\sf weak:} & .
    \end{tabular}
\end{center}

\begin{center}
    \begin{tabular}{ r p{9cm}}
    {\bf exceptional rules:} & 73, 204. \\
    & \\
    {\sf strong:} & - \\
    {\sf moderate:} & 73. \\
    {\sf weak:} & 204.
    \end{tabular}
\end{center}

\subsubsection{Morphological diversity classification (2010)}

Morphological diversity classification in ``On Generative Morphological Diversity of Elementary Cellular Automata'', establishes five ECA classes:

\begin{center}
\{{\sf chaotic, complex, periodic, two-cycle, fixed point, null}\}.
\end{center}

See details in [Adamatzky \& Mart{\'i}nez, 2010].

\begin{center}
    \begin{tabular}{ r p{9cm}}
    {\bf chaotic:} & 2, 10, 18, 22, 24, 26, 30, 34, 42, 45, 56, 60, 73, 74, 90, 94, 105, 106, 126, 130, 138, 150, 152, 154, 161(122), 162, 170, 184. \\
    & \\
    {\sf strong:} & 2, 10, 18, 22, 24, 26, 30, 34, 42, 56, 94, 106, 122, 126, 130, 138, 152, 154, 162, 170, 184. \\
    {\sf moderate:} & 73, 74. \\
    {\sf weak:} & 60, 90, 105, 150.
    \end{tabular}
\end{center}

\begin{center}
    \begin{tabular}{ r p{9cm}}
    {\bf complex:} & 54, 110. \\
    & \\
    {\sf strong:} & 54, 110. \\
    {\sf moderate:} & - \\
    {\sf weak:} & -
    \end{tabular}
\end{center}

\begin{center}
    \begin{tabular}{ r p{9cm}}
    {\bf periodic:} & 18, 26, 60, 90, 94, 154. \\
    & \\
    {\sf strong:} & 18, 26, 94, 154. \\
    {\sf moderate:} & - \\
    {\sf weak:} & 60, 90.
    \end{tabular}
\end{center}

\begin{center}
    \begin{tabular}{ r p{9cm}}
    {\bf two-cycle:} & 1, 2, 3, 4, 5, 6, 7, 9, 10, 11, 12, 13, 14, 15, 19, 23, 24, 25, 27, 28, 29, 33, 34, 35, 36, 37, 38, 42, 43, 44, 46, 50, 51, 56, 58, 74, 76, 106, 108, 130, 132, 134, 138, 140, 142, 152, 156, 162, 164, 170, 172, 178, 184, 204. \\
    & \\
    {\sf strong:} & 2, 7, 9, 10, 11, 15, 24, 25, 34, 35, 42, 46, 56, 58, 106, 108, 130, 138, 152, 162, 170, 178, 184. \\
    {\sf moderate:} & 1, 3, 4, 5, 6, 13, 14, 27, 28, 29, 33, 37, 38, 43, 44, 74, 132, 134, 140, 142, 156, 164, 172. \\
    {\sf weak:} & 12, 19, 23, 36, 50, 51, 76, 204.
    \end{tabular}
\end{center}

\begin{center}
    \begin{tabular}{ r p{9cm}}
    {\bf fixed point:} & 0, 2, 4, 8, 10, 11, 12, 13, 14, 24, 32, 34, 36, 40, 42, 43, 44, 46, 50, 56, 57, 58, 72, 74, 76, 77, 78, 104, 106, 108, 128, 130, 132, 136, 138, 140, 142, 152, 160, 162, 164, 168, 170, 172, 178, 184, 200, 204, 232. \\
    & \\
    {\sf strong:} & 2, 10, 11, 24, 34, 42, 46, 56, 57, 58, 106, 108, 128, 130, 138, 152, 162, 170, 178, 184. \\
    {\sf moderate:} & 4, 8, 13, 14, 32, 40, 43, 44, 72, 74, 77, 78, 104, 132, 136, 140, 142, 160, 164, 168, 172. \\
    {\sf weak:} & 0, 12, 36, 50, 76, 200, 204, 232.
    \end{tabular}
\end{center}

\begin{center}
    \begin{tabular}{ r p{9cm}}
    {\bf null:} & 0, 8, 32, 40, 72, 104, 128, 136, 160, 168, 200, 232. \\
    & \\
    {\sf strong:} & 128. \\
    {\sf moderate:} & 8, 32, 40, 72, 104, 136, 160, 168. \\
    {\sf weak:} & 0, 200, 232.
    \end{tabular}
\end{center}

\subsubsection{Distributive and non-distributive lattices classification (2010)}

Distributive and non-distributive lattices classification in ``Inducing Class 4 Behavior on the Basis of Lattice Analysis'', establishes four ECA classes:

\begin{center}
\{{\sf class 1, class 2, class 3, class 4}\}.
\end{center}

See details in [Gunji, 2010].

\begin{center}
    \begin{tabular}{ r p{9cm}}
    {\bf class 1:} & 0, 32, 128, 160, 250(160), 254(128). \\
    & \\
    {\sf strong:} & 128. \\
    {\sf moderate:} & 32, 160. \\
    {\sf weak:} & 0.
    \end{tabular}
\end{center}

\begin{center}
    \begin{tabular}{ r p{9cm}}
    {\bf class 2:} & 4, 36, 50, 72, 76, 94, 104, 108, 132, 164, 178, 200, 204, 218(164), 232, 236(200). \\
    & \\
    {\sf strong:} & 94, 108, 178. \\
    {\sf moderate:} & 4, 72, 104, 132, 164. \\
    {\sf weak:} & 36, 50, 76, 200, 204, 232.
    \end{tabular}
\end{center}

\begin{center}
    \begin{tabular}{ r p{9cm}}
    {\bf class 3:} & 18, 22, 54, 122, 126, 146, 150, 182(146). \\
    & \\
    {\sf strong:} & 18, 22, 54, 122, 126, 146. \\
    {\sf moderate:} & - \\
    {\sf weak:} & 150.
    \end{tabular}
\end{center}

\begin{center}
    \begin{tabular}{ r p{9cm}}
    {\bf class 4:} & 110, 124(110), 137(110), 193(110). \\
    & \\
    {\sf strong:} & 110. \\
    {\sf moderate:} & . \\
    {\sf weak:} & .
    \end{tabular}
\end{center}

\subsubsection{Topological dynamics classification (2012)}

Topological classification in ``A Full Computation-Relevant Topological Dynamics Classification of Elementary Cellular Automata'', establishes four ECA classes: 

\begin{center}
\{{\sf equicontinuous, almost-equicontinuous, sensitive, sensitive positively expansive}\}.
\end{center}

See details in [Sch\"ule \& Stoop, 2012], [Cattaneo et al., 2000].

\begin{center}
    \begin{tabular}{ r p{9cm}}
    {\bf equicontinuous:} & 0, 1, 4, 5, 8, 12, 19, 29, 36, 51, 72, 76, 108, 200, 204. \\
    & \\
    {\sf strong:} & 108. \\
    {\sf moderate:} & 1, 4, 5, 8, 29, 72. \\
    {\sf weak:} & 0, 12, 19, 36, 51, 76, 200, 204.
    \end{tabular}
\end{center}

\begin{center}
    \begin{tabular}{ r p{6cm}}
    {\bf almost-equicontinuous:} & 13, 23, 28, 32, 33, 40, 44, 50, 73, 77, 78, 94, 104, 128, 132, 136, 140, 156, 160, 164, 168, 172, 178, 232. \\
    & \\
    {\sf strong:} & 94, 128, 178. \\
    {\sf moderate:} & 13, 28, 32, 40, 73, 77, 78, 104, 132, 136, 140, 156, 160, 164, 168, 172. \\
    {\sf weak:} & 23, 50, 232.
    \end{tabular}
\end{center}

\begin{center}
    \begin{tabular}{ r p{9cm}}
    {\bf sensitive:} & 2, 3, 6, 7, 9, 10, 11, 14, 15, 18, 22, 24, 25, 26, 27, 30, 34, 35, 37, 38, 41, 42, 43, 45, 46, 54, 56, 57, 58, 60, 62, 74, 106, 110, 122, 126, 130, 134, 138, 142, 146, 152, 154, 162, 170, 184. \\
    & \\
    {\sf strong:} & 2, 7, 9, 10, 11, 15, 18, 22, 24, 25, 26, 30, 34, 35, 41, 42, 45, 46, 54, 56, 57, 58, 62, 106, 110, 122, 126, 130, 138, 146, 152, 154, 162, 170, 184. \\
    {\sf moderate:} & 3, 6, 14, 27, 37, 38, 43, 74, 134, 142. \\
    {\sf weak:} & 60.
    \end{tabular}
\end{center}

\begin{center}
    \begin{tabular}{ r p{9cm}}
    {\bf sensitive positively expansive:} & 90, 105, 150. \\
    & \\
    {\sf strong:} & - \\
    {\sf moderate:} & - \\
    {\sf weak:} & 90, 105, 150.
    \end{tabular}
\end{center}

Also, this classification can be refined into three sub-classes: weakly periodic, surjective, and chaotic (in the sense of Denavey).

\begin{center}
    \begin{tabular}{ r p{9cm}}
    {\bf weakly periodic:} & 2, 3, 10, 15, 24, 34, 38, 42, 46, 138, 170. \\
    & \\
    {\sf strong:} & 2, 10, 15, 24, 34, 42, 46, 138, 170. \\
    {\sf moderate:} & 3, 38. \\
    {\sf weak:} & -
    \end{tabular}
\end{center}

\begin{center}
    \begin{tabular}{ r p{9cm}}
    {\bf surjective:} & 15, 30, 45, 51, 60, 90, 105, 106, 150, 154, 170, 204. \\
    & \\
    {\sf strong:} & 15, 30, 45, 154, 170. \\
    {\sf moderate:} & -  \\
    {\sf weak:} & 51, 60, 90, 105, 150, 204.
    \end{tabular}
\end{center}

\begin{center}
    \begin{tabular}{ r p{5cm}}
    {\bf chaotic (in the sense of Denavey):} & 15, 30, 45, 60, 90, 105, 106, 150, 154, 170. \\
    & \\
    {\sf strong:} & 15, 30, 45, 106, 154, 170. \\
    {\sf moderate:} & - \\
    {\sf weak:} & 60, 90, 105, 150.
    \end{tabular}
\end{center}

\subsubsection{Expressivity analysis (2013)}

This is a classification by the evolution of a configuration consisting of an isolated one surrounded by zeros, that is a bit different from conventional ECA classifications previously displayed. In ``Expressiveness of Elementary Cellular Automata'', we can see five ECA kinds of expressivity:

\begin{center}
\{{\sf 0, periodic patterns, complex, Sierpinski patterns, finite growth}\}.
\end{center}

See details in [Redeker et al., 2013].

\begin{center}
    \begin{tabular}{ r p{9cm}}
    {\bf 0:} & 0, 7, 8, 19, 23, 31, 32, 40, 55, 63, 72, 104, 127, 128, 136, 160, 168, 200, 232. \\
    & \\
    {\sf strong:} & 7, 128. \\
    {\sf moderate:} & 8, 32, 40, 72, 104, 136, 160, 168. \\
    {\sf weak:} & 0, 19, 23, 200, 232.
    \end{tabular}
\end{center}

\begin{center}
    \begin{tabular}{ r p{8cm}}
    {\bf periodic patterns:} & 13, 28, 50, 54, 57, 58, 62, 77, 78, 94, 99, 109, 122, 156, 178. \\
    & \\
    {\sf strong:} & 54, 57, 58, 62, 94, 122, 178. \\
    {\sf moderate:} & 13, 28, 73, 77, 78, 156. \\
    {\sf weak:} & 50.
    \end{tabular}
\end{center}

\begin{center}
    \begin{tabular}{ r p{9cm}}
    {\bf complex:} & 30, 45, 73, 75, 110. \\
    & \\
    {\sf strong:} & 30, 45, 110. \\
    {\sf moderate:} & 73. \\
    {\sf weak:} & -
    \end{tabular}
\end{center}

\begin{center}
    \begin{tabular}{ r p{9cm}}
    {\bf Sierpinski patterns:} & 18, 22, 26, 60, 90, 105, 126, 146, 150, 154. \\
    & \\
    {\sf strong:} & 18, 22, 26, 126, 146, 154. \\
    {\sf moderate:} & - \\
    {\sf weak:} & 60, 90, 105, 150.
    \end{tabular}
\end{center}

\begin{center}
    \begin{tabular}{ r p{9cm}}
    {\bf finite growth:} & 1, 2, 3, 4, 5, 6, 9, 10, 11, 12, 14, 15, 24, 25, 27, 29, 33, 34, 35, 36, 37, 38, 39, 41, 42, 43, 44, 46, 47, 51, 56, 59, 71, 74, 76, 103, 106, 107, 108, 111, 130, 132, 134, 138, 140, 142, 152, 162, 164, 170, 172, 184, 204. \\
    & \\
    {\sf strong:} & 2, 9, 10, 11, 15, 24, 25, 34, 35, 41, 42, 46, 56, 106, 108, 130, 152, 162, 170, 184. \\
    {\sf moderate:} & 1, 3, 4, 5, 6, 14, 27, 29, 33, 37, 38, 43, 44, 74, 140, 142, 164, 172. \\
    {\sf weak:} & 12, 36, 51, 76, 204.
    \end{tabular}
\end{center}

\subsubsection{Normalised compression classification (2013)}

Normalised compression classification in ``Asymptotic Behaviour and Ratios of Complexity in Cellular Automata Rule Spaces'', establishes two ECA classes:

\begin{center}
\{{\sf $C_{1,2}$, $C_{3,4}$}\}.
\end{center}

See details in [Zenil \& Zapata].

\begin{center}
    \begin{tabular}{ r p{9cm}}
    {\bf $C_{1,2}$:} & 0, 1, 2, 3, 4, 5, 6, 7, 8, 9, 10, 11, 12, 13, 14, 15, 19, 23, 24, 25, 26, 27, 28, 29, 32, 33, 34, 35, 36, 37, 38, 40, 42, 43, 44, 46, 50, 51, 56, 57, 58, 72, 74, 76, 77, 78, 104, 108, 128, 130, 132, 134, 136, 138, 140, 142, 152, 154, 156, 160, 162, 164, 168, 170, 172, 178, 184, 200, 204, 232. \\
    & \\
    {\sf strong:} & 2, 7, 9, 10, 11, 15, 24, 25, 26, 34, 35, 42, 46, 56, 57, 58, 108, 128, 130, 138, 152, 154, 170, 178, 184. \\
    {\sf moderate:} & 1, 3, 4, 5, 6, 8, 13, 14, 27, 28, 29, 32, 33, 37, 38, 40, 43, 44, 72, 74, 77, 78, 104, 132, 134, 136, 140, 142, 156, 160. \\
    {\sf weak:} & 0, 12, 19, 23, 36, 50, 51, 76, 200, 204, 232.
    \end{tabular}
\end{center}

\begin{center}
    \begin{tabular}{ r p{9cm}}
    {\bf $C_{3,4}$:} & 18, 22, 30, 41, 45, 54, 60, 62, 73, 90, 94, 105, 106, 110, 122, 126, 146, 150. \\
    & \\
    {\sf strong:} & 18, 22, 30, 41, 45, 54, 62, 94, 106, 110, 122, 126, 146. \\
    {\sf moderate:} & 73. \\
    {\sf weak:} & 60, 90, 105, 150.
    \end{tabular}
\end{center}

\subsubsection{Surface dynamics classification (2013)}

Expressivity classification in ``Emergence of Surface Dynamics in Elementary Cellular Automata'', establishes three ECA classes: 

\begin{center}
\{{\sf type A, type B, type C}\}.
\end{center}

See details in [Mora et al.].

\begin{center}
    \begin{tabular}{ r p{9cm}}
    {\bf type A:} & 0, 1, 2, 3, 4, 5, 6, 7, 8, 9, 10, 11, 12, 13, 14, 15, 19, 23, 24, 25, 27, 28, 29, 32, 33, 34, 35, 36, 37, 38, 40, 42, 43, 44, 46, 50, 51, 56, 57, 58, 72, 74, 76, 77, 78, 94, 104, 108, 128, 130, 132, 134, 136, 138, 140, 142, 152, 156, 160, 162, 164, 168, 170, 172, 178, 184, 200, 204, 232. \\
    & \\
    {\sf strong:} & 2, 7, 9, 10, 11, 15, 24, 25, 34, 35, 42, 46, 56, 57, 58, 128, 130, 152, 170, 178, 184. \\
    {\sf moderate:} & 1, 3, 4, 5, 6, 8, 13, 14, 27, 28, 29, 32, 33, 37, 38, 40, 43, 44, 72, 74, 77, 78, 104, 108, 132, 134, 136, 140, 142, 156, 160, 164, 168, 172. \\
    {\sf weak:} & 0, 12, 19, 23, 36, 50, 51, 76, 200, 204, 232.
    \end{tabular}
\end{center}

\begin{center}
    \begin{tabular}{ r p{9cm}}
    {\bf type B:} & 18, 22, 26, 30, 41, 45, 60, 90, 105, 106, 122, 126, 146, 150, 154. \\
    & \\
    {\sf strong:} & 18, 22, 26, 30, 45, 106, 122, 126, 146, 154. \\
    {\sf moderate:} & - \\
    {\sf weak:} & 18, 22, 26, 30, 45, 106, 122, 126, 146, 154.
    \end{tabular}
\end{center}

\begin{center}
    \begin{tabular}{ r p{9cm}}
    {\bf type C:} & 54, 62, 73, 110. \\
    & \\
    {\sf strong:} & 54, 62, 110. \\
    {\sf moderate:} & 73. \\
    {\sf weak:} & -
    \end{tabular}
\end{center}

\subsubsection{Spectral classification (2013)}

Spectral classification in ``A Spectral Portrait of the Elementary Cellular Automata Rule Space'', establishes four ECA classes: 

\begin{center}
\{{\sf DE/SFC, DE/SFC SFC, EB, S}\}.
\end{center}

See details in [Ruivo \& de Oliveira, 2013].

\begin{center}
    \begin{tabular}{ r p{9cm}}
    {\bf DE/SFC:} & 0, 1, 2, 5, 6, 7, 8, 9, 10, 11, 12, 14, 19, 22, 23, 24, 25, 26, 27, 29, 32, 33, 34, 35, 36, 37, 38, 40, 41, 42, 43, 44, 46, 50, 54, 56, 57, 58, 62, 72, 73, 74, 76, 77, 94, 104, 108, 110, 128, 130, 132, 134, 136, 138, 140, 142, 152, 160, 162, 164, 168, 172, 178, 184, 200, 232. \\
    & \\
    {\sf strong:} & 2, 7, 9, 10, 11, 22, 24, 25, 26, 29, 34, 35, 41, 42, 46, 54, 56, 57, 58, 62, 94, 108, 110, 128, 130, 138, 152, 162, 178, 184. \\
    {\sf moderate:} & 1, 5, 6, 8, 14, 27, 32, 33, 37, 38, 40, 43, 44, 72, 73, 74, 77, 104, 132, 134, 136, 140, 142, 160, 164, 168, 172. \\
    {\sf weak:} & 0, 12, 19, 23, 36, 50, 76, 200, 232.
    \end{tabular}
\end{center}

\begin{center}
    \begin{tabular}{ r p{9cm}}
    {\bf DE/SFC SFC:} & 3, 4. \\
    & \\
    {\sf strong:} & - \\
    {\sf moderate:} & 3, 4. \\
    {\sf weak:} & -
    \end{tabular}
\end{center}

\begin{center}
    \begin{tabular}{ r p{9cm}}
    {\bf EB:} & 13, 18, 28, 78, 122, 126, 146, 156. \\
    & \\
    {\sf strong:} & 18, 122, 126, 146. \\
    {\sf moderate:} & 13, 28, 78, 156. \\
    {\sf weak:} & -
    \end{tabular}
\end{center}

\begin{center}
    \begin{tabular}{ r p{9cm}}
    {\bf S:} & 15, 30, 45, 51, 60, 90, 105, 106, 150, 154, 170, 204. \\
    & \\
    {\sf strong:} & 15, 30, 45, 106, 154, 170. \\
    {\sf moderate:} & - \\
    {\sf weak:} & 51, 60, 90, 105, 150, 204.
    \end{tabular}
\end{center}

\subsubsection{Bijective and surjective classification (2013)}

In this section, we have just bijective and surjective classification (personal communication, Harold V. McIntosh and Juan C. Seck Tuoh Mora): 

\begin{center}
\{{\sf bijective, surjective}\}.
\end{center}

See details in [McIntosh, 1990], [McIntosh, 2009].

\begin{center}
    \begin{tabular}{ r p{9cm}}
    {\bf bijective:} & 15, 51, 170, 204. \\
    & \\
    {\sf strong:} & 15, 170. \\
    {\sf moderate:} & - \\
    {\sf weak:} & 51, 204.
    \end{tabular}
\end{center}

\begin{center}
    \begin{tabular}{ r p{9cm}}
    {\bf surjective:} & 30, 45, 60, 90, 105, 106, 150, 154. \\
    & \\
    {\sf strong:} & 30, 45, 106, 154. \\
    {\sf moderate:} & - \\
    {\sf weak:} & 60, 90, 105, 150.
    \end{tabular}
\end{center}

\subsubsection{Creativity classification (2013)}

Creativity classification in ``On Creativity of Elementary Cellular Automata'', establishes four ECA classes: 

\begin{center}
\{{\sf creative, schizophrenic, autistic savants, severely autistic}\}.
\end{center}

See details in [Adamatzky \& Wuensche].

\begin{center}
    \begin{tabular}{ r p{9cm}}
    {\bf creative:} & 3, 5, 11, 13, 15, 35. \\
    & \\
    {\sf strong:} & 11, 15, 35. \\
    {\sf moderate:} & 3, 5, 13. \\
    {\sf weak:} & -
    \end{tabular}
\end{center}

\begin{center}
    \begin{tabular}{ r p{9cm}}
    {\bf schizophrenic:} & 9, 18, 22, 25, 26, 28, 30, 37, 41, 43, 45, 54, 57, 60, 62, 73, 77, 78, 90, 94, 105, 110, 122, 126, 146, 150, 154, 156. \\
    & \\
    {\sf strong:} & 9, 18, 22, 25, 26, 30, 41, 45, 54, 57, 62, 110, 122, 126, 146, 152, 154. \\
    {\sf moderate:} & 28, 37, 43, 73, 77, 78, 156. \\
    {\sf weak:} & 60, 90, 105.
    \end{tabular}
\end{center}

\begin{center}
    \begin{tabular}{ r p{8cm}}
    {\bf autistic savants:} & 1, 2, 4, 7, 8, 10, 12, 14, 19, 32, 34, 42, 50, 51, 76, 128, 136, 138, 140, 160, 162, 168, 170, 200, 204. \\
    & \\
    {\sf strong:} & 2, 7, 10, 34, 42, 128, 138, 162, 170. \\
    {\sf moderate:} & 1, 4, 8, 14, 32, 136, 140, 160, 168. \\
    {\sf weak:} & 12, 19, 50, 51, 76, 200, 204.
    \end{tabular}
\end{center}

\begin{center}
    \begin{tabular}{ r p{8cm}}
    {\bf severely autistic:} & 23, 24, 27, 29, 33, 36, 40, 44, 46, 56, 58, 72, 74, 104, 106, 108, 130, 132, 142, 152, 164, 172, 178, 184, 232. \\
    & \\
    {\sf strong:} & 24, 46, 56, 58, 106, 108, 130, 152, 178, 184. \\
    {\sf moderate:} & 27, 29, 33, 40, 44, 72, 74, 104, 132, 142, 164, 172. \\
    {\sf weak:} & 23, 36, 232.
    \end{tabular}
\end{center}

\subsection{Universal relations in ECAM classes}

After checking that memory has similar effect for every rule in the same  equivalence class (please see a full description of them in [Wuensche \& Lesser, 1992]), we will deal here for simplicity with the canonical representative rule of every one of the 88 equivalence classes, and not explicitly with the 256 rules. 

In what follows, we enumerate the most important relations.

\begin{itemize}
\item Transition of {\it uniform} to {\it uniform}.

\begin{equation}
uniform \xrightarrow{\phi_{CAm:\tau}} uniform
\label{un-un_UNI}
\end{equation}

this is transition from ECA $\varphi_{R32}$ to ECAM $\phi_{R32maj:3}$.
\item Transition of {\it uniform} to {\it periodic}.

\begin{equation}
uniform \xrightarrow{\phi_{CAm:\tau}} periodic
\label{un-per_UNI}
\end{equation}

this is transition rom ECA $\varphi_{R160}$ to ECAM $\phi_{R160par:5}$.
\item Transition of {\it uniform} to {\it chaos}.

\begin{equation}
uniform \xrightarrow{\phi_{CAm:\tau}} chaos
\label{un-ch_UNI}
\end{equation}

this is transition from ECA $\varphi_{R40}$ to ECAM $\phi_{R40par:2}$.
\item Transition of {\it uniform} to {\it complex}.

\begin{equation}
uniform \xrightarrow{\phi_{CAm:\tau}} complex
\label{un-com_UNI}
\end{equation}

this is transition from ECA $\varphi_{R40}$ to ECAM $\phi_{R40par:4}$.
\item Transition of {\it periodic} to {\it uniform}.

\begin{equation}
periodic \xrightarrow{\phi_{CAm:\tau}} uniform
\label{per-un_PER}
\end{equation}

this is transition from ECA $\varphi_{R130}$ to ECAM $\phi_{R130maj:4}$.
\item Transition of {\it periodic} to {\it periodic}.

\begin{equation}
periodic \xrightarrow{\phi_{CAm:\tau}} periodic
\label{per-per_PER}
\end{equation}

this is transition from ECA $\varphi_{R130}$ to ECAM $\phi_{R130maj:3}$.
\item Transition of {\it periodic} to {\it chaos}.

\begin{equation}
periodic \xrightarrow{\phi_{CAm:\tau}} chaos
\label{per-ch_PER}
\end{equation}

this is transition from ECA $\varphi_{R130}$ to ECAM $\phi_{R130par:3}$.
\item Transition of {\it periodic} to {\it complex}.

\begin{equation}
periodic \xrightarrow{\phi_{CAm:\tau}} complex
\label{per-com_PER}
\end{equation}

this is transition from ECA $\varphi_{R94}$ to ECAM $\phi_{R94par:2}$.
\item Transition of {\it chaos} to {\it uniform}.

\begin{equation}
chaos \xrightarrow{\phi_{CAm:\tau}} uniform
\label{ch-un_CH}
\end{equation}

this is transition from ECA $\varphi_{R18}$ to ECAM $\phi_{R18maj:10}$.
\item Transition of {\it chaos} to {\it periodic}.

\begin{equation}
chaos \xrightarrow{\phi_{CAm:\tau}} periodic
\label{ch-per_CH}
\end{equation}

this is transition from ECA $\varphi_{R30}$ to ECAM $\phi_{R30maj:4}$.
\item Transition of {\it chaos} to {\it chaos}.

\begin{equation}
chaos \xrightarrow{\phi_{CAm:\tau}} chaos
\label{ch-ch_CH}
\end{equation}

this is transition from ECA $\varphi_{R30}$ to ECAM $\phi_{R30par:2}$.
\item Transition of {\it chaos} to {\it complex}.

\begin{equation}
chaos \xrightarrow{\phi_{CAm:\tau}} complex
\label{ch-com_CH}
\end{equation}

this is transition from ECA $\varphi_{R126}$ to ECAM $\phi_{R126maj:4}$.
\item Transition of {\it complex} to {\it uniform}.

\begin{equation}
complex \xrightarrow{\phi_{CAm:\tau}} uniform
\label{com-un_COM}
\end{equation}

this is v from ECA $\varphi_{R54}$ to ECAM $\phi_{R54maj:6}$.
\item Transition of {\it complex} to {\it periodic}.

\begin{equation}
complex \xrightarrow{\phi_{CAm:\tau}} periodic
\label{com-per_COM}
\end{equation}

this is transition from ECA $\varphi_{R54}$ to ECAM $\phi_{R54par:2}$.
\item Transition of {\it complex} to {\it chaos}.

\begin{equation}
complex \xrightarrow{\phi_{CAm:\tau}} chaos
\label{com-ch_COM}
\end{equation}

this is transition from ECA $\varphi_{R110}$ to ECAM $\phi_{R110min:3}$.
\item Transition of {\it complex} to {\it complex}.

\begin{equation}
complex \xrightarrow{\phi_{CAm:\tau}} complex
\label{com-com_COM}
\end{equation}

this is transition from ECA $\varphi_{R54}$ to ECAM $\phi_{R54maj:8}$.

\end{itemize}

\begin{figure}
\centerline{\includegraphics[width=4.2in]{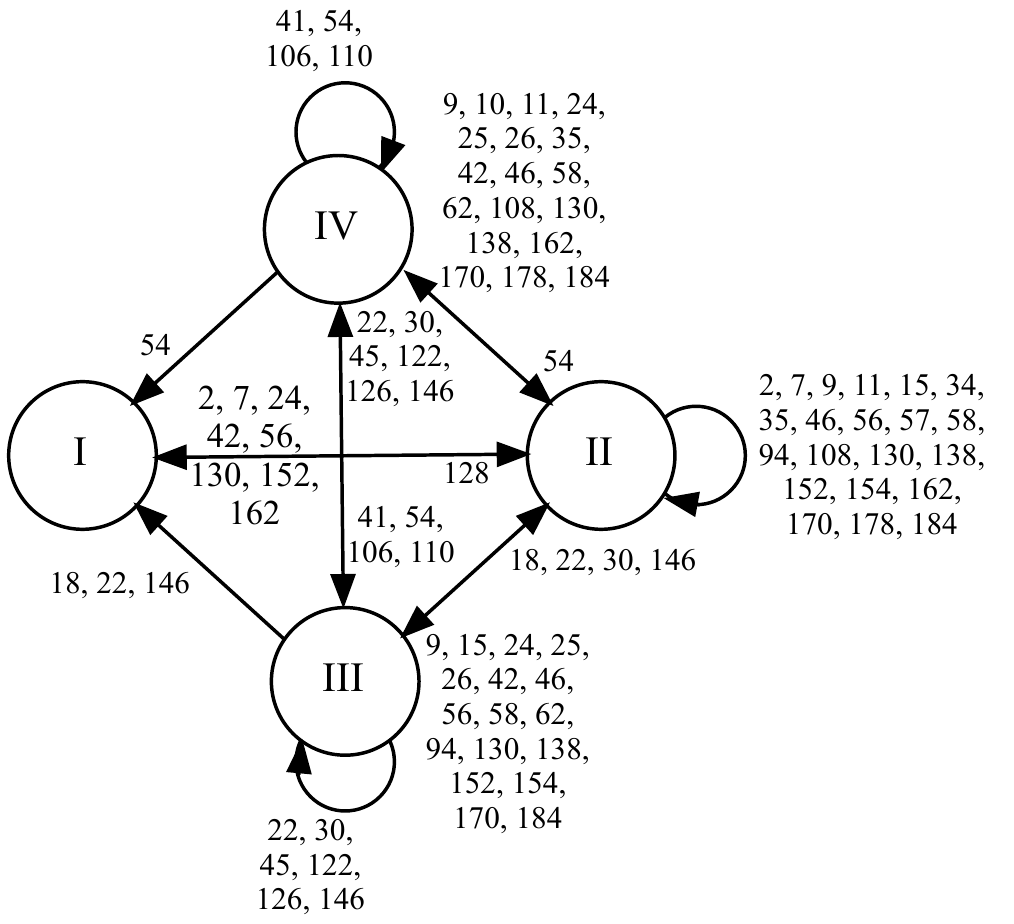}}
\caption{``Strong'' ECAM class is able to reach some other classes. Starting from a Wolfram's class (rule) and selecting some kind of memory inside {\it strong} ECAM class, one can reach some other class with such a rule.}
\label{classes-1}
\end{figure}

\begin{figure}
\centerline{\includegraphics[width=5.2in]{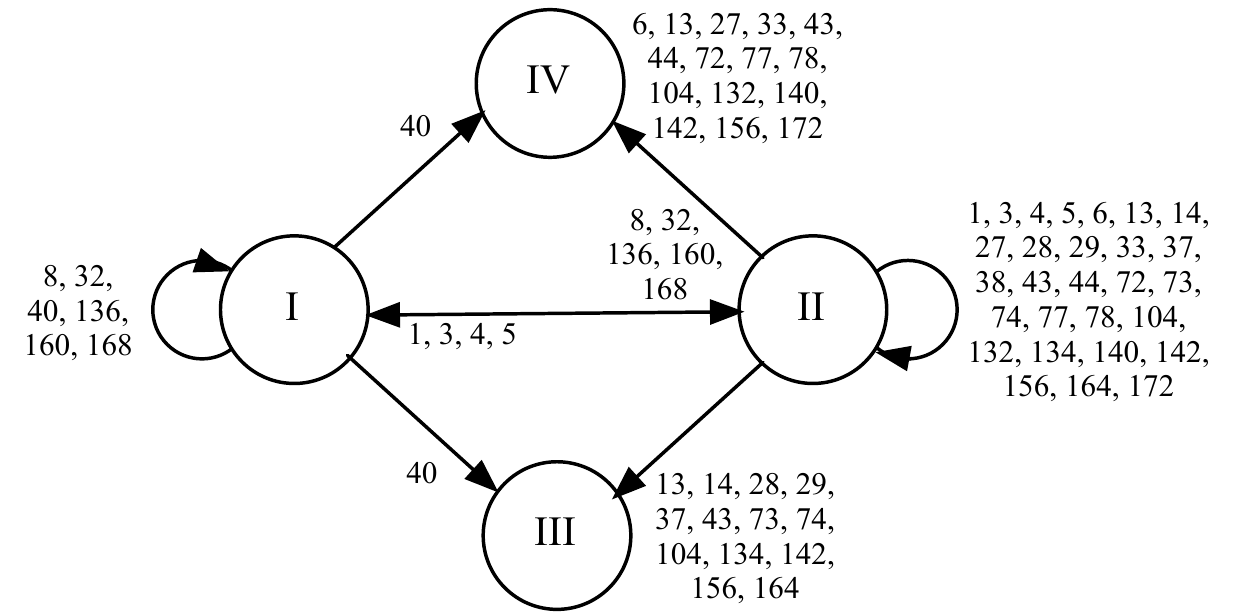}}
\caption{``Moderate'' ECAM class is able to reach some other classes. Starting from a Wolfram's class (rule) and selecting some kind of memory inside {\it moderate} ECAM class, one each some other class with such a rule.}
\label{classes-2}
\end{figure}

Therefore, from transitions \ref{un-un_UNI}--\ref{com-com_COM} we  we can reach a class from any other class with some kind of memory at least once.

ECAM preserves main characteristics of the original evolution rule and they can be found in both ECA and ECAM rules. As was detailed in ECA rule 126, a glider that is found in ECAM $\phi_{R126maj:4}$ already there is in the conventional ahistoric formulation rule (section~\ref{ECAM126section}). This way, dynamics in ECA move around of the memory effect in ECAM. As a consequence from this systematic analysis, we have that: \\

\begin{figure}[th]
\centerline{\includegraphics[width=4.7in]{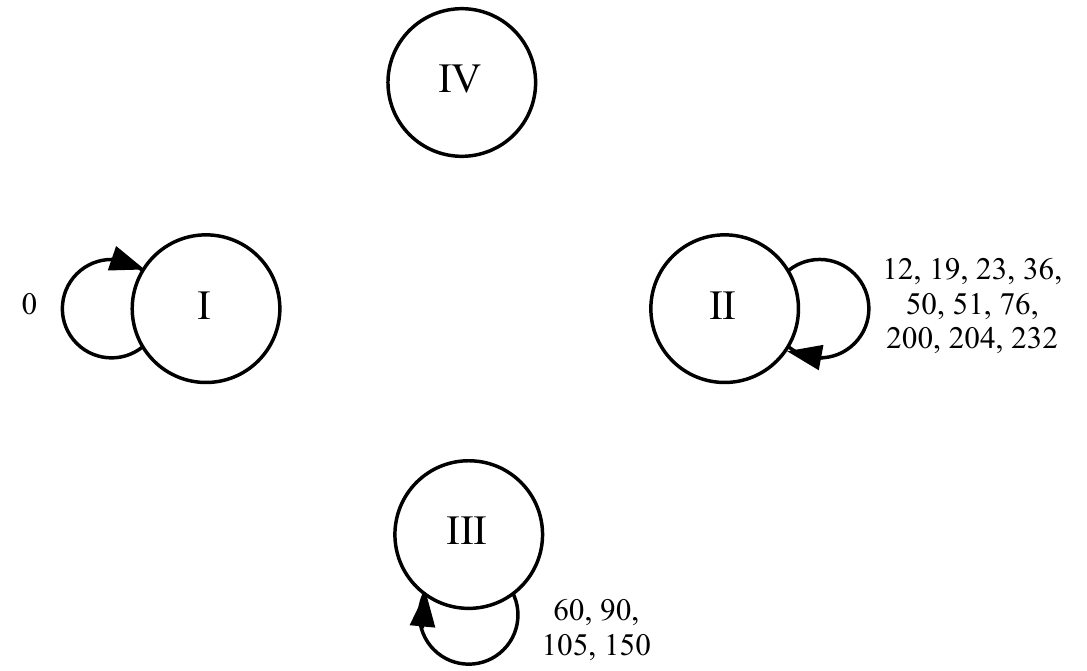}}
\caption{``Weak'' ECAM class is not able to reach other classes. Starting from a Wolfram's class (rule) and selecting some kind of memory inside {\it weak} ECAM class, one cannot reach some classes with such a rule.}
\label{classes-3}
\end{figure}

\noindent {\bf Proposition 4.} Dynamics in ECAM also cannot be induced from some previous ECA. \\

If you have selected a ECA class I, II, III, or IV; you could obtain a ECAM class I, II, III, or IV without some prefix which determines exactly the result. Diagrams displayed in Fig.~\ref{classes-1},~\ref{classes-2},~\ref{classes-3} show how move between classes. If you choice a specific ECA rule (that is in some Wolfram's class) hence with a kind of memory you can `move' to another class if it is the case. You can see these finite machines with respect to ECAM classification, Fig.~\ref{classes-1} for strong class, Fig.~\ref{classes-2} for moderate class, and Fig.~\ref{classes-3} for the weak class.

\begin{figure}
\centerline{\includegraphics[width=6.5in]{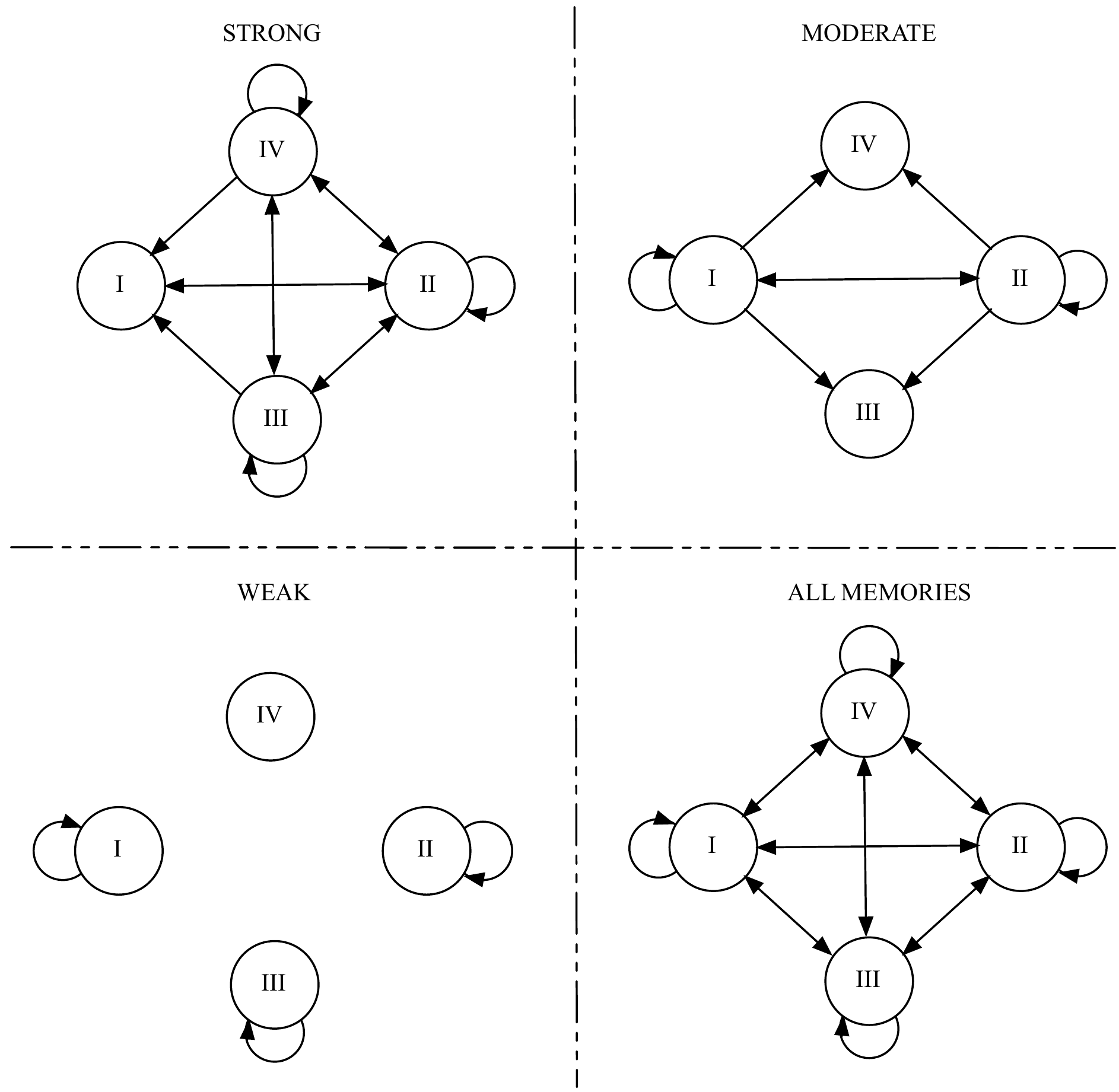}}
\caption{Every ECAM class has rules with behaviour class I, II, III, or IV. If you take one ECA rule with a kind of memory hence you can change to another class. ``All memories'' diagram show that it is possible to reach any class from some ECA enriched  with memory, thus some ECAM is able to reach any class.}
\label{classes-Mem}
\end{figure}

Finally, diagram in Fig.~\ref{classes-Mem} (all memories) shows a directed graph strongly connected due to the transitions \ref{un-un_UNI}--\ref{com-com_COM}. That means than you can reach any class from any class including them self (loops).

As outlined in [Culik II \& Yu, 1988] in the conventional ahistoric context, it is not possible to determine the behaviour of a ECAM from that of its conventional ahistoric ECA. This way, it is undecidable determines the behaviour of a CAM from any CA. Of course, memory can be selected on any dynamical systems useful mainly for discover hidden information, such as was studied in excitable CA [Alonso-Sanz \& Adamatzky, 2008].

\section{Unconventional computing with ECAM}
\label{unconvComputing}

In this section, we present a kind of complex CA derived since ECA rule 22 with memory. Again, we have selected the majority memory and we 
focus on $\tau=4$, generating a new ECAM rule, $\phi_{R22maj:4}$.

\begin{figure}[th]
\centerline{\includegraphics[width=6.4in]{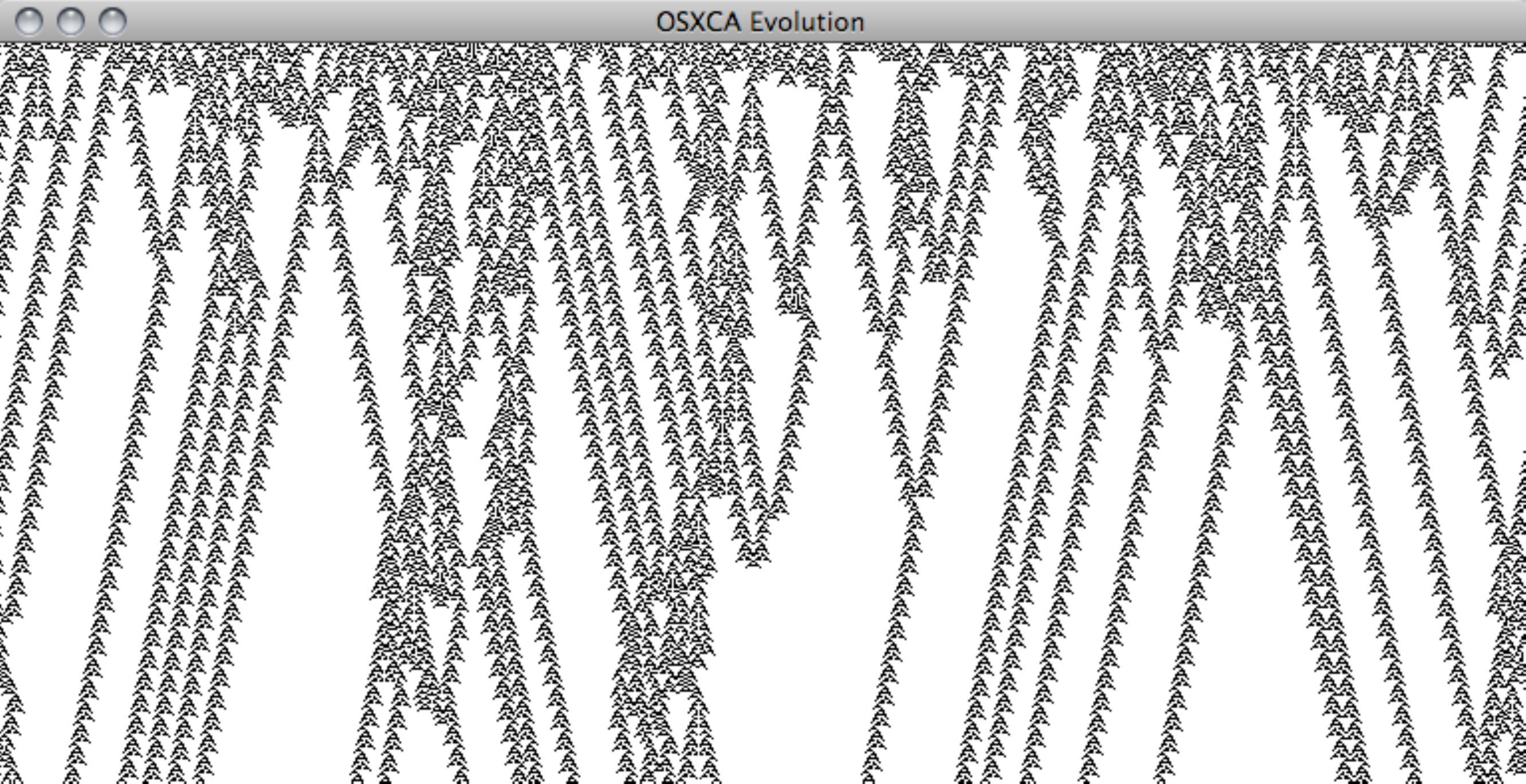}}
\caption{Typical random evolution of $\phi_{R22maj:4}$ from an initial configuration where 37\% of cells take state '1'. The automaton is a ring of 767 cells. Evolution is recorded for 372 generations.}
\label{evolR22maj4}
\end{figure}

Figure~\ref{evolR22maj4} displays a typical random evolution of ECAM $\phi_{R22maj:4}$. There we witness emergence of non-trivial travelling patterns and outcomes of their collisions. 

\begin{figure}
\begin{center}
\subfigure[]{\scalebox{0.54}{\includegraphics{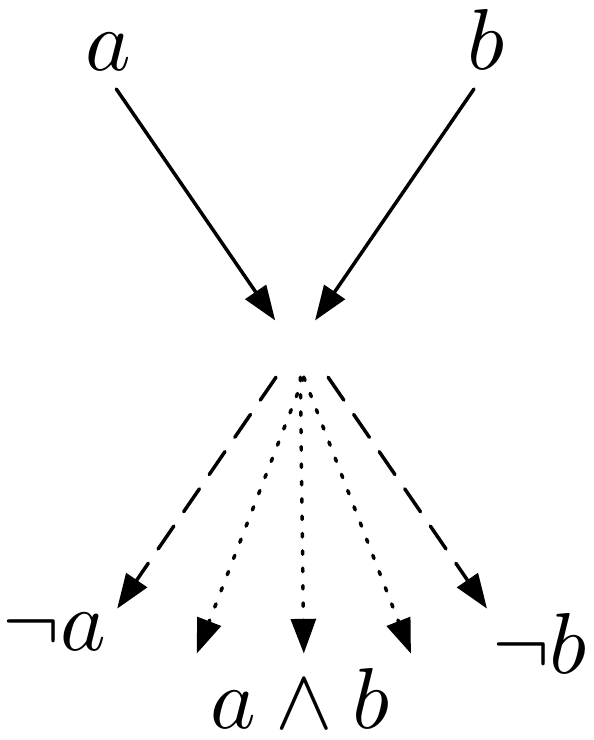}}} 
\subfigure[]{\scalebox{0.54}{\includegraphics{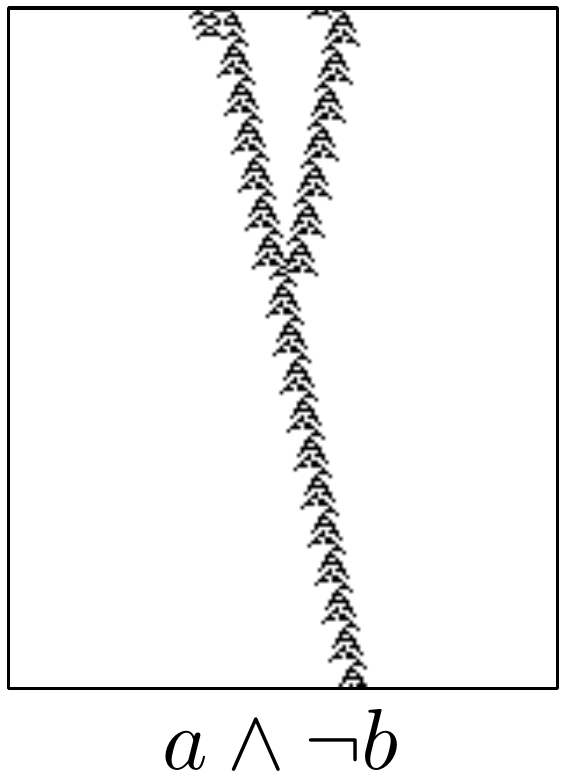}}} 
\subfigure[]{\scalebox{0.54}{\includegraphics{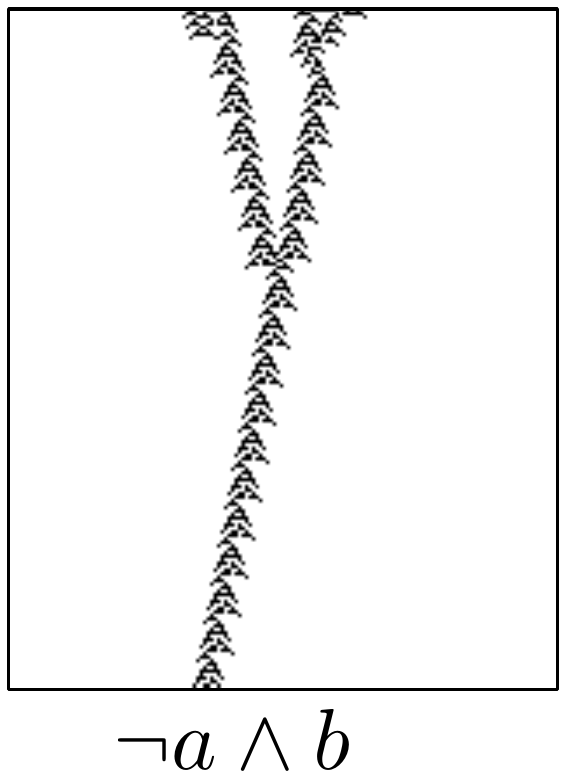}}} 
\subfigure[]{\scalebox{0.54}{\includegraphics{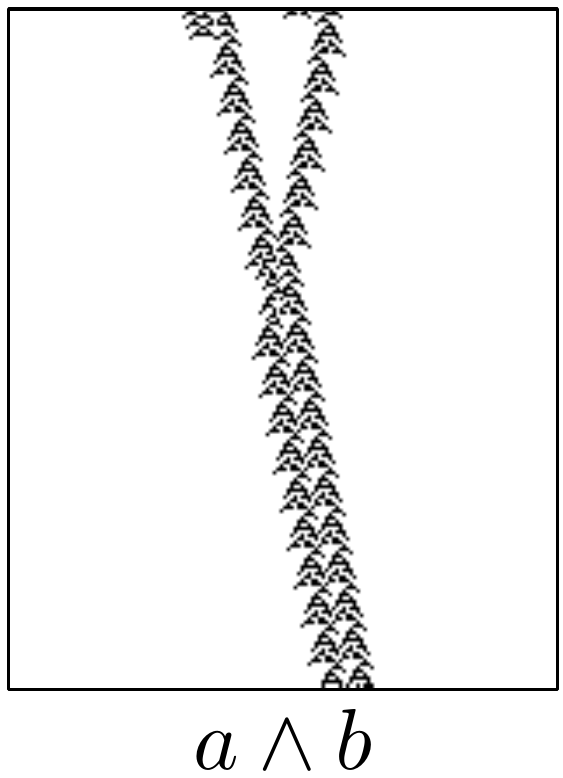}}} 
\subfigure[]{\scalebox{0.53}{\includegraphics{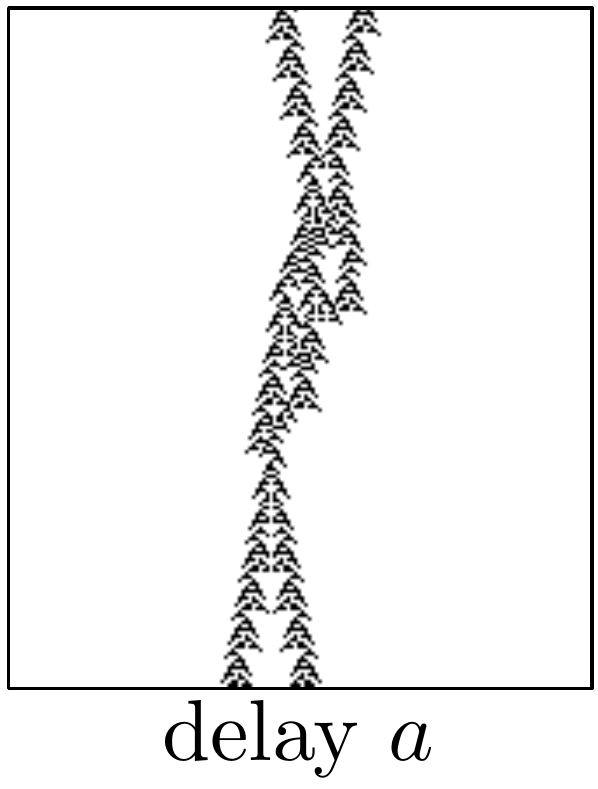}}}
\end{center}
\caption{We implement basic logic functions as {\sc not} and {\sc and} gates via collisions of gliders and  a {\sc delay} element. 
Single or pair of particles represent bits 0's or 1's respectively.}
\label{basicGates}
\end{figure}

\begin{figure}
\begin{center}
\subfigure[]{\scalebox{1.2}{\includegraphics{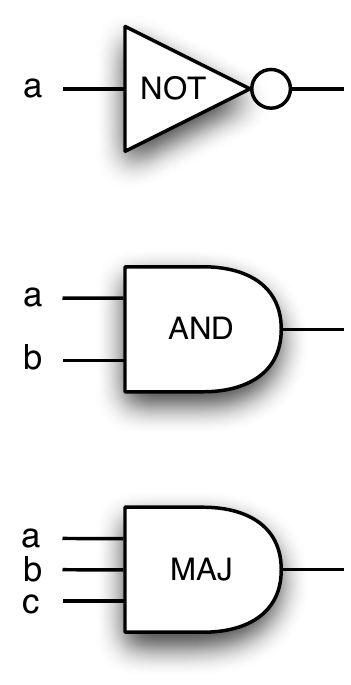}}} \hspace{2cm}
\subfigure[]{\scalebox{1.2}{\includegraphics{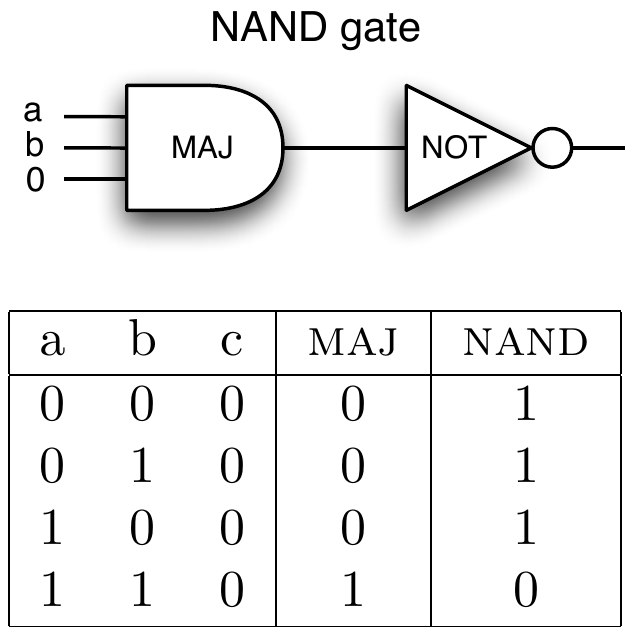}}}
\end{center}
\caption{A {\sc nand} gate based in {\sc majority} and {\sc not} gates.}
\label{majorityGate-0}
\end{figure}

The main and most interesting characteristic is that this complex ECA with memory has only two gliders, maybe we can tell only one with its respective reflection. With these gliders  $\cal G$$_{\phi_{R22maj:4}} = \{g_{L},g_{R}\}$ we can design computing circuits (this is a partial result 
of our research detailed in [Mart{\'i}nez et al. a]).

We should start with basic logic gates derived from simple binary collisions. A {\it logic gate} performs a logic operation on one or more logic inputs yielding just one logic output. Normally a logic gate is a Boolean function, such that for some positive integer $n$ we have that $f: \Sigma^n \rightarrow \Sigma$ for $\Sigma=\{0,1\}$, and therefore it can be represented by a truth table that describes the behaviour of a logic gate [Minsky, 1967].

Figure~\ref{basicGates} displays implementation of {\sc not} and {\sc and} gates with gliders  $\cal G$$_{\phi_{R22maj:4}}$ and a symmetric {\sc delay} element.

Generally a problem to implement computations in injective CA is related to the synchronisation of collisions between gliders and accurate positioning of gliders in initial configuration. 

\begin{figure}
\begin{center}
\subfigure[]{\scalebox{1.2}{\includegraphics{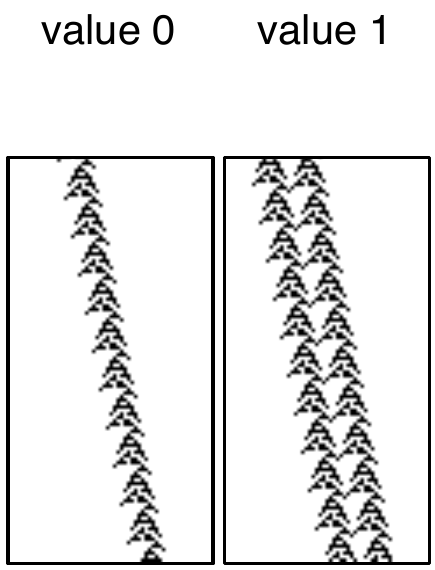}}} \hspace{2.4cm}
\subfigure[]{\scalebox{1.2}{\includegraphics{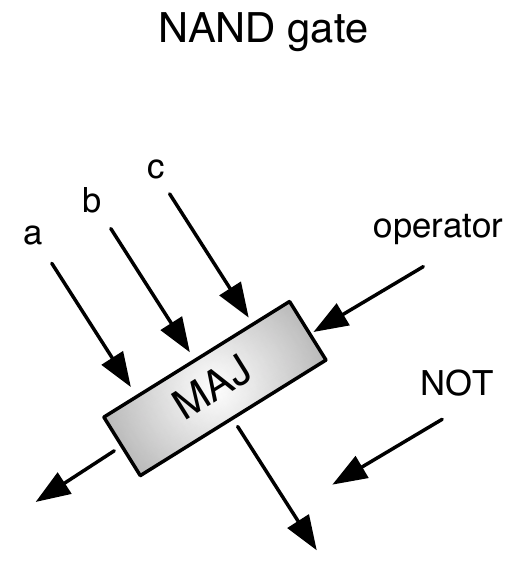}}} \hspace{0.5cm}
\subfigure[]{\scalebox{1}{\includegraphics{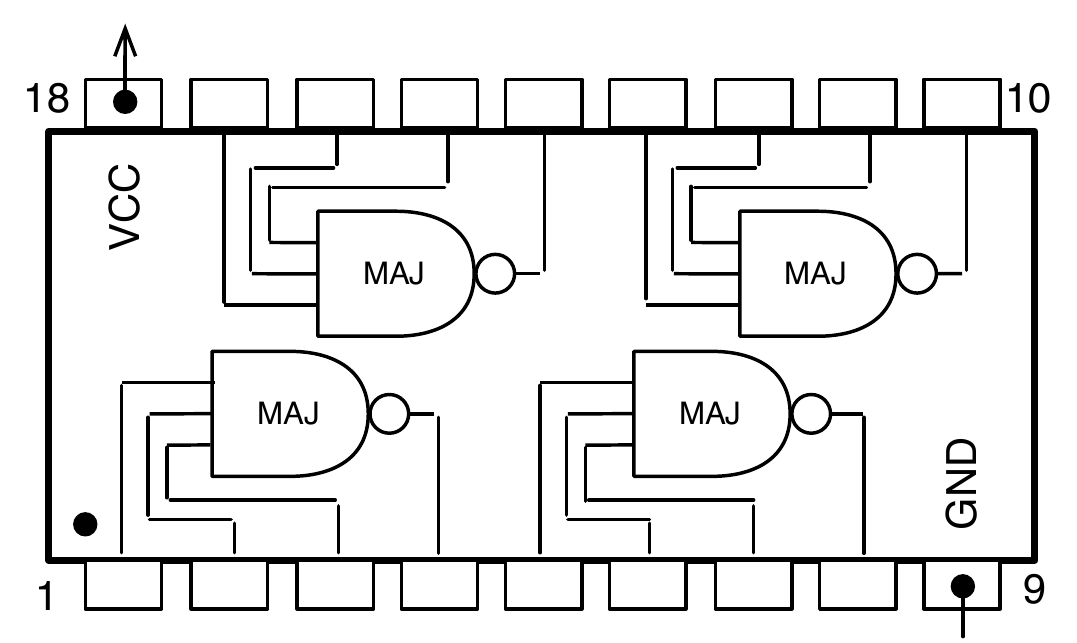}}}
\end{center}
\caption{(a) binary values by gliders codification, (b) scheme of a {\sc nand} gate from {\sc majority} and {\sc not} gates with glider reaction, and (c)circuit based on four {\sc nand} gates like a modified 7400 chip but now with 18 pins (for the {\sc majority} gate).}
\label{majorityGate-1}
\end{figure}

A {\sc majority} gate and {\sc and} gate as is represented in the Fig.~\ref{majorityGate-0}. A {\sc not} gate is aggregated to get a {\sc nand} 
gate (Fig.~\ref{majorityGate-0}b).

\begin{figure}
\begin{center}
\subfigure[]{\scalebox{1.1}{\includegraphics{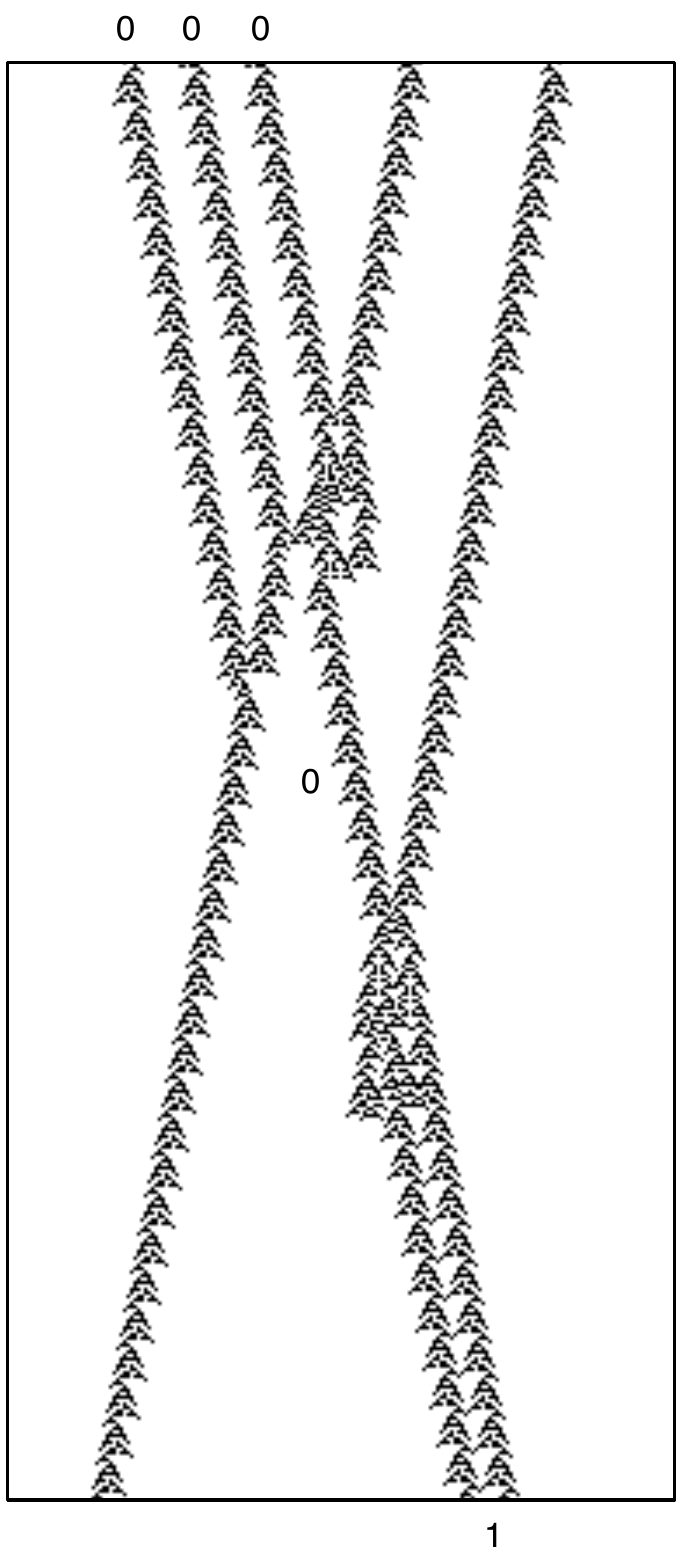}}}  \hspace{0.2cm}
\subfigure[]{\scalebox{1.1}{\includegraphics{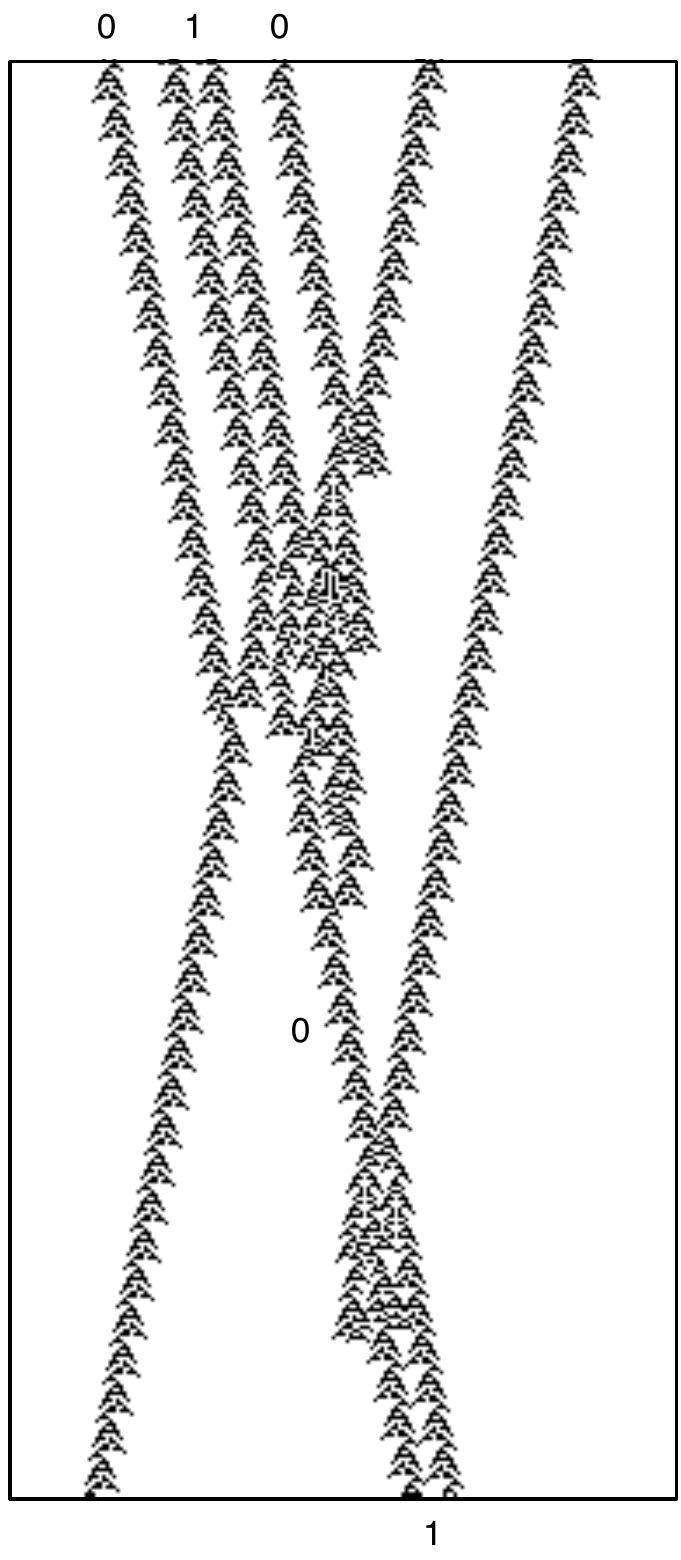}}}
\end{center}
\caption{{\sc nand} gate implemented from {\sc majority} and {\sc not} gates in $\phi_{R22maj:4}$. Inputs (a) 000 and (b) 010.}
\label{majorityGate-2}
\end{figure}

\begin{figure}
\begin{center}
\subfigure[]{\scalebox{1.1}{\includegraphics{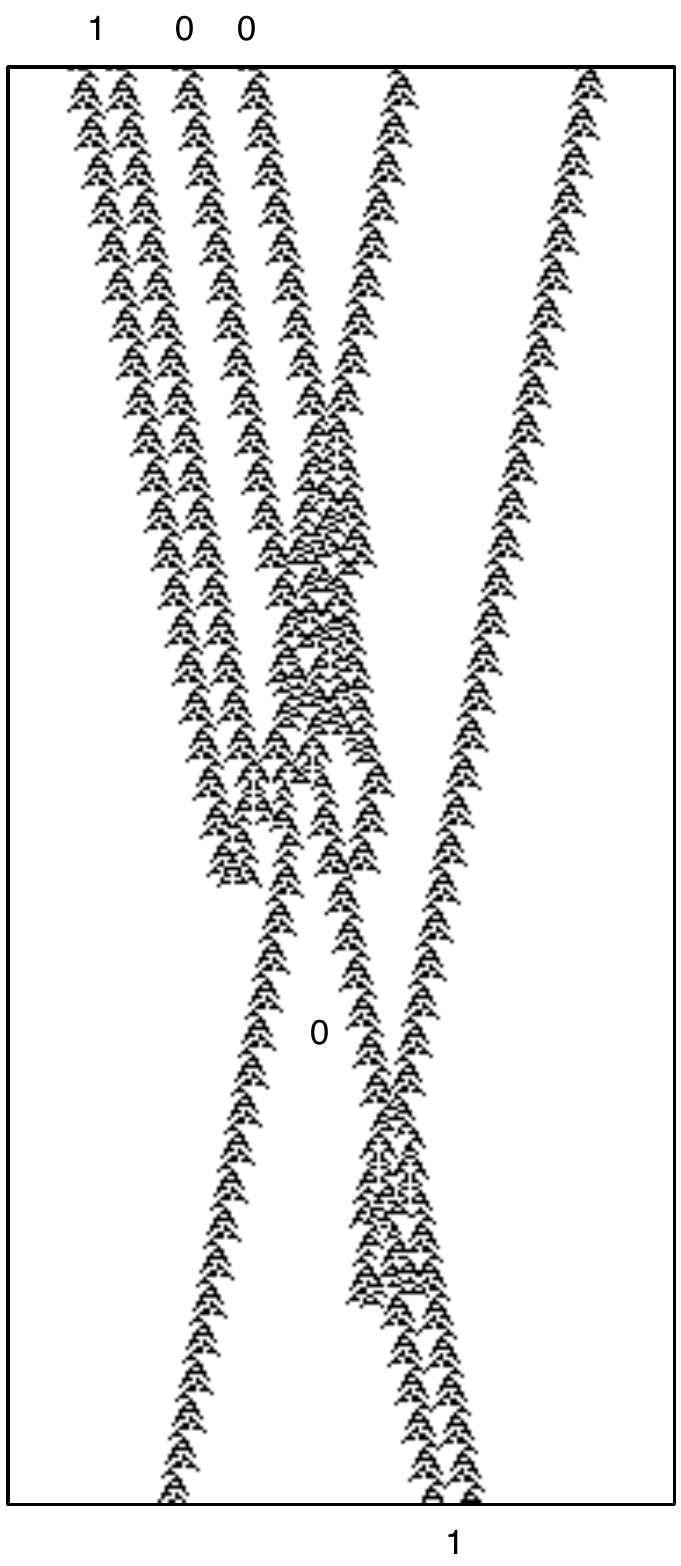}}} \hspace{0.2cm}
\subfigure[]{\scalebox{1.1}{\includegraphics{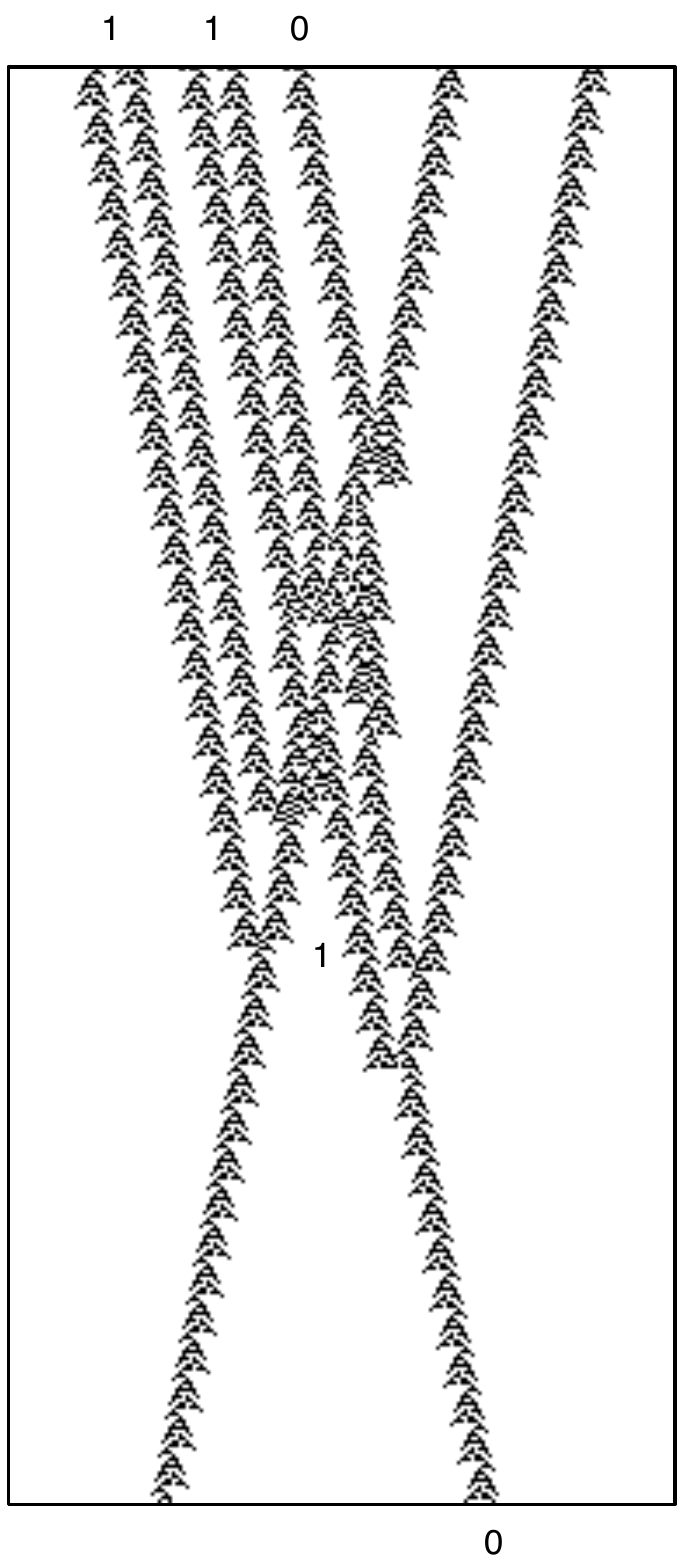}}}
\end{center}
\caption{{\sc nand} gate implemented from {\sc majority} and {\sc not} gates in $\phi_{R22maj:4}$. Inputs (a) 100 and (b) 110.}
\label{majorityGate-2A}
\end{figure}

To implement a majority gate we must represent binary values across of gliders (Fig.~\ref{majorityGate-1}(a)). Later, let use this construction to implement a {\sc nand} gate by gliders collisions as Fig.~\ref{majorityGate-1}(b) display. This way, we use a $g_R$ glider that works as an operator processing three input values at the same time. A {\sc not} gate is represented by a second $g_R$ glider inverting the final result. Also, we can utilise this scheme to design a modified chip related to 7400 chip\footnote{National semiconductor web site. Device 5400/DM5400/DM7400 Quad 2-Input {\sc NAND} Gates \url{http://www.national.com/ds/54/5400.pdf}} but with four {\sc majority} and {\sc not} gates instead of four {\sc nand} gates, working with three independent inputs per gate on 18 pins as in the Fig.~\ref{majorityGate-1}(c).

Figures~\ref{majorityGate-2} and \ref{majorityGate-2A} show the implementation of {\sc nand} gate with $\phi_{R22maj:4}$. As illustrated in diagram (Fig.~\ref{majorityGate-1}c) a glider works as an operator of the {\sc majority} gate and this operator is reused in the next {\sc majority} gate. We present all stages where the {\sc nand} gate works, thus proving functionality of the design.

\section{Final remarks}
\label{FinalRemarks}
We demonstrated that a memory is a `universal' switch which allows us to change dynamics of a complex spatially extended systems and to guide the system in a `labyrinth' of complexity classes. Memory allows us to make complex systems simple and to simple ones complex.

The memory implementation mechanism studied here constitute a simple extension (of straightforward computer codification) of the basic CA paradigm allowing for an easy systematic study of the effect of memory in cellular automata (and other discrete dynamical systems). This may  inspire some useful ideas in using cellular automata  as a tool for modeling phenomena with memory. This task has been traditionally attacked by means of differential, or finite-difference, equations, with some (or all) continuous component. In contrast, full discrete models are ideally suited to digital computers. Thus, it seems plausible that further study on cellular automata  with memory should prove profitable, and may be possible to paraphrase T. Toffoli [Toffoli, 1984] in presenting cellular automata with memory {\it as an alternative to (rather than an approximation of)} integro-differential equations in modeling phenomena with memory. Besides their potential applications, cellular automata with memory have an aesthetic and mathematical interest on their own, so that we believe that the subject is worth to studying.

Last but not least, other memories are possible. In this study we have implemented an explicit dependence in the dynamics of the past states in the manner: first summary then rule. But the order summary-rule may be inverted, i.e., the rule is first applied and a summary is then presented as new state (for details see [Alonso-Sanz, 2013]. This alternative memory implemention enriches the potential use of memory in discrete systems as a tool for modeling, and, again, in our opinion deserves attention on its own.


\section{Appendix A}

Appendix and full paper is available in \url{http://eprints.uwe.ac.uk/21980/}.

\end{document}